\documentclass[fleqn,usenatbib]{mnras}

\usepackage{newtxtext,newtxmath}
\usepackage[T1]{fontenc}

\DeclareRobustCommand{\VAN}[3]{#2}
\let\VANthebibliography\thebibliography
\def\thebibliography{\DeclareRobustCommand{\VAN}[3]{##3}\VANthebibliography}

\usepackage{graphicx}	% Including figure files
\usepackage{amsmath,bm}	% Advanced maths commands
\usepackage{pdflscape}

\usepackage{color, soul}

\newcommand\poet{\texttt{POET}}
\newcommand\feh{\left[\frac{Fe}{H}\right]}

\graphicspath{{figures_and_tables/}}

\title[Tidal Dissipation in Open Cluster Binaries I]{Comprehensive Bayesian
Modeling of Tidal Circularization in Open Cluster Binaries part I: M 35, NGC
6819, NGC 188}

\author[K. Penev et al.]{
    Kaloyan M. Penev,$^{1}$\thanks{E-mail: kaloyan.penev@utdallas.edu (KMP)}
    Joshua A. Schussler$^{1}$
\\
$^{1}$Department of Physics, University of Texas at Dallas, 800 W. Campbell Rd.,
Richardson, TX 75080, USA\\
}

\date{Accepted XXX. Received YYY; in original form ZZZ}

\pubyear{2015}

\begin{document}
\label{firstpage}
\pagerange{\pageref{firstpage}--\pageref{lastpage}}
\maketitle

\begin{abstract}
    Tidal friction has long been recognized to circularize the orbits of binary
    stars over time. In this study, we use the observed distribution of orbital
    eccentricities in populations of binary stars to probe tidal dissipation. In
    contrast to previous studies, we incorporate a host of physical effects
    often neglected in other analyses, provide a much more general description
    of tides, model individual systems in detail (in lieu of population
    statistics), and account for all observational uncertainties. The goal is to
    provide a reliable measurement of the properties of tidal dissipation that
    is fully supported by the data, properly accounts for different dissipation
    affecting each tidal wave on each object separately, and evolves with the
    internal structure of the stars. We extract high precision measurements of
    tidal dissipation in short period binaries of Sun-like stars in three open
    clusters. We find that the tidal quality factor on the main sequence falls
    in the range $5.7 < \log_{10}Q_\star' < 6$ for tidal periods between 3 and
    7.5 days. In contrast, the observed circularization in the 150\,Myr old
    M\,35 cluster requires that pre-main sequence stars are much more
    dissipative: $Q_\star' < 4\times10^4$.  We test for frequency dependence of
    the tidal dissipation, finding that for tidal periods between 3 and 7.5
    days, if a dependence exists, it is sub-linear for main-sequence stars.
    Furthermore, by using a more complete physical model for the evolution, and
    by accounting for the particular properties of each system, we alleviate
    previously observed tensions in the circularization in the open clusters
    analyzed.
\end{abstract}


%%%%%%%%%%%%%%%%%%%%%%%%%%%%%%%%%%%%%%%%%%%%%%%%%%

%%%%%%%%%%%%%%%%% BODY OF PAPER %%%%%%%%%%%%%%%%%%

\section{Introduction}

The friction associated with tidally-induced fluid flow leads to long-term
energy dissipation, with profound consequences throughout all of astrophysics.
At present however, many gaps remain in our understanding of the complex fluid
mechanical processes responsible for tidal dissipation. Furthermore, analyses of
observations have not been able to sufficiently narrow down the realm of
possible theories.

Even if we focus on the narrow subset of sun-like stars, we find numerous
astrophysical problems for which progress is blocked by our limited
understanding of tidal dissipation. For example, the formation of all short
period giant planets and ultra-short period planets of any kind will be strongly
affected by tides \citep{Rasio_Ford_96, Fabrycky_Tremaine_07,
Beauge_Nesvorny_12, Vick_Lai_Anderson_19}. In particular, two of the three
formation models for short period gas giant planets (disk migration and in-situ
formation) predict the planet will be close to its star when the star is still
young and large, and thus experience enhanced tides. If the star spins slower
than the orbit, the stellar tides will quickly shrink the orbit, threatening the
survival of the planet. Alternatively, \citet{Lin_et_al_96} suggest tides on a
super-synchronously spinning star may be what stops the inward migration of
planets, thus potentially playing a key role in the formation mechanism of hot
Jupiters. The third class of formation models, high-eccentricity migration,
explicitly invokes tides as the mechanism that drives migration, but is
dominated by the tides on the planet.  There is also observational evidence for
stellar tides re-aligning exoplanet orbits \citep{Winn_et_al_10,
Valsecchi_Rasio_14}, driving changes in the orbital period
\citep{Maciejewski_et_al_16, Patra_et_al_17}, and destroying planets through
orbital decay \citep{Debes_Jackson_10, Hamer_Schlaufman_19}.

The orbits of short period binary stars also show clear signs of being shaped by
tides. Our current understanding suggests these systems start in a wide range of
states, and over time tides partially or completely circularize their orbits and
synchronize the spins of their components to the orbit.

More generally, reliable measurements of tidal dissipation can provide a unique
window into the physical processes going on inside stars, complimentary to what
can be probed by theoretical models and simulations.

The importance of tidal dissipation is evident by the huge effort invested in
attempting to understand the processes involved. Unfortunately, this has
resulted in many theoretical models, giving contradictory predictions for the
rate of energy loss for tidally perturbed stars. These models can be broadly
grouped in two classes: equilibrium tide models \citep[][and others]{Zahn_66,
Zahn_89, Zahn_Bouchet_89, Goldreich_Keeley_77, Goldreich_Nicholson_77,
Ogilvie_Lin_04, Penev_Barranco_Sasselov_09, Penev_et_al_09, Barker_20}, in which
the dominant dissipation mechanism is a turbulent cascade in the convective
zone; and dynamical tide models \citep[][and many others]{Ogilvie_Lin_04,
    Ogilvie_13, Barker_20, Goodman_Dickson_98, Barker_Ogilvie_10,
Essick_Weinberg_16, Savonije_Witte_02}, in which the tidal perturbations
interact with various waves supported in the perturbed body, leading to
potentially much more efficient dissipation relative to the equilibrium tide.

Under the equilibrium tide assumption, turbulent eddies within the convective
zone siphon kinetic energy from the large scale of the tides to smaller and
smaller scales, until eventually the molecular viscosity turns that energy into
heat. The usual treatment is to fold all the complications of this eddy cascade
into an effective eddy viscosity, many orders of magnitude larger than the usual
molecular viscosity.  For tidal periods much longer than the longest eddy
turnover times, dimensional analysis predicts the order of magnitude of the
effective viscosity. At shorter tidal periods, however, dissipation is less
efficient and a number of different arguments predict different period scalings
based on different physical reasoning.

For example, \citet{Zahn_66, Zahn_89} and \citet{Zahn_Bouchet_89} predict linear
scaling by arguing that the largest eddies will dominate the dissipation and
that the rate they remove energy from a large scale high frequency flow will be
proportional to the fraction of the eddy turnover that is completed in a single
tidal period. On the other hand, \citet{Goldreich_Keeley_77} and
\citet{Goldreich_Nicholson_77} argue that if the eddy turnover time is longer
than the tidal period, its contribution to the dissipation is exponentially
suppressed. As a result, the largest eddies with turnover times shorter than the
tidal period dominate the dissipation, which combined with the assumption of a
Kolmogorov inertial range predict that the effective eddy viscosity falls off as
the square of the tidal period. The difference between these prescriptions can
be as much as two orders of magnitude at periods of order days, appropriate for
short-period detached binary star and exoplanet systems, and about five orders
of magnitude at the periods typical of solar p--modes ($\sim5$\,min).

More recently, equilibrium tidal dissipation has been addressed with numerical
simulations.  \citet{Penev_Barranco_Sasselov_09, Penev_et_al_09} find effective
viscosity that scales linearly with the tidal period at long periods and
quadratically at short ones (without being able to discern the physical
mechanism).  \citet{Ogilvie_Lesur_12} argued analytically and confirmed with
numerical simulations that large eddies (with long turnover times) provide an
effective viscosity that decreases with the square of the period at short tidal
periods, matching the quadratic scaling predicted by Goldreich and collaborators
but for a different physical reason. The latter paper even finds (both
analytically and numerically) that the effective viscosity can in principle be
negative (i.e.  anti-dissipation), though perhaps not in physically relevant
situations.  \citet{Duguid_Barker_Jones_20} carry out an extensive parameter
survey of simulated convective dissipation, confirming that the largest eddies
dominate the dissipation even at high frequencies, confirming the quadratic
scaling and interpretation of \citet{Ogilvie_Lesur_12}, and further argue for
the existence of an intermediate regime at moderate tidal frequencies, where the
dissipation scales as the square root of the period.  \citet{Vidal_Barker_20}
also confirm the quadratic scaling for high frequency tides in idealized global
simulations, as well as reproduce negative effective viscosities in certain
circumstances. These authors also find an intermediate regime where the
effective viscosity scales linearly with the tidal period for tidal frequencies
below but comparable to the dominant eddy turnover time.  Finally,
\citet{Terquem_21, Terquem_Martin_21} suggest that previous analyses ignore the
potentially dominant contribution to tidal dissipation proportional to
correlations of the tidal flow velocities. Relying on strong assumptions, they
demonstrate that the implied dissipation is roughly consistent with the
circularization of Hot Jupiter systems and binary stars, as well as with the
tidal migration rates of Jupiter and Saturn moon systems.

While an exhaustive review of the various dynamical tide models is not
practical, we can provide a general feeling for the range of possibilities. See
\citet{Ogilvie_14} and \citet{Mathis_19} for in-depth, recent reviews of tidal
dissipation theory.

One scenario involves the excitation of g--modes at the
core--envelope boundary, which for surface radiative zone stars travel outwards
and are radiatively dissipated \citep{Zahn_75}. For stars with convective
envelopes, g-modes travel inwards, get focused near the center and may break if
the amplitude grows large enough \citep{Goodman_Dickson_98}. For intermediate
amplitudes, not enough for wave breaking but sufficient to exhibit
non-linearity, \citet{Essick_Weinberg_16} argue that resonant mode--mode
coupling can lead to strongly enhanced dissipation.  Lastly,
\citet{Ma_Fuller_21} argue that if the tidal excitation is weak enough for
g-modes to remain linear (e.g. if companion is a planet, or waves are excluded
from the center of the star by a convective core), tides may resonantly lock to
a gravity mode, also leading to enhanced dissipation.

Another prominent class of models suggest that, in the correct frequency range,
tides may excite rotationally supported inertial waves, leading to strongly
enhanced dissipation if the tidal frequency is less than twice the spin
frequency of the star \citep[cf.][]{Ogilvie_Lin_07, Ogilvie_09,
Rieutord_Valdettaro_10, Barker_20} with highly irregular frequency dependence.
The usual assumption is that perhaps as yet unmodeled physical processes will
act to smooth out the tidal dissipation. As a result, the enhanced inertial mode
dissipation is usually modeled as an appropriate frequency average of the
theoretical predictions. Recently, \citet{Barker_22} showed that properly
accounting for stellar evolution in combination with the enhanced dissipation
due to inertial waves is capable of approximately reproducing at least the
overall circularization period ($P_{circ}$) for a number of open clusters. That
study also demonstrates the inadequacy of using a single statistic, like
$P_{circ}$, to characterize a population of objects. Their figures show large
variations of that quantity with the masses and spins of the binary stars, with
the mass dependence being especially pronounced for ages of several Gyrs. As the
definition of $P_{circ}$ for each cluster necessarily includes a range of masses
and the spin periods are presumably pseudo-syncronized with the orbital period,
it is expected, and indeed observed, that in each cluster there should be a
broad range of periods where some systems are circularized and others are not.
Thus, proper treatment of tidal circularization requires modeling the combined
tidal and spin evolution of individual binaries separately as we do here.

Attempts to constrain tidal dissipation rates from observations have provided
some guidance, but so far fail to uniquely identify which of the models operates
under which conditions or if perhaps new models are needed. The tidal
dissipation efficiency is often parameterized by the dimensionless number $Q'
\equiv Q/k_2$, where $1/Q$ is the fraction of the energy in a tidal wave lost
for each radian the wave travels, and $k_2$ is the tidal Love number (the ratio
    of the quadrupole component of the gravitational potential of a tidally
distorted object to the quadrupole moment of the external potential driving the
deformation). At present, empirical constraints often disagree by orders of
magnitude, even when based on the same class of objects
\citep[c.f.][]{Husnoo_et_al_12, Jackson_Greenberg_Barnes_08a}.

These inconsistencies are the result of simplifications made in order to make
the problem more tractable. First, $Q_\star'$ is expected to depend on tidal
frequency and amplitude, as well as the spin and internal structure of the
dissipating body. The only studies that account for all of these complexities
\citep[e.g.][]{Bolmont_Mathis_16, Benbakoura_et_al_19} assume a particular
theoretical model for the dissipation and only treat circular orbits with no
spin-orbit misalignment. Other studies ignore many of these expected
dependencies, accounting for at most one possible dependency, or assume
$Q_\star'=const$. Second, the parameters that affect tidal dissipation (e.g.
object radius, moment of inertia, and structure) often change on a similar
timescale as the effects of tides are felt. Hence, the combined evolution of all
these parameters must be included in the tidal calculations. Many prior efforts
ignore stellar evolution in their modeling. Finally, except for circular orbits
with zero obliquity, there are a number of different tidal deformations acting
simultaneously, each with its own frequency and amplitude, and hence dissipation
rate \citep[see for example][]{Alexander_73, Zahn_89, Lai_12}. This is also
something often ignored in the literature.

Our research group has embarked on a systematic effort
\citep[including][]{Penev_et_al_18, Anderson_et_al_21, Patel_Penev_21} to relax
as many of the above simplifications as possible and provide empirical
constraints on $Q_\star'$, and its dependence on the properties of the object
subjected to tides, as well as the particular tidal wave.

In this study, we analyze the eccentricities of short period single line
spectroscopic binaries (SB1) in the 150 Myr old M 35
\citep{Meibom_Mathieu_Stassun_09}, 2.4 Gyr old NGC 6819 \citep{Basu_et_al_11},
and the 7 Gyr old NGC 188 \citep{Sarajedini_et_al_99} open clusters. We model
each individual system's evolution in detail, accounting for stellar evolution,
loss of angular momentum to stellar winds, non-solid body rotation of the
component stars, and allowing for frequency dependent tidal dissipation that
evolves as the orbit and the structure of the stars in the binary evolve.
Notably, we split the tidal potential in spatial spherical harmonics and
temporal Fourier series, evaluating our tidal dissipation model separately for
each term in this expansion, resulting in different dissipation efficiency.
At this stage we assume that the tidal coupling is exclusively within the
convective zones of the stars, in line with some theoretical models, but not
others. Exploring the hypothesis that tides couple to radiative regions is
subject to future articles.

The rest of this article is organized as follows: in Section \ref{sec:methods}
we describe the details of the method for using the observed eccentricities of
SB1 to measure tidal dissipation, our model for computing the tidal evolution,
and the Bayesian analysis applied that allows us to fully incorporate
observational and model uncertainties in the results; in Section
\ref{sec:input_data} we describe the collection of observations we use in the
analysis; in Section \ref{sec:results} we report the empirical constraints our
analysis yields; and in Section \ref{sec:conclusions} we discuss the
implications of our results, caveats in our analysis, and prospects for the
future.

\section{Methods}
\label{sec:methods}

Tidal dissipation acts to circularize orbits over time. The observed
eccentricities of binary star systems show clear signs of being shaped by tides,
with all of the shortest period systems in circular orbits, a gradual increase
in the eccentricity envelope as the period increases, and a broad
period-independent eccentricity distribution at long periods
\citep{Milliman_et_al_14, Windemuth_Agol_Kiefer_19} (see Fig.
    \ref{fig:m35_period_eccentricity}, \ref{fig:ngc6819_period_eccentricity},
and \ref{fig:ngc188_period_eccentricity}). As tides rapidly lose influence with
increasing orbital period, this observed trend is precisely what is predicted if
all systems started with a broad range of initial eccentricities and tides were
the dominant mechanism shaping their orbits.  Furthermore, ongoing
circularization is detectable over the lifetime of stars
\citep{Meibom_Mathieu_05}, hence the late-time eccentricity envelope is not set
during binary formation or early evolution, leaving tides as the most likely
explanation. Assuming this picture is correct, an upper limit for the tidal
dissipation (lower limit on $Q_\star'$) in an individual system can be derived
by requiring that there exist an initial period and eccentricity such that tidal
evolution reproduces the present day eccentricity of the system at the current
age of the system. A lower limit to the dissipation (upper limit on $Q_\star'$)
follows by requiring that no matter how large the  initial eccentricity of the
system is, tides must circularize it to lie below the period--eccentricity
envelope by the present system age. In other words, the dissipation should be
enough to make it impossible for a binary containing exactly the stars of the
binary being analyzed to end up above the period-eccentricity envelope, no
matter the initial eccentricity, but not so much as to make it impossible to
reproduce the observed eccentricities of the systems.

When applying the Bayesian analysis described in \ref{sec:bayesian_analysis},
the evolution will start with high initial eccentricity ($e_i=0.8$), and an
initial orbital period such that the present day orbital period is reproduced at
the present system age. The likelihood function then assigns equal probability
for the final eccentricity to lie anywhere above the observed eccentricity for
the particular system and below the period-eccentricity envelope for the given
collection of systems being analyzed.  Uncertainties in the measured
eccentricity are handled by marginalizing over the corresponding distribution.
The remaining parameters are sampled during the Bayesian analysis.

\subsection{Orbital Evolution}
\label{sec:orbital_evolution}

At the core of this work is the ability to simulate the evolution of binary star
systems under the combined effects of tides (with a separate prescription for
the dissipation in the two stars), evolving stellar structure, loss of angular
momentum to stellar winds, and the internal redistribution of angular momentum
between the surface and the interior of stars. Furthermore, orbital evolution
must handle high eccentricities, and allow for variable tidal dissipation. All
of those requirements are met by the latest version of a publicly available tool
called \poet{} \footnote{\url{https://github.com/kpenev/poet}}.
\citet{Penev_Zhang_Jackson_14} presents a preliminary version of the code
describing the handling of stellar evolution, stellar winds, and the
redistribution of angular momentum within stars.  Subsequent development already
allows for:
\begin{itemize}
    \item planet--star, star--star, planet--planet, and single star systems
    \item splitting objects into an arbitrary number of zones, with imperfect
        rotational coupling between neighboring zones, evolving zone boundaries
        in both mass and radius, and separate tidal dissipation for each zone
    \item evolving the obliquity of each zone of each object, under zone--zone
        coupling and tidal dissipation
    \item eccentric orbits evolving under the effects of tidal dissipation, with
        dynamically adjusted eccentricity expansion that accurately handles even
        extreme eccentricities.
    \item tracking the dissipation for each tidal wave in each zone separately,
        under a fully general prescription for the dissipation as an arbitrary
        user-specified function of internal structure and spin of the object
        being tidally distorted, as well as the frequency and amplitude of the
        tidal wave.
\end{itemize}

In the work presented here, only the star-star regime of \poet{} was used. Stars
are modeled as objects consisting of two zones: a radiative core and a
convective envelope, with only the convective zone assumed to be subject to
tidal dissipation. The evolution of the radius of the star, the moments of
inertia of the two zones, and the mass and radius boundary between the two zones
is calculated by interpolating within a grid of stellar evolution tracks
computed following exactly the MESA Isochrones \& Stellar Tracks
\citep{Dotter_16, Choi_et_al_16} prescription for generating solar calibrated
stellar evolution tracks using the Modules for Experiments in Stellar
Astrophysics code \citep{Paxton_et_al_11, Paxton_et_al_13, Paxton_et_al_15}. The
angular momentum vector of each zone is evolved separately. The exchange of mass
between the two zones of the star implies exchange of angular momentum according
to:

\begin{equation}
    \frac{d\vec{L}_\mathrm{rad}}{dt}
    =
    -\frac{d\vec{L}_\mathrm{conv}}{dt}
    =
    \frac{2}{3} R_\mathrm{rad}^2 \frac{dM_\mathrm{rad}}{dt}
    \vec{\omega}_\mathrm{conv}
\end{equation}

Where $\vec{L}_\mathrm{rad/conv}$ are the angular momentum vectors of the
convective and radiative zones respectively, $R_\mathrm{rad}$ and
$M_\mathrm{rad}$ are the mass and radius boundary between the core and the
envelope, and $\vec{\omega}_\mathrm{conv}$ is the convective zone angular
velocity vector.

Differential rotation between the core and the envelope is assumed to decay
exponentially with a timescale of $\tau_{c-e}$ \citep{Irwin_et_al_07}:

\begin{equation}
    \frac{d\vec{L}_\mathrm{rad}}{dt}
    =
    -\frac{d\vec{L}_\mathrm{conv}}{dt}
    =
    \frac{I_\mathrm{conv}\vec{L}_\mathrm{rad}-
    I_\mathrm{rad}\vec{L}_\mathrm{conv}}{(I_\mathrm{conv}+I_\mathrm{rad})\tau_{c-e}}
\end{equation}

Where $I_\mathrm{conv/rad}$ are the moments of inertia of the convective and
radiative zone respectively.

In addition, the envelope loses angular momentum to stellar winds according to
\citep{Barnes_Sofia_96,Irwin_et_al_07}:

\begin{equation}
    \frac{d\vec{L}_\mathrm{conv}}{dt}
    =
    -K_w\vec{\omega}_\mathrm{conv} \min(\omega_{\mathrm{conv}},
    \omega_\mathrm{sat})^2
		\left(
			\frac{R_*}{R_\odot}\right)^{1/2} \left( \frac{M_*}{M_\odot}
		\right)^{-1/2}
\end{equation}

Where $K_w$ parametrizes how efficiently the stellar wind removes angular
momentum and $\omega_{sat}$ is a frequency cutoff, above which the mass loss rate
due to stellar wind is assumed to saturate, as required to explain the spin-down
observed in stars near the zero age main-sequence.

Finally, tidal evolution follows the formalism of \citet{Lai_12}, but
generalized for eccentric orbits, and applied to each zone of each star
separately. Briefly, the tidal potential, $U_{tide}(\vec{r}, t)$, experienced by
a star due to its companion is expanded in a series of spatial spherical
harmonics, and temporal Fourier terms:

\begin{align}
    U_{tide}(\vec{r}, t)
    \equiv & {}
    \frac{GM'}{|\vec{r}_{M'}|}\left(
        \frac{\vec{r}\cdot\vec{r}_{M'}}{\left|\vec{r}_{M'}\right|^2}
        -
        \frac{\left|\vec{r_{M'}}\right|}{\left|\vec{r} -
        \vec{r}_{M'}\right|}
    \right)\nonumber \\
    = & {}
    \sum_{m=-2}^2 \sum_{m'=0}^{\infty} U_{m,m'} e^{-i\Omega_{m,m'} t}
\end{align}

With:

\begin{align}
    \Omega_{m,m'} & \equiv m\Omega_\star-m'\Omega_{orb}\\
    U_{m,m'} & \equiv -\frac{GM'}{a^3}
    \sum_{s\in\{-2, 0, 2\}}
    \rho^2 Y_{2,m}(\theta,\phi)
    W_{2,s}D_{m,s}(\Theta) p_{s,m'}
\end{align}

Where $M$ is the mass of the tidally distorted star, $M'$ is the mass of the
companion, $\vec{r}$ is a position within the tidally distorted object (with
spherical coordinates $\rho$, $\theta$, $\phi$ in a frame with origin the center
of $M$ and rotating with the tidally distorted zone of $M$), $\vec{r}_{M'}$ is
the position of the companion, $a$ is the semi-major axis of the orbit,
$Y_{2,m}(\theta,\phi)$ is the second degree spherical harmonic of order $m$,
$W_{2,s}D_{m,s}$ are given in \citet{Lai_12}, $\Omega_{orb}$ is the orbital
angular velocity, $\Omega_\star$ is the spin angular velocity of the tidally
distorted zone, and $p_{s,m'}$ are expansion coefficients that depend solely on
the eccentricity of the orbit. The summation is over $m=-2\ldots2$, $s\in\{-2,
0, 2\}$, and $m'=0 \ldots \infty$. In \poet{}, the $p_{s,m'}$ coefficients were
calculated numerically for $s\leq400$, on a sufficiently dense grid of
eccentricities to ensure precise and accurate interpolation. The summation on
$m'$ is truncated to a dynamically adjusted order to guarantee a user-specified
precision target in the tidal potential.

Again, following \citet{Lai_12}, we assume the object responds to each ($m$,
$m'$) term in the above series independently, with a fluid displacement
$\vec{\xi}_{m,m'}$ and density perturbation $\delta\rho_{m,m'}$, given by:

\begin{align}
    \vec{\xi}_{m,m'}(\vec{r},t) & =
    \frac{U_{m,m'}}{\omega_0^2}\vec{\bar{\xi}}_{m,m'}(\vec{r})
    \exp(-i\Omega_{m,m'} t + i\Delta_{m,m'})\\
    \delta\rho_{m,m'}(\vec{r},t) & =\frac{U_{m,m'}}{\omega_0^2}
    \delta\bar{\rho}_{m,s}(\vec{r}) \exp(-i\Omega_{m,m'} t + i\Delta_{m,m'})
\end{align}

Where $\omega_0 \equiv \sqrt{\frac{G M}{R^3}}$ is the dynamical frequency of the
tidally distorted star, and $\Delta_{m,m'}$ is a phase lag that parametrizes the
tidal energy loss for this particular wave.  While this prescription is based on
linear equilibrium tide, it is capable of capturing the dynamics produced by any
tidal model (equilibrium or dynamic) as long as the principle of superposition
applies (i.e. the effect of each wave is independent of the other waves
present). What is required is that $\Delta_{m,m'}$ be specified in such a way as
to reproduce the rate at which energy is converted to heat. This will in general
imply that $\Delta_{m,m'}$ will be different for different waves and will evolve
as the star or the orbit evolves.  The phase lag has a one-to-one relationship
with the often used tidal quality factor defined in the introduction:
$\Delta_{m,m'}\propto Q_{m,m'}^{-1}$.  Note that by allowing $\Delta_{m,m'}$ to
depend on $m$, $m'$, the structure and spin of the star, and the orbit, we can
accommodate a wide range of equilibrium and dynamic tidal theories.

Finally, the rate of tidal energy dissipation $\dot{E}$ and the torque $\vec{T}$
experienced by the tidally distorted stellar zone are calculated as:

\begin{align}
    \dot{E} &= -\int d^3 x \rho(\vec{r})
    \frac{\partial \vec{\xi}(\vec{r}, t)}{\partial t}
    \cdot\nabla U_{tide}(\vec{r}, t)\\
    \vec{T}&= -\int d^3 x \delta\rho(\vec{r}, t) \vec{r}
    \times \nabla U_{tide}(\vec{r}, t)
\end{align}

expanded to linear order in $\Delta_{m,m'}$. In the resulting expressions, the
equilibrium tidal response and the phase lag for a particular wave always appear
together as $\kappa_{m,m'} \Delta_{m,m'}$, where:

\begin{equation}
    \kappa_{m,m'} \equiv \frac{1}{M_\star R_\star^2} \int d^3x
    \delta\bar{\rho}_{m,s}(\vec{r}) r^2 Y_{2,m}^*(\theta, \phi)
\end{equation}

Thus we fold the complicated equilibrium response of the star to tides into a
re-defined modified phase lag $\Delta'_{m,m'} \equiv \kappa_{m,m'}
\Delta_{m,m'}$, or a corresponding modified tidal quality factor,
$Q'\equiv\frac{Q}{k_2}$ ($k_2$ is the usual tidal Love number for the particular
star).

The procedure above describes the differential equations that control the
evolution. To fully specify the problem, we must also specify initial
conditions. In particular, we need values for the initial orbital period
($P_{orb,i}$), initial eccentricity ($e_i$), and initial angular momenta for the
four zones comprising our stars (each star is split in a radiative core and a
convective envelope).

As discussed in the beginning of the section, our method relies on
over-estimating the initial eccentricity, starting all our evolutions with
$e_i=0.8$. The initial orbital period is then determined by requiring that after
running the evolution to present day, the observed present day orbital period is
reproduced. Finally, the initial angular momenta of all zones are determined by
starting all stars at young enough age so they are fully convective and
calculating the evolution of the core angular momentum while holding the
envelope spin period fixed at $P_{\star,0}=10\,days$ until an age of $t_0$
($t_0=20\,Myr$ for NGC 6819 and NGC 188 and $t_0=2\,Myr$ for M 35). This
prescription seems to work for single star spin evolution models
\citep[e.g.][]{Gallet_Bouvier_15}, which posit that the surface of the star
remains locked to the inner edge of the circum-stellar disk until the disk
evaporates. For single stars, this locked spin appears to last between a few
and 10\,Myrs. Short period binaries are likely also affected by the presence of
circum-stellar material at very early ages, but modeling such evolution is
beyond the scope of this paper.

We start the binary evolution at later ages for the two older clusters because
that improves the numerical stability without affecting the results. Appendix
\ref{sec:disk_lifetime_effect} shows that by ages of few Gyrs, the difference
between evolutions started at 20Myrs vs all earlier ages is negligible. It is
worth pointing out that if tidal dissipation were strongly age dependent the
effect of the starting age may be more pronounced, but for the tidal model
assumed here that is not the case.

On the other hand, the M 35 cluster is young enough that delaying the start of
tidal circularization could potentially have a significant effect. The choice of
starting age of 2\,Myrs in that case is a compromise between numerical stability
requirements and the desire to start as early as possible.  In Appendix
\ref{sec:disk_lifetime_effect} we show that the effect of the starting age is
entirely negligible for NGC\,6819 and NGC\,188, and less than 0.3\,dex in the
inferred value of $\log_{10}Q_\star'$ for M\,35.

The initial angular momenta have very little effect on the subsequent orbital
evolution for two reasons. First, within just a few hundred Myrs, differences in
initial spin are erased due to the strong spin dependence of the rate at which
angular momentum is lost to wind.  Second, the orbit dominates the angular
momentum budget by several orders of magnitude, leading to virtually identical
evolution regardless of the initial spin (see Appendix
\ref{sec:disk_period_effect} ). This also implies that the influence of the
$K_w$, $\omega_{sat}$, and $\tau_{c-e}$ parameters on the predicted final
eccentricity is going to be rather limited.  For that reason, we do not attempt
to account for uncertainty in these parameters, and instead fix their values for
all systems analyzed (see Table \ref{tab:parameter_distributions} for values).

\subsection{Tidal Dissipation Model}
\label{sec:tidal_dissipation_model}

Given the broad range of models for tidal dissipation and its dependence on many
system parameters, a parametric prescription flexible enough to accommodate all
the major theoretical possibilities would involve an impractical number of free
parameters. Luckily, analyzing one system at a time means only dependence on
parameters that change significantly during the evolution needs to be explicitly
included in the prescription of $Q'_{m,m'}$, while dependencies on fixed
parameters (e.g.  stellar mass or metallicity) will be captured as differences
between systems. In this work, we assume that tidal dissipation occurs only in
the convective zones of the stars. Furthermore, we assume the same tidal
dissipation prescription applies to the convective zones of both stars, with the
tidal quality factor (or phase lag) following a saturating powerlaw dependence
on the tidal period $P_{m,m'}\equiv2\pi/\Omega_{m,m'}$:

\begin{equation}
    Q'_{m,m'} = Q_0
    \max\left[
        1,
        \left(\frac{P_{m,m'}}{P_0}\right)^\alpha
    \right]
    \label{eq:Q_prescription}
\end{equation}

The $Q_0$, $P_0$, and $\alpha$ parameters are then constrained using Bayesian
analysis (see Section \ref{sec:bayesian_analysis}) separately for each binary
system, but are assumed to be constant and the same for all $m$, $m'$ tidal
terms experienced by a particular binary.

This prescription is much more flexible than it appears, because each system is
analyzed independently. As a result, deviations from powerlaw behavior, or
dependencies on stellar mass and metallicity, will be detected as differences
between systems. Furthermore, having independent constraints from each system
allows us to check if systems with similar properties produce similar
constraints.

\subsection{Bayesian Likelihood}
\label{sec:bayesian_analysis}

Our analysis is designed to incorporate as broad a range of observational or
model uncertainties in the result as possible in a manner that accounts for
correlations between parameters. In particular, for single line spectroscopic
binaries (SB1), the masses of the two components are highly correlated, because
their values are constrained by multi-band photometric observations and radial
velocity (RV) measurements, combined with knowledge of the age and metallicity
of the cluster the binary is a member of. Furthermore, the inclination of the
orbit relative to the line of sight for our binaries is unknown, and so is the
actual initial eccentricity the binary started out with. In our analysis we wish to
properly account for these correlations and unknowns. Here we describe the
procedure we use to accomplish this task.

Given the final eccentricity $e_{max}$ found by an orbital evolution calculated
as described in Section \ref{sec:orbital_evolution}, the value of the
period-eccentricity envelope evaluated at the orbital period of the system
($e_{env}$), metallicity ($\feh$), masses of the primary and secondary
components ($M_1$, and $M_2$ respectively), inclination of the orbit relative to
the line of sight ($i$), present day eccentricity ($e$), system age ($t$), and
orbital period ($P_{orb}$), the posterior we wish to sample from can be written
as:

\begin{align}
    \mathcal{P}
    \propto &
    {} \mathcal{L}_{max}(e_{max}, e, e_{env})
    \tau(t)
    \phi\left(\feh\right)
    f_{phot}\left(M_1, M_2, t, \feh\right)
    \nonumber\\
    & \epsilon(e)
    \kappa\left[K_{RV}\left(M_1, M_2, P_{orb}, e, i\right)\right]
    \pi(Q_0, P_0, \alpha, i)
    \label{eq:raw_posterior}
\end{align}

Where:
\begin{itemize}
    \item $\mathcal{L}_{max}$ is 1 if $e < e_{max} < e_{env}$ and 0 otherwise.
        It implements the requirement of our method that over-estimating the
        initial eccentricity should result in a predicted present day
        eccentricity somewhere between the observed value and the envelope.
    \item $\tau$ and $\phi$ are the likelihoods that the cluster the binary is
        a member of has the specified age and metallicity. We approximate both
        as independent Gaussians with the appropriate mean and standard
        deviation taken from the literature (see Table
        \ref{tab:cluster_properties}).
    \item $f_{phot}$ is the likelihood of the observed multi-band photometry
        available for the system given that the components have the specified
        masses, age and $\feh$. We approximate the available measurements in
        a collection of photometry bands or colors as following an independent
        Gaussian distribution with the appropriate mean and standard deviation.
        Theoretical predictions for the magnitude of a star with given mass and
        $\feh$ from a given cluster were calculated by interpolating among the
        CMD 3.4 \footnote{\url{http://stev.oapd.inaf.it/cmd}} stellar isochrones
        \citep{Bressan_et_al_12, Chen_et_al_14, Chen_et_al_15, Tang_et_al_14,
        Marigo_et_al_17, Pastorelli_et_al_19} queried for the age and extinction
        appropriate for each cluster (Table \ref{tab:cluster_properties}). The
        brightnesses of the two stars are added and compared to the measured
        values. This is the dominant observational constraint for the mass of the
        brighter star in the system. Furthermore, many SB1 systems are found to
        lie significantly above the single star color-magnitude sequence for the
        cluster. In those cases, photometry ends up constraining the mass of the
        secondary tighter than RV observations (though we always combine both
        constraints). We verified that running the MCMC analysis with
        $\mathcal{L}_{max} = 1$ reproduces primary and secondary masses when
        those are quoted in the literature, even when using different
        photometry datasets than the literature (e.g. for NGC 188).
    \item $\epsilon$ and $\kappa$ approximate the likelihood of the observed RV
        measurements given the expected RV semi-amplitude $K_{RV}$ and
        eccentricity $e$. In particular we approximate $K_{RV}$ and $e$ as
        independently distributed, with $K_{RV}$ following a Rice distribution.
        The eccentricity cannot be negative, but frequently circular orbits
        cannot be excluded. In those cases we assign a finite value for the
        probability that the eccentricity is exactly zero and a Gaussian for all
        positive values.
    \item $\pi$ are prior distributions we impose on the dissipation parameters
        ($Q_0$, $P_0$, and $\alpha$) and the inclination of the orbit relative
        to the line of sight ($i$). These priors are assumed independent of each
        other.
\end{itemize}

Note that we ignore the uncertainty in the orbital period. For all our binaries
$P_{orb}$ is measured to very high precision (few parts per $10^5$ or better) so it
contributes only negligibly to the final uncertainty in the inferred dissipation
parameters.

The posterior in Eq. \ref{eq:raw_posterior} contains a very large number of
``nuisance'' parameters, and it is highly peaked for many of them. This will
dramatically reduce the performance of most Bayesian analysis algorithms. We
take several steps to alleviate these challenges.

First, we marginalize Eq. \ref{eq:raw_posterior} over $e$ (actual present day
eccentricity of the system) and $i$ (inclination of the orbit relative to the
line of sight), under the assumption that directions from which we are observing
the orbit are uniformly distributed on the sphere (i.e. $\zeta\equiv\cos i$ is
distributed uniformly between 0 and 1). The component of Eq.
\ref{eq:raw_posterior} that depends on $e$ or $i$ is:

\begin{equation}
    \lambda(e_{max}, K_0)
    \equiv
    \int_0^{e_{max}} \epsilon(e)\int_0^1 d\zeta
    \kappa\left(K_0\sqrt\frac{1-\zeta^2}{1-e^2}\right)
\end{equation}

where $K_0\equiv\sqrt[3]{\frac{2\pi G M_2^3}{P_{orb}(M_1 + M_2)^2}}$ is the RV
semi-amplitude for a circular aligned orbit viewed edge-on. We calculate
$\lambda$ numerically on a grid of $K_0$ and $e_{max}$ values with an adaptively
increased resolution until linear interpolation among the tabulated values is
accurate to a part per million.

Additionally, $\lambda$ simultaneously contains a sharp peak, highly correlated
between $M_1$ and $M_2$, and a very long low probability tail, again leading to
inefficient Bayesian sampling. To improve the efficiency, we rewrite the
posterior as:

\begin{align}
    \mathcal{P}
    \propto &
    \tau\Big(t\Big)
    \phi\Big(\feh\Big)
    f_{phot}\Big(M_1, M_2, t, \feh \Big)
    \pi\Big(Q_0, P_0, \alpha\Big)
    \lambda\Big(e_{env}, K_0\Big)
    \nonumber\\
    & \frac{\lambda\Big(e_{max}, K_0\Big)}{\lambda\Big(e_{env}, K_0\Big)}
    \label{eq:final_posterior}
\end{align}

Notice that, since nothing on the first line above depends on $e_{max}$, it does
not require calculating the orbital evolution to evaluate. We treat that as our
new prior and the bottom line becomes our likelihood function.

The final modification we introduce to the Bayesian analysis is that we define a
prior transform function that maps a set of independent random variables,
uniformly distributed between zero and one, to our parameters of interest,
such that the transformed variables follow the re-defined prior (top line in Eq.
\ref{eq:final_posterior}). That way the acceptance probability of any proposed
steps is only determined by the new likelihood (bottom line of Eq.
\ref{eq:final_posterior}), which is dominated by how well the assumed
dissipation parameters reproduce the observed and envelope eccentricity;
satisfying RV and photometry data, as well as cluster measurements, is handled
by the prior transform.

Table \ref{tab:parameter_distributions} lists the assumed distributions for all
model and observational parameters needed for calculating the orbital evolution,
the prior transform, and the likelihood.

For practical reasons we impose lower limits on $\log_{10}Q_\star'$ that are
larger than some theoretical models predict. This is necessary because
calculating the evolution under the assumption of very efficient dissipation
becomes very computationally intensive and not practical as part of a Bayesian
analysis. That assumption does not affect the results we report in this paper.
In particular, the final constraints we find for the dissipation in NGC\,6819
and NGC\,188 are well away from the assumed boundaries, hence not influenced by
them. For M\,35 we find that even our smallest assumed $Q_\star'$ is consistent
with the data, so we only report an upper limit.

As Appendix \ref{sec:disk_period_effect} demonstrates, the effect of the stellar
spin on the evolution is small, because it accounts for only a very small
fraction of the total angular momentum available in the system. This means that
the exact values we assume for the spin parameters ($P_{\star,0}$, $K_w$,
$w_{sat}$, and $\tau_{c-e}$) will not have a significant influence on the final
results. Note that initial stellar spin may be important if tidal dissipation
increases rapidly with stellar spin or at young ages, but it is not important
for the parametrization of $Q_\star'$ assumed here (Eq. \ref{eq:Q_prescription}).

\begin{table*}
    \centering
    \caption{
        The values/distributions for all model and observational parameters
        needed for calculating the orbital evolution or the posterior
        distribution.  See text for detailed descriptions of the parameters. $U$
        denotes uniform distribution with the given bounds, $N$ denotes
        Gaussian distribution with the given mean and standard deviation,
        and $R$ denotes Rice distribution with the given parameters.
    }
    \label{tab:parameter_distributions}
    \begin{tabular}{l@{\quad}l@{\quad}l}
        \hline
        \multicolumn{1}{c}{\textbf{Parameter}} &
        \multicolumn{1}{c}{\textbf{Description}} &
        \multicolumn{1}{c}{\textbf{Distribution/Value}}\\
        \hline \hline
        $\pi(\log_{10}Q_0)$ & Prior of normalization of tidal dissipation &
        $U[5, 12]$ for NGC 6819 and NGC 188, $U[4, 12]$ for M 35\\
        \hline
        $\pi(\log_{10}P_0)$ & Prior of Saturation period of tidal dissipation &
        $U[\log_{10}(0.5\,d), \log_{10}(20\,d)]$\\
        \hline
        $\pi(\alpha)$ & Prior of tidal dissipation powerlaw index & U[-5, 5]\\
        \hline
        $P_{\star,0}$ & Constant surface spin period of stars before binary
        evolution starts & $10\,d$\\
        \hline
        $t_0$ & Age at which binary evolution starts & $20\,Myr$ for NGC 6819
        and NGC 188, $2\,Myr$ for M 35\\
        \hline
        $K_w$ & Normalization of stellar angular momentum loss to wind & $0.17\,
        M_\odot\,R_\odot^2\,day^2\,rad^{-2}\,Gyr^{-1}$\\
        \hline
        $\omega_{sat}$ & Convective zone angular velocity of wind saturation &
        $2.45\,rad\,day^{-1}$\\
        \hline
        $\tau_{c-e}$ & Core-envelope coupling timescale & $5\,Myr$\\
        \hline
        $\tau(t)$ & Observational constraint for the age of the system &
        $N\left(\mu_{t,cl}, \sigma_{t,cl}\right)$\\
        \hline
        $\phi(t)$ & Observational constraint for the $\feh$ of the system &
        $N\left(\mu_{\feh,cl}, \sigma_{\feh,cl}\right)$\\
        \hline
        $\epsilon(e)$ & Observational constraint for the present day
        eccentricity & $f N\left(\mu_{e,sys}, \sigma_{e, sys}\right) + (1-f)
        \delta(e=0)$\\
        \hline
        $\kappa(K_{RV})$ & Observational constraint for the RV semi-amplitude &
        $R\left(\frac{\mu_{K_{RV},sys}}{\sigma_{K_{RV}}, sys}, \sigma_{K_{RV},
        sys}\right)$\\
    \end{tabular}
\end{table*}

\begin{figure}
    \includegraphics[width=\columnwidth]{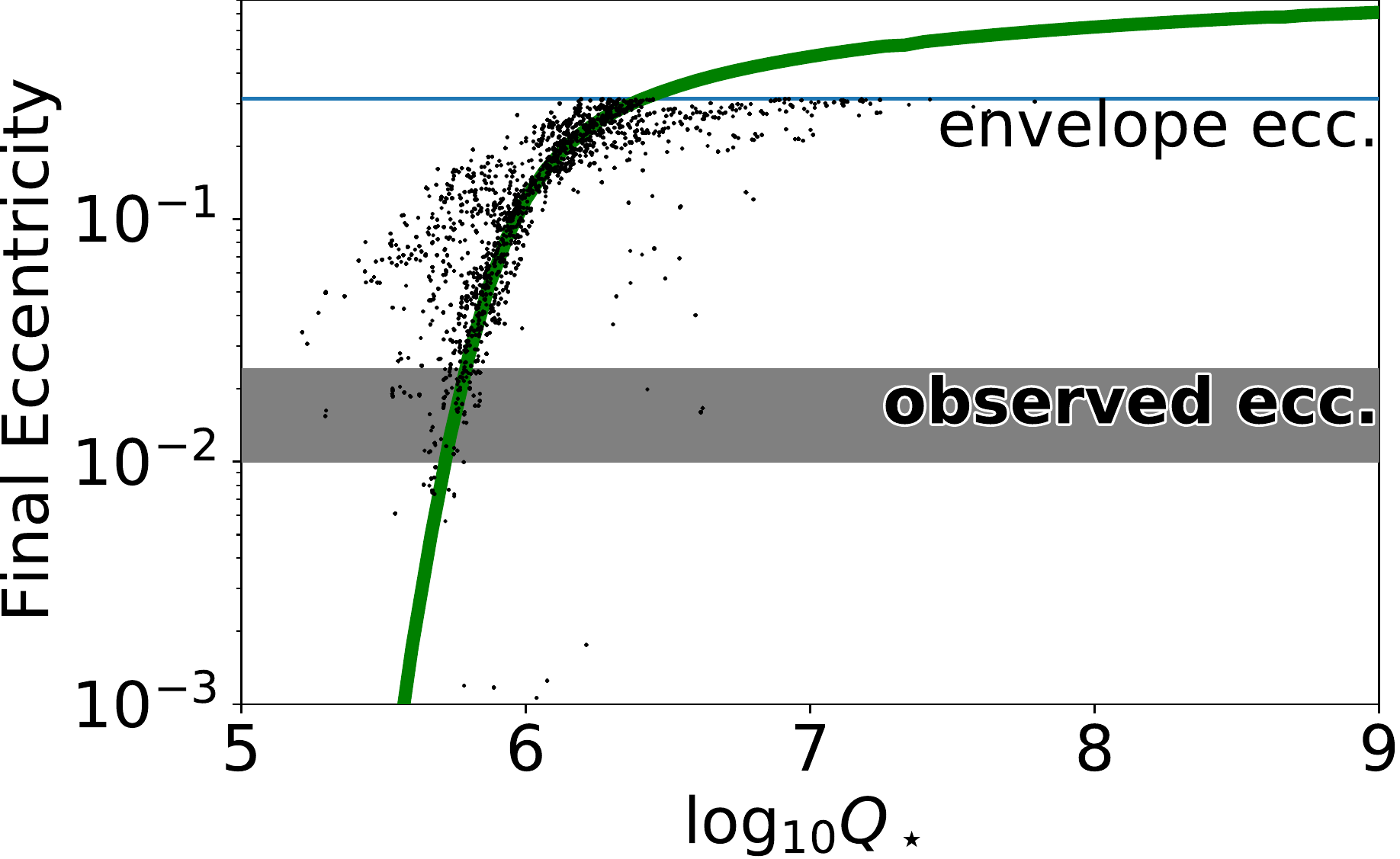}
    \caption{
        Example of the Bayesian analysis applied to a single binary in our
        sample (NGC 188 binary PKM 4618). Green line: final eccentricity as a
        function of  $Q_\star'$ (assumed constant), ignoring observational
        uncertainties (assuming median values for all binary properties). Cloud
        of black points: MCMC samples of $\log_{10}Q_\star'(P_{tide}=5\,d)$
        accounting for all observational uncertainties and variable $Q_\star'$.
    }
    \label{fig:analysis_demo}
\end{figure}

Fig. \ref{fig:analysis_demo} shows this method applied to a single binary (NGC
188 binary PKM 4618). The horizontal line near the top is the value of the
period-eccentricity envelope evaluated at the orbital period of the binary. The
grey area slightly below the middle shows the measured eccentricity of the
system (1-sigma error bar), and the green line shows the final eccentricity of a
binary containing the same stars as PKM 4618 (median parameters), if it started
with an initial eccentricity of 0.8, and with an orbital period that evolves to
the observed orbital period at the present age. Finally, the points are the
Bayesian analysis samples, following the posterior distribution defined above.
At low $Q_\star'$ (high dissipation), no matter the initial eccentricity, the
system is circularized before its present age, so no initial conditions exist
that would evolve to the observed present orbit. At high $Q_\star'$, a system of
identical stars starting with high initial eccentricity would end up above the
envelope, also inconsistent with observations.

\subsection{Sampling and Convergence Diagnostics}
\label{sec:convergence_diagnostics}

In order to carry out the Bayesian analysis, using the likelihood function
defined above, we rely on the \texttt{emcee} python package
\citep{Foreman-Mackey_et_al_13} based on the \citet{Goodman_Weare_10} affine
invariant sampling algorithm. At each MCMC step, the algorithm simultaneously
advances a collection of multiple individual, \emph{not independent}, chains,
called walkers. We chose to use 64 walkers. The combined samples of all walkers
obey all standard requirements of the MCMC algorithm, and hence as the number of
samples increases the distribution of samples approaches the desired posterior.

We initialize the sampler with 64 initial positions chosen to be over-dispersed
compared to the posterior. To generate these samples, we draw uniform random
values between 0 and 1, apply the prior transform described above, run the
orbital evolution to the present system age, and keep any samples that land below
the period-eccentricity envelope. For samples that land above the envelope, we
re-draw  new independent values and repeat the process. The reason to avoid
starting walkers in configurations that evolve to above the period-eccentricity
envelope is that per our definition, the likelihood function there is exactly
zero, so the acceptance probability for most proposed updates for such values is
undefined.

One of the key issue any MCMC analysis needs to address is demonstrating that
sampling has run long enough for the samples to follow the target posterior
distribution. One common approach is to split sampling in two stages. First, an
initial burn-in period is introduced, during which samples potentially started
from very low likelihood regions find their way to higher likelihoods. These
initial samples have the potential to bias any estimates of quantities of
interest away from their true values for finite length chains, so the usual
practice is to discard those from further analysis. After the burn-in period,
samples are used to extract estimates of the desired quantities. In addition to
the need to determine the burn-in cut off, the often highly correlated nature of
MCMC chains presents a challenge to estimating uncertainties in reported
quantities.

In this effort we wish to find confidence intervals for the value of $\log_{10}
Q_\star'$ as a function of tidal period, which implies measuring quantiles of
its distribution. \citet{Raftery_Lewis_91} provide a procedure to find a
required burn-in period, and a formalism for calculating the variance of
quantile estimates. We follow their method, but adapt it to the peculiar emcee
sampling involving many walkers. Appendix \ref{sec:convergence_derivation} gives
the details of how the method was adapted to this analysis. As a result of
following the procedure of Appendix \ref{sec:convergence_derivation}, we get an
estimate of the burn-in period (the number of samples to be discarded before our
    estimates of the 2.3\%, 15.9\%, 84.1\%, 97.7\% quantiles of the posterior
distribution have converged) and an estimate of the uncertainty of the CDF at
each of those quantiles using only steps after the burn-in period.

\section{Input Data}
\label{sec:input_data}

In order to carry out the analysis described in Section \ref{sec:methods}, we
need observational constraints of the age, metallicity, extinction, and distance
to the three open clusters targeted by this study, as well as RV and photometric
measurements for the short period SB1 systems contained in these clusters.

The cluster properties assumed (along with references) are presented in Table
\ref{tab:cluster_properties}. For each cluster, we analyze single line
spectroscopic binaries flagged as open cluster members with orbital periods
shorter than 50 days. The binaries which meet these criteria but for which no
circularization analysis is reported here, along with the reason for their
exclusion, are described separately for each cluster below.

\begin{table}
    {
        \centering
        \caption{The observationally constrained properties of the three open
        clusters analyzed.}
        \label{tab:cluster_properties}
        \begin{tabular}{r@{\quad}l@{\quad}}
            \hline \hline
            \multicolumn{2}{c}{\textbf{M 35}} \\
            \hline
            Age $\Big(\mu_{t,cl} \pm \sigma_{t,cl}\Big)$ &
            $150 \pm 30\,Myr^f$ \\
            $\feh$ $\Big(\mu_{\feh,cl} \pm \sigma_{\feh,cl}\Big)$ &
            $-0.18 \pm 0.03^{i,j}$ \\
            Extinction ($E(B-V)$) & $0.2^h$ \\
            Distance modulus & $9.6^g$\\
            \hline \hline
            \multicolumn{2}{c}{\textbf{NGC 6819}} \\
            \hline
            Age $\Big(\mu_{t,cl} \pm \sigma_{t,cl}\Big)$ &
            $2.4 \pm 0.3\,Gyr^a$ \\
            $\feh$ $\Big(\mu_{\feh,cl} \pm \sigma_{\feh,cl}\Big)$ &
            $0.09 \pm 0.03^{b}$ \\
            $\mathrm{(B-V)}$ & $0.15^b$ \\
            Distance modulus & $11.85^a$\\
            \hline \hline
            \multicolumn{2}{c}{\textbf{NGC 188}}\\
            \hline
            Age $\Big(\mu_{t,cl} \pm \sigma_{t,cl}\Big)$ &
            $7.0 \pm 0.5\,Gyr^c$ \\
            $\feh$ $\Big(\mu_{\feh,cl} \pm \sigma_{\feh,cl}\Big)$ &
            $0.10 \pm 0.05^e$ \\
            Extinction ($E(B-V)$) & $0.08^c$ \\
            Distance modulus & $11.3^{c,d}$\\
            \hline
        \end{tabular}\\
    }
    \begin{itemize}
        \item[$^a$] \cite{Basu_et_al_11}

        \item[$^b$] \citet{Bragaglia_et_al_01}
        \item[$^c$] \citet{Sarajedini_et_al_99}
        \item[$^d$] \citet{Gao_18}
        \item[$^e$] \citet{Worthey_Jowett_03}
        \item[$^f$] \citet{Meibom_Mathieu_Stassun_09}
        \item[$^g$] \citet{Abdelaziz_et_al_22}
        \item[$^h$] \citet{Kalirai_et_al_03}
        \item[$^i$] \citet{Netopil_et_al_16}
        \item[$^j$] \citet{Barrado_et_al_01}
    \end{itemize}
\end{table}

\subsection{M 35}
\label{sec:m35_data}

We use the \citet{Leiner_et_al_15} $V$ and $B-V$ photometry and RV measurements
for SB1s in M\,35. We select only systems with orbital period less than 50 days.
The authors collected photometry from two sources: observations by T. von
Hippel using the Burrell Schmidt telescope at KPNO and observations by
Deliyannis using the WIYN 0.9m telescope (no cite-able sources were given and we
were unable to locate such ourselves). The second source is deemed higher
precision by the authors, and \citet{Geller_et_al_10} quote that the $V$
magnitude differences between the two sources is approximately Gaussian with
$\sigma=0.06\,mag$. Based on this information, we model the photometry
likelihood function ($f_{phot}$ from Section \ref{sec:bayesian_analysis}) as a
product of a Gaussian distribution of the $V$ magnitude with $\sigma=0.06/0.03$,
and an independent Gaussian distribution of the $B$ magnitude with
$\sigma=0.08/0.04$ if the photometry for the particular binary came from the
low/high precision source. In addition, since we only use SB1 systems, only the
spectrum of the brighter star in the binary was detectable. We conservatively
require that the difference in $V$ magnitudes between the primary and the
secondary star in the system should be at least 1, otherwise presumably this
would have been a double lined binary.

We use the values given in Table 5 of \citet{Leiner_et_al_15} for the orbital
periods and the parameters for the RV semi-amplitude ($\kappa$) and eccentricity
($\epsilon$) distributions. As described in Section \ref{sec:bayesian_analysis}
we approximate $\kappa$ by a Rice distribution with shape parameter given by the
ratio of the measured value to the uncertainty and scale given by the
uncertainty from \citet{Leiner_et_al_15} Table 5, and $\epsilon$ is assumed
Gaussian with the given mean and $\sigma$.

\begin{figure}

    \includegraphics[width=\columnwidth]{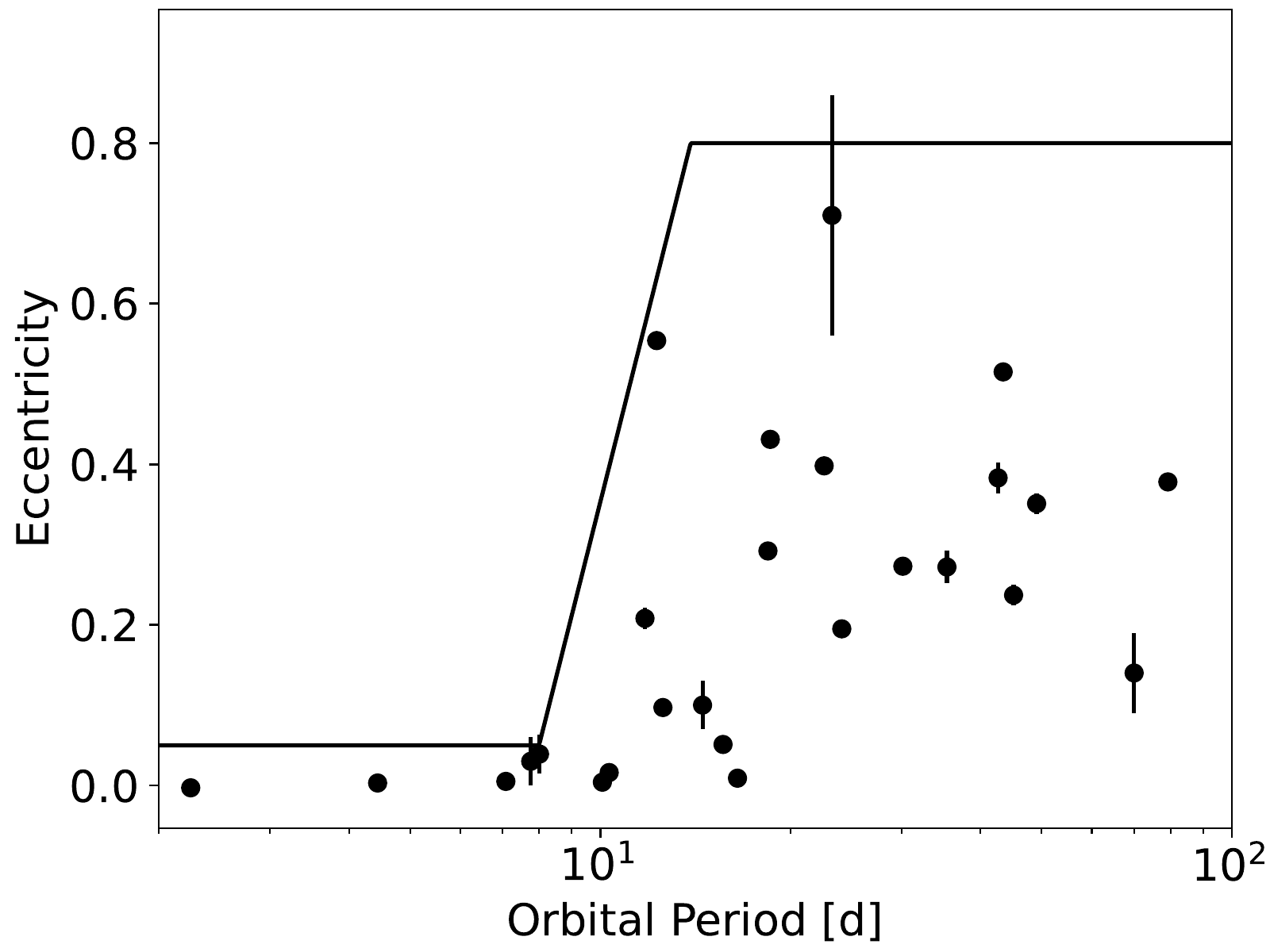}

    \caption{
        Period eccentricity envelope (black line) assumed for M
        35. The circles show all binaries flagged as members by
        \citet{Leiner_et_al_15}. At large periods we show the envelope
        saturating at $e_{env}=0.8$ because our orbital evolution with
        $e_0=0.8$. Hence, the highest eccentricity our model is capable of
        reproducing is $e=0.8$
    }
    \label{fig:m35_period_eccentricity}
\end{figure}

We also use the \citet{Leiner_et_al_15} binaries deemed to be members of M 35 to
construct a period-eccentricity envelope (see Fig.
\ref{fig:m35_period_eccentricity}):

\begin{equation}
    e_{env}(P_{orb})
    =
    0.05
    +
    \max\left[0, 1.356 \log_{10} \left(\frac{P_{orb}}{8\,days}) \right)\right]
    \label{eq:m35_eccentricity_envelope}
\end{equation}

Two SB1 systems with orbital periods less than 50 days were not included in this
analysis: WOCS 49043 and WOCS 15012, because for rare cases of initial
conditions, calculating their orbital evolution encountered numerical
instabilities at very young ages. We worried that simply rejecting such samples
could in theory bias the results, as it is in effect imposing another,
ill-defined, prior. Work is on-going on a follow-up article that will expand the
analysis presented here to many more clusters, and include double-lined binaries
in addition to SB1s, where we hope to resolve these issues and include these
systems. The observational data for all M 35 binaries used by this study is
presented in Table \ref{tab:m35_sb1_data}.

\begin{table}
    \centering
    \caption{
        Observationally determined properties of the M 35 binary systems used to
        constrain tidal dissipation.
    }
    \label{tab:m35_sb1_data}
    \begin{tabular}{rrrrrr}
    \multicolumn{1}{c}{\textbf{WOCS}} &
    \multicolumn{1}{c}{\textbf{$\bm{P_{orb}}$}} &
    \multicolumn{1}{c}{\textbf{$\bm{e}$}} &
    \multicolumn{1}{c}{\textbf{$\bm{K_{RV}}$}} &
    \multicolumn{1}{c}{\textbf{V}} &
    \multicolumn{1}{c}{\textbf{B-V}}\\
    \hline
    \hline
    54027 &
    2.247 &
    $-0.003\pm0.004$ &
    $46.13\pm0.21$ &
    15.95 &
    1.02\\
    16016 &
    7.089 &
    $0.005\pm0.006$ &
    $30.18\pm0.15$ &
    14.25 &
    0.63\\
    23043 &
    7.761 &
    $0.030\pm0.030$ &
    $10.00\pm0.40$ &
    14.75 &
    0.82\\
    54054 &
    8.013 &
    $0.039\pm0.024$ &
    $8.26\pm0.21$ &
    15.00 &
    0.78\\
    35045 &
    10.077 &
    $0.004\pm0.005$ &
    $49.86\pm0.24$ &
    14.50 &
    0.82\\
    40015 &
    10.330 &
    $0.016\pm0.007$ &
    $40.50\pm0.30$ &
    15.34 &
    0.84\\
    33054 &
    11.771 &
    $0.208\pm0.013$ &
    $45.50\pm0.60$ &
    14.88 &
    0.73\\
    59018 &
    18.427 &
    $0.292\pm0.010$ &
    $37.40\pm0.40$ &
    16.10 &
    1.10\\
    14025 &
    22.619 &
    $0.398\pm0.010$ &
    $17.47\pm0.21$ &
    14.32 &
    0.64\\
    24023 &
    30.134 &
    $0.273\pm0.004$ &
    $28.19\pm0.22$ &
    15.02 &
    0.85\\
    9016 &
    41.201 &
    $0.372\pm0.013$ &
    $16.23\pm0.22$ &
    13.55 &
    0.63\\
    37029 &
    45.121 &
    $0.237\pm0.013$ &
    $22.70\pm0.30$ &
    15.14 &
    0.87\\
    20016 &
    49.073 &
    $0.351\pm0.013$ &
    $16.81\pm0.18$ &
    14.86 &
    0.77\\
\end{tabular}

\end{table}

\subsection{NGC 6819}

We use the orbital period, RV semi-amplitude, and eccentricity for NGC 6819 SB1
systems reported in Table 5 of \citet{Milliman_et_al_14}, together with the $V$
and $I$ photometry of \citet{Yang_et_al_13} as reported in table 2 of
\citet{Milliman_et_al_14}. \citet{Milliman_et_al_14} also provide error
estimates for the photometric measurements by comparing to earlier photometry
collected by \citet{Sarrazine_et_al_03}, estimating $\sigma_V=0.036\,mag$ and
$\sigma_I=0.039\,mag$.

\begin{figure}

    \includegraphics[width=\columnwidth]{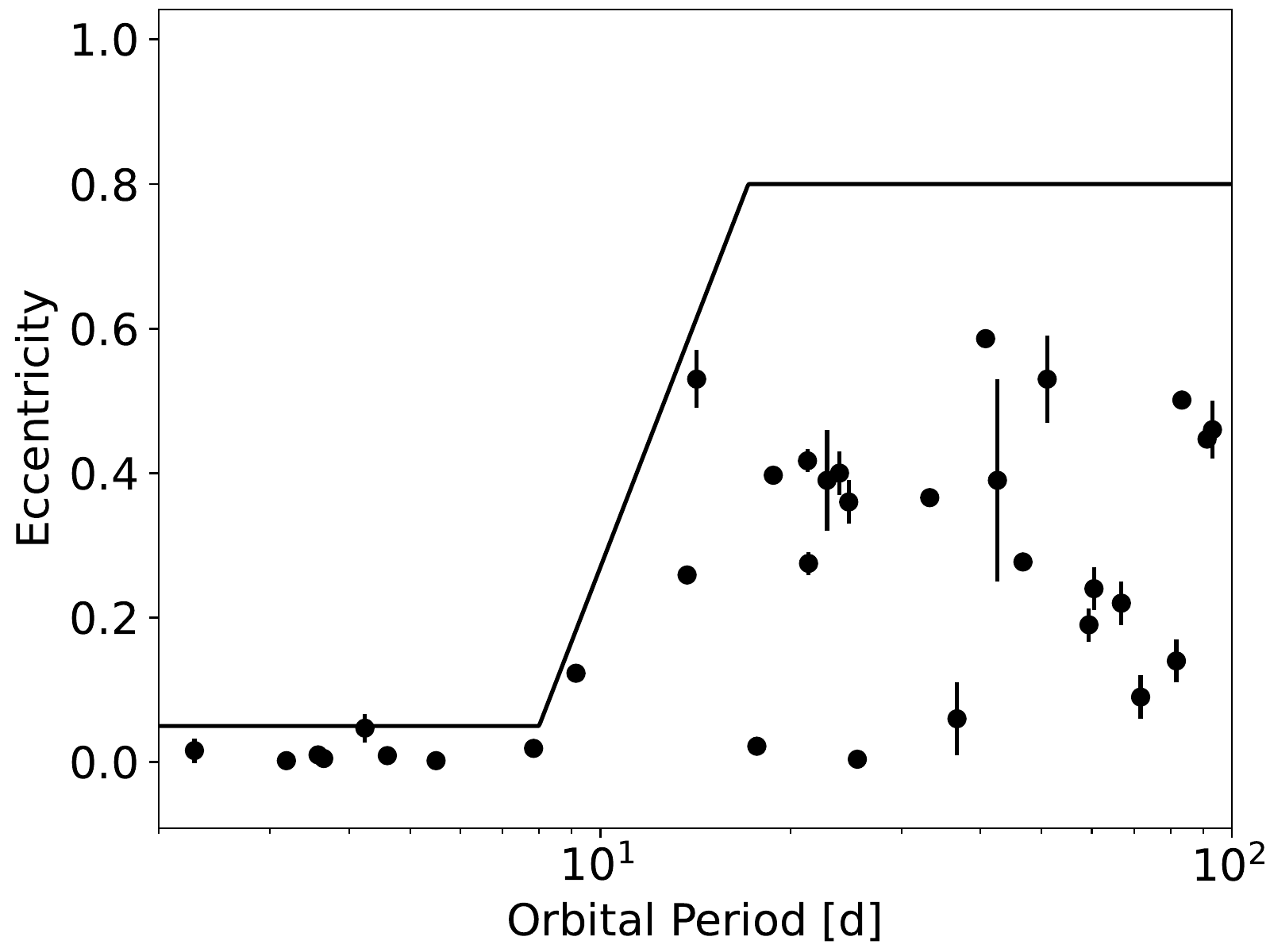}

    \caption{
        Same as Fig. \ref{fig:m35_period_eccentricity} but for NGC 6819. Binary
        data from \citet{Milliman_et_al_14}.
    }

    \label{fig:ngc6819_period_eccentricity}
\end{figure}

Similarly to M 35 we define a period-eccentricity envelope for NGC 6819, shown
in Fig. \ref{fig:ngc6819_period_eccentricity}:

\begin{equation}
    e_{env}(P_{orb})
    =
    0.05
    +
    \max\left[0, 1.340 \log_{10} \left(\frac{P_{orb}}{8\,days}) \right)\right]
    \label{eq:ngc6819_eccentricity_envelope}
\end{equation}

In the case of NGC 6819, a significant number of binaries with $P_{orb}<50\,d$
were excluded from this analysis (listed below).

\textbf{WOCS 57004:} We found that the posterior likelihood (Eq.
\ref{eq:final_posterior}) for this system is non-negligible only for a very
narrow range of parameters. This is mostly due to the fact that the observed
eccentricity ($e=0.123\pm0.006$) is very close to the envelope eccentricity of
Eq.  \ref{eq:ngc6819_eccentricity_envelope} at the orbital period of this system
($e_{env}=0.128$). While in principle this is good news, since it implies this
binary will provide very stringent constraints on tidal dissipation, it also
makes the inferred posterior very sensitive to the exact shape of the
distribution assumed for the present day eccentricity. Extracting reliable
results for this binary thus requires an in-depth study of the impact of
different assumed $\epsilon(e)$ and period-eccentricity envelopes, which will be
addressed in an upcoming study that will expand the analysis presented here to
a significantly larger sample of binaries.

\textbf{Binaries with WOCS identifiers: 33002, 3002, 25004, 31004, 21007, 24012,
39013, 8012, 26007:} The primary mass in these systems is near or above the
threshold where the surface convective zone becomes negligible. Most theoretical
models predict that such stars will be subject to very different tidal
dissipation compared to stars with significant surface convective zones. Since
our goal is to construct a combined constraint based on all binaries in a
cluster we exclude these systems in order to avoid erroneously combining
measurements of potentially different physical quantities.

The observational data for all NGC 6819 binaries used by this study is
presented in Table \ref{tab:ngc6819_sb1_data}.

\begin{table}
    \centering
    \caption{
        Observationally determined properties of the NGC 6819 binary systems
        used to constrain tidal dissipation.
    }
    \label{tab:ngc6819_sb1_data}
    \begin{tabular}{rrrrrr}
    \multicolumn{1}{c}{\textbf{WOCS}} &
    \multicolumn{1}{c}{\textbf{$\bm{P_{orb}}$}} &
    \multicolumn{1}{c}{\textbf{$\bm{e}$}} &
    \multicolumn{1}{c}{\textbf{$\bm{K_{RV}}$}} &
    \multicolumn{1}{c}{\textbf{V}} &
    \multicolumn{1}{c}{\textbf{V-I}}\\
    \hline
    \hline
    49002 &
    1.616 &
    $0.022\pm0.020$ &
    $53.70\pm1.20$ &
    16.23 &
    0.80\\
    66004 &
    2.278 &
    $0.016\pm0.017$ &
    $38.60\pm0.70$ &
    16.10 &
    0.80\\
    13001 &
    4.241 &
    $0.047\pm0.020$ &
    $55.10\pm1.00$ &
    15.87 &
    0.85\\
    60006 &
    7.843 &
    $0.019\pm0.011$ &
    $33.14\pm0.22$ &
    16.11 &
    0.83\\
    46013 &
    14.211 &
    $0.530\pm0.040$ &
    $10.80\pm0.50$ &
    15.95 &
    0.76\\
    59003 &
    21.368 &
    $0.275\pm0.016$ &
    $31.60\pm0.40$ &
    16.40 &
    0.83\\
    35020 &
    24.737 &
    $0.360\pm0.030$ &
    $12.40\pm0.50$ &
    15.80 &
    0.77\\
    53003 &
    42.540 &
    $0.390\pm0.140$ &
    $5.40\pm1.00$ &
    16.29 &
    0.81\\
\end{tabular}

\end{table}

\subsection{NGC 188}
\label{sec:ngc188_data}

Orbital period, RV semi-amplitude, and eccentricity information for NGC 188
binaries used in this analysis were taken from Table 2 of
\citet{Geller_et_al_09}. Photometric measurements in $u'$, $g'$, $r'$, $i'$, and
$z'$ bands were taken from \citet{Fornal_et_al_07} and $U$, $B$, $V$, $R$, and
$I$ measurements were taken from \citet{Stetson_et_al_04}.

\begin{figure}

    \includegraphics[width=\columnwidth]{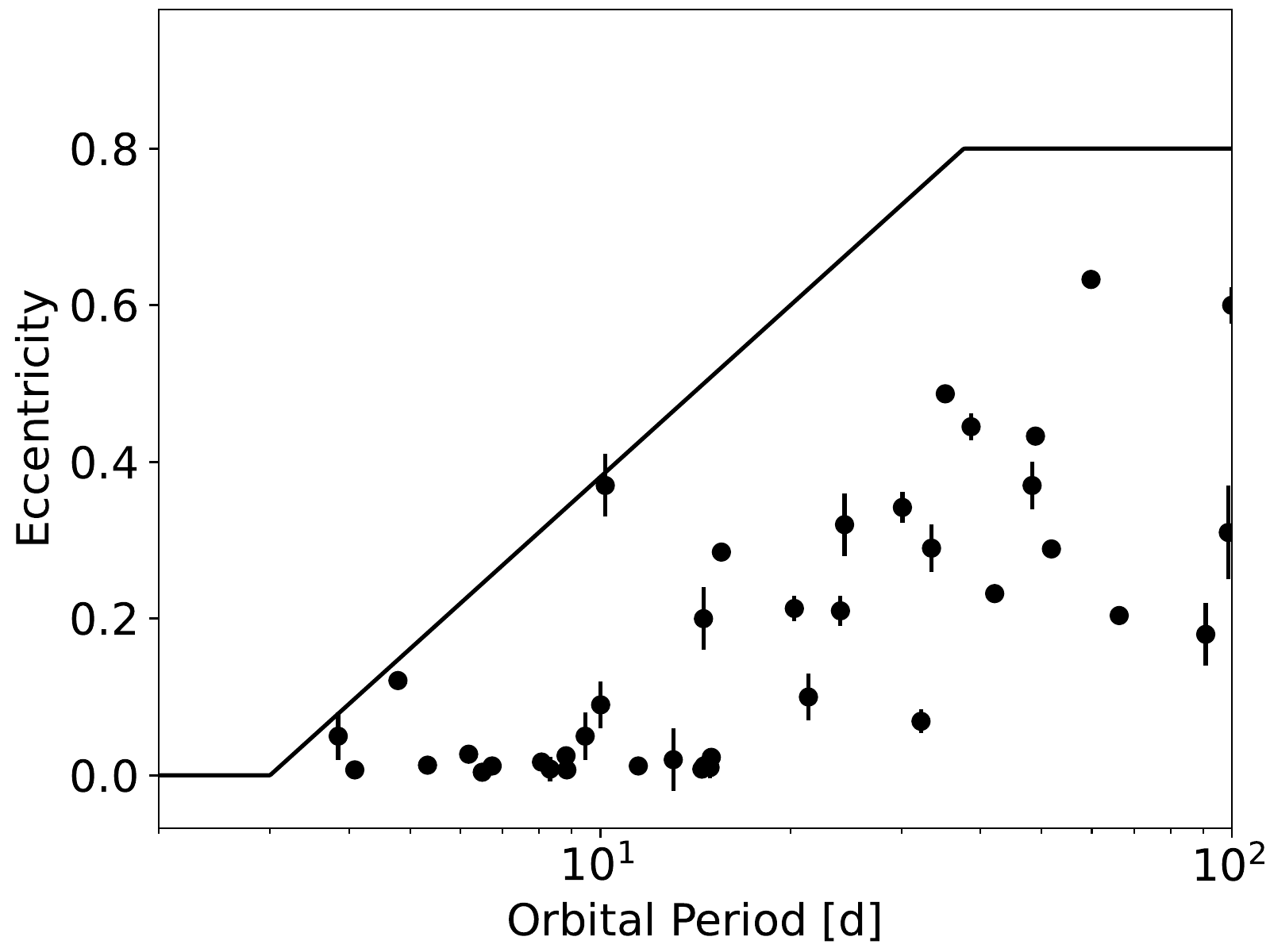}

    \caption{
        Same as Fig. \ref{fig:m35_period_eccentricity} but for NGC 188. Binary
        data from \citet{Geller_et_al_09}.
    }

    \label{fig:ngc188_period_eccentricity}
\end{figure}

Like for the other clusters, we define a period-eccentricity envelope based on
the observed binary orbits (see Fig. \ref{fig:ngc188_period_eccentricity}):

\begin{equation}
    e_{env}(P_{orb})
    =
    \max\left[0, 0.316 \log_{10} \left(\frac{P_{orb}}{3\,days}) \right)\right]
    \label{eq:ngc188_eccentricity_envelope}
\end{equation}

Note that the period eccentricity envelope we assumed for NGC 188 is pushed to
significantly higher eccentricities by just two systems, with PKM IDs 5078
($P_{orb} = 4.78303 \pm 0.00012$, $e = 0.121 \pm 0.006$) and 4904 ($P_{orb} =
10.185 \pm 0.003$, $e=0.37 \pm 0.04$). There is a clear pile-up of circular
systems for orbital periods all the way to 14 days or so, indicating these
systems are likely outliers. There is very little doubt either binary is a
member of the cluster with both their systemic radial velocity and proper motion
closely matching that of the cluster. One possibility \citet{Geller_et_al_09}
point out is that tertiary companions are not infrequent for short period
binary systems of solar type stars. As a result, these systems are potential
exceptions to the envelope for binaries shaped only by tides. Regardless, we
choose to define the envelope in a way that accommodates these systems, since
per our method that is the conservative assumption, increasing the uncertainty
in the inferred tidal dissipation parameters, while still including the values
that would be inferred with a lower period-eccentricity envelope.

Of the \citet{Geller_et_al_09} SB1 binaries with $P_{orb}<50\,d$ we exclude 2
systems:

\textbf{PKM 4710:} This system is flagged as a cluster member in
\citet{Geller_et_al_09} but at the same time the same authors quote its radial
velocity membership probability to be 1\%, though it appears to be a member
based on its proper motion. To be safe, we exclude the binary from our analysis.

\textbf{PKM 4999:} The distribution of the secondary star's mass in this system
extends to very small values (as low as $\sim0.1M_\odot$). Thus, there is a
reasonable chance that this is a fully convective star. At least some tidal
dissipation models predict very different dissipation for fully convective stars
vs stars with a significant radiative core. Since tidal circularization is
dominated by the dissipation in the secondary star, we don't necessarily expect
the dissipation measured for PKM 4999 to be similar to all other binaries
analyzed in this effort so we exclude the system from the analysis.

Table \ref{tab:ngc188_sb1_data} presents the observational data for all NGC 188
binaries included in this study.

\begin{table*}
    \centering
    \caption{
        Observationally determined properties of the NGC 188 binary systems
        used to constrain tidal dissipation.
    }
    \label{tab:ngc188_sb1_data}
    \begin{tabular}{rrrrrrrrrrrrrr}
    \multicolumn{1}{c}{\textbf{PKM}} &
    \multicolumn{1}{c}{\textbf{$\bm{P_{orb}}$}} &
    \multicolumn{1}{c}{\textbf{$\bm{e}$}} &
    \multicolumn{1}{c}{\textbf{$\bm{K_{RV}}$}} &
    \multicolumn{1}{c}{\textbf{U}} &
    \multicolumn{1}{c}{\textbf{B}} &
    \multicolumn{1}{c}{\textbf{V}} &
    \multicolumn{1}{c}{\textbf{R}} &
    \multicolumn{1}{c}{\textbf{I}} &
    \multicolumn{1}{c}{\textbf{u}} &
    \multicolumn{1}{c}{\textbf{g}} &
    \multicolumn{1}{c}{\textbf{r}} &
    \multicolumn{1}{c}{\textbf{i}} &
    \multicolumn{1}{c}{\textbf{z}}\\
    \hline
    \hline
    5052 &
    3.847 &
    $0.050\pm0.030$ &
    $8.80\pm0.30$ &
    16.82 &
    16.63 &
    15.92 &
    15.47 &
    15.08 &
    17.77 &
    16.30 &
    15.75 &
    15.58 &
    15.60\\
    4618 &
    8.073 &
    $0.017\pm0.007$ &
    $55.10\pm0.40$ &
    16.72 &
    16.48 &
    15.77 &
    15.33 &
    14.90 &
    17.57 &
    16.13 &
    15.57 &
    15.36 &
    15.27\\
    5015 &
    8.329 &
    $0.008\pm0.016$ &
    $45.60\pm0.70$ &
    17.20 &
    16.83 &
    16.03 &
    15.55 &
    15.10 &
    18.05 &
    16.37 &
    15.76 &
    15.55 &
    15.40\\
    6171 &
    8.828 &
    $0.025\pm0.011$ &
    $35.90\pm0.40$ &
    17.32 &
    17.05 &
    16.31 &
    15.83 &
    15.46 &
    --- &
    --- &
    --- &
    --- &
    ---\\
    5463 &
    9.465 &
    $0.050\pm0.030$ &
    $6.24\pm0.22$ &
    15.80 &
    15.64 &
    14.98 &
    14.58 &
    14.20 &
    16.61 &
    15.29 &
    14.83 &
    14.66 &
    14.55\\
    5601 &
    10.014 &
    $0.090\pm0.030$ &
    $5.67\pm0.10$ &
    16.45 &
    16.28 &
    15.60 &
    15.16 &
    14.79 &
    --- &
    --- &
    --- &
    --- &
    ---\\
    4904 &
    10.185 &
    $0.370\pm0.040$ &
    $11.20\pm0.50$ &
    16.91 &
    16.66 &
    15.93 &
    15.49 &
    15.07 &
    --- &
    --- &
    --- &
    --- &
    ---\\
    4289 &
    11.488 &
    $0.012\pm0.010$ &
    $40.80\pm0.30$ &
    16.94 &
    16.24 &
    15.30 &
    14.77 &
    14.20 &
    --- &
    --- &
    --- &
    --- &
    ---\\
    5738 &
    13.049 &
    $0.020\pm0.040$ &
    $5.92\pm0.20$ &
    16.10 &
    15.99 &
    15.31 &
    14.91 &
    14.55 &
    16.97 &
    15.62 &
    15.16 &
    14.98 &
    14.86\\
    6292 &
    14.599 &
    $0.012\pm0.007$ &
    $30.99\pm0.19$ &
    --- &
    16.21 &
    15.52 &
    --- &
    --- &
    --- &
    --- &
    --- &
    --- &
    ---\\
    4965 &
    14.922 &
    $0.010\pm0.014$ &
    $17.38\pm0.20$ &
    16.14 &
    15.97 &
    15.28 &
    14.86 &
    14.45 &
    16.80 &
    15.35 &
    14.83 &
    14.65 &
    14.60\\
    5797 &
    14.991 &
    $0.023\pm0.011$ &
    $19.53\pm0.23$ &
    16.76 &
    16.54 &
    15.85 &
    15.41 &
    15.03 &
    --- &
    --- &
    --- &
    --- &
    ---\\
    5647 &
    15.553 &
    $0.285\pm0.011$ &
    $45.50\pm0.70$ &
    17.00 &
    16.74 &
    16.00 &
    15.54 &
    15.13 &
    17.82 &
    16.36 &
    15.80 &
    15.60 &
    15.55\\
    880 &
    20.292 &
    $0.213\pm0.016$ &
    $15.40\pm0.30$ &
    --- &
    16.07 &
    15.38 &
    --- &
    --- &
    --- &
    --- &
    --- &
    --- &
    ---\\
    4080 &
    32.197 &
    $0.069\pm0.015$ &
    $20.70\pm0.30$ &
    16.38 &
    16.20 &
    15.52 &
    15.11 &
    14.73 &
    --- &
    --- &
    --- &
    --- &
    ---\\
    5040 &
    33.455 &
    $0.290\pm0.030$ &
    $6.91\pm0.24$ &
    15.86 &
    15.68 &
    15.01 &
    14.60 &
    14.23 &
    16.72 &
    15.31 &
    14.82 &
    14.65 &
    14.59\\
    4673 &
    48.270 &
    $0.370\pm0.030$ &
    $8.40\pm0.30$ &
    15.88 &
    15.60 &
    14.88 &
    14.45 &
    14.05 &
    16.72 &
    15.21 &
    14.68 &
    14.50 &
    14.41\\
\end{tabular}

\end{table*}

\section{Results}
\label{sec:results}

\subsection{System-by-System Constraints}
\label{sec:individual_results}

\begin{figure*}
    \includegraphics[width=\textwidth]{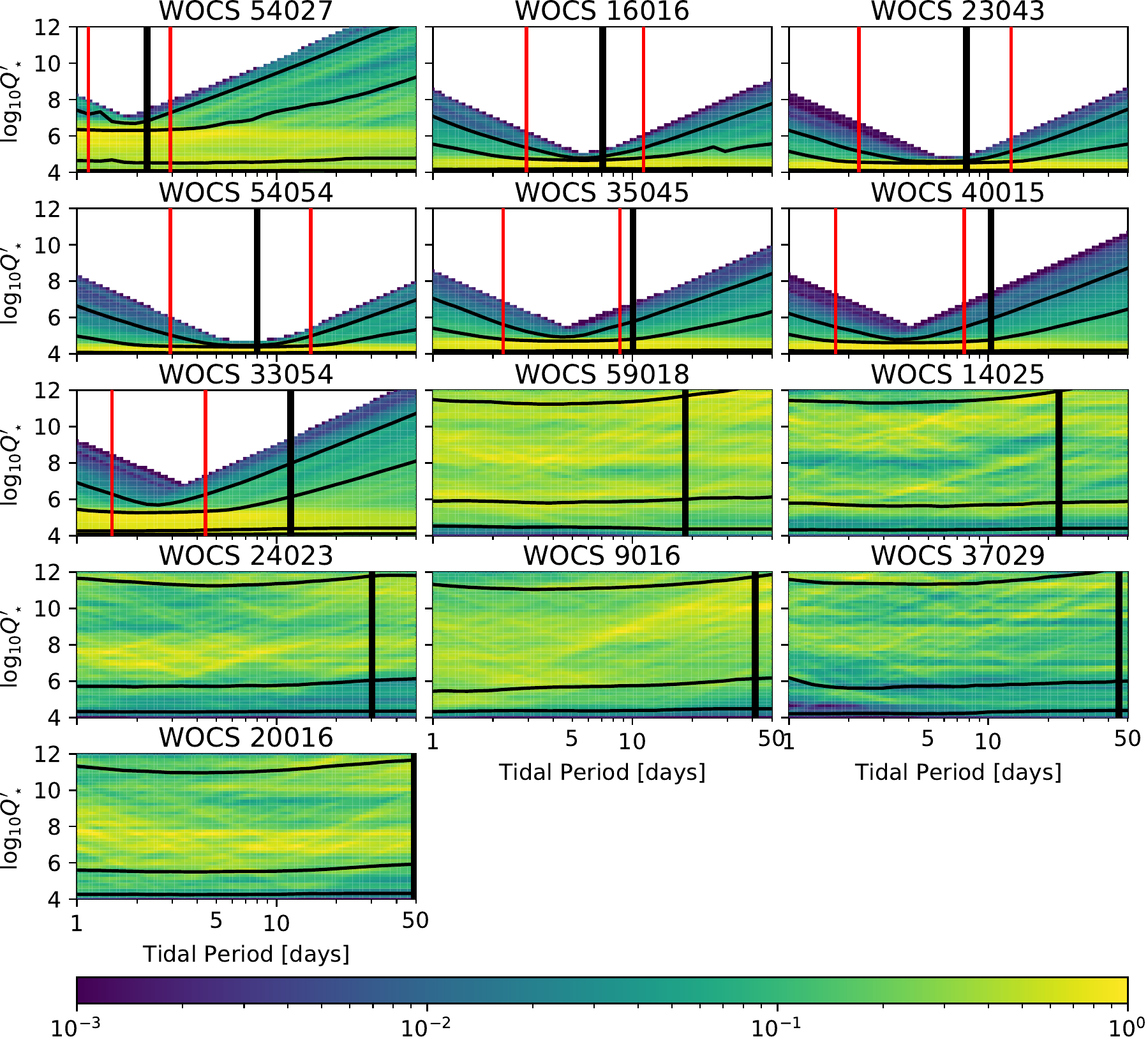}
    \caption{
        Constraints for $log_{10}Q_\star'$ obtained for the individual M\,35
        binaries. The color scale in each plot shows the posterior probability
        density function ($\log_{10}$ scale) of the tidal dissipation parameter
        as a function of tidal period.  The vertical red lines demarcate the
        range of tidal periods where the constraints are dominated by
        observations rather than model assumptions and priors (see text). The
        thin black curves show 2.3\%, 15.9\%, 84.1\%, and 97.7\% quantiles
        respectively.
    }
    \label{fig:m35_individual_constraints}
    \label{fig:first_individual_constraints}
\end{figure*}

\begin{figure*}
    \includegraphics[width=\textwidth]{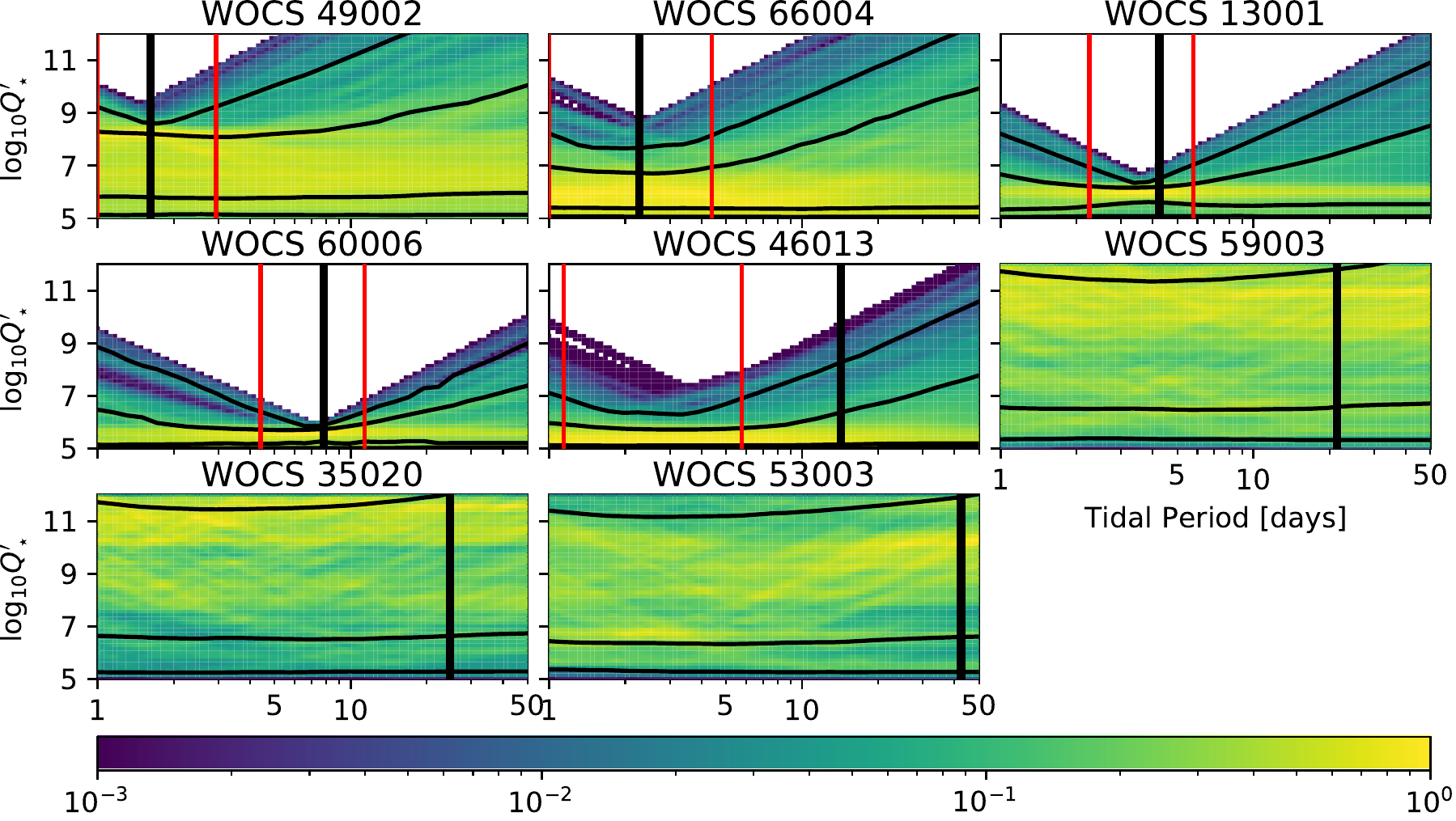}
    \caption{
        Same as Fig. \ref{fig:m35_individual_constraints} but for the NGC 6819
        binaries analyzed.
    }
    \label{fig:ngc6819_individual_constraints}
\end{figure*}

\begin{figure*}
    \includegraphics[width=\textwidth]{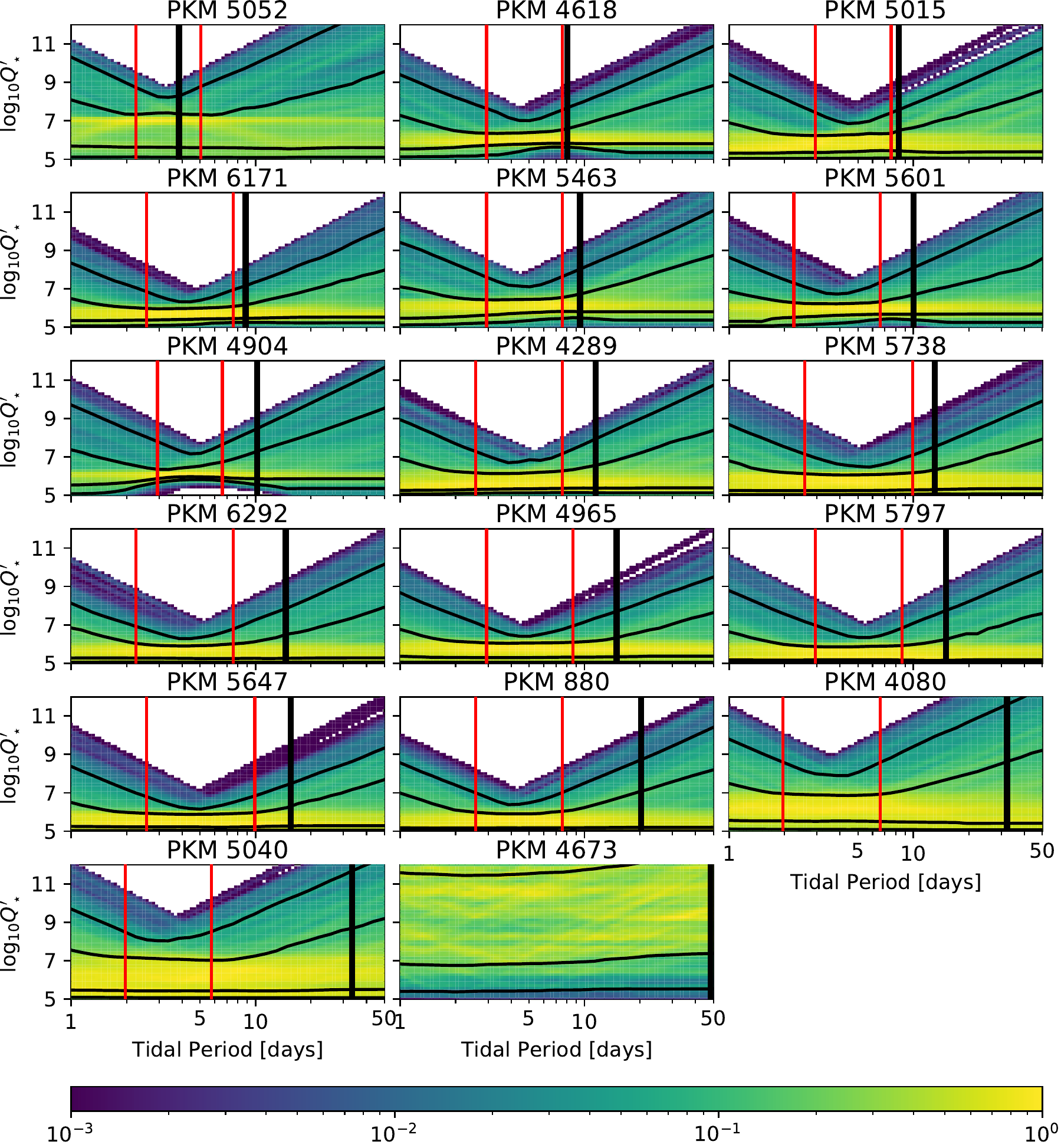}
    \caption{
        Same as Fig. \ref{fig:m35_individual_constraints} but for the NGC 188
        binaries analyzed.
    }
    \label{fig:ngc188_individual_constraints}
    \label{fig:last_individual_constraints}
\end{figure*}

For all binaries included in this study, we generated MCMC samples following the
procedure described in Sec. \ref{sec:methods}. For each MCMC sample generated,
we evaluate the tidal dissipation model (Eq. \ref{eq:Q_prescription}) at each
tidal period to produce samples of $\log_{10}Q_\star'$ as a function of
$P_{tide}$. We then convolve these samples with an Epanechnikov (quadratic)
kernel of width 0.2 to approximate the posterior probability density function of
$\log_{10}Q_\star'$ at each tidal period. Figures
\ref{fig:first_individual_constraints} --- \ref{fig:last_individual_constraints}
display the decimal logarithm of the inferred probability density for each
system scaled such that it has a maximum value of 1 (the color scale spans 3
orders of magnitude in likelihood). The vertical black line marks the orbital
period of the system, and the 4 black curves from bottom to top show the 1- and
2-sigma equivalent quantiles ($\log_{10}Q_\star'$ where the cumulative
distribution is 2.3\%, 15.9\%, 84.1\%, and 97.7\% respectively).

The v-shaped appearance of Fig. \ref{fig:first_individual_constraints} ---
\ref{fig:last_individual_constraints} upper quantiles and the period of the most
stringent upper limit (bottom of the V) for each system can intuitively be
understood to be produced by the period-eccentricity envelope. Per our method,
the highest value $Q_\star'$ can take is such that the system should land on the
period-eccentricity envelope. For orbital periods where significant
circularization is expected (envelope is significantly below the starting value
of $e=0.8$) there is a narrow range of tidal terms that dominate the final
stages of the circularization.

At short orbital periods, when the envelope only allows for very close to
circular orbits, the limiting stage of the circularization process is at the
very end, when the eccentricity is small.  In that case, the higher frequency
tidal potential terms have very small amplitudes, so they are unable to support
a spin-orbit lock at the corresponding frequency. As a result, the stellar spin
period is very close to the orbital period (star spins pseudo-synchronously).
The dominant tidal perturbation in those cases has a frequency equal to the
orbital period, because in a reference frame rotating with the star, the
companion moves closer and further away with each orbit but stays in
approximately the same direction.  Hence, this is the term that is most
constrained by our analysis.

At longer orbital periods, $Q_\star'$ values at the highest quantiles of the
distribution place the system on the envelope, which has significant
eccentricity, and can sustain higher order spin-orbit locks: $2\Omega_\star = k
\Omega_{orb}$ for k > 2. Since stars get spun-up as they contract to the main
sequence to rates far exceeding the orbital period, their spins generally get
stuck at the highest $k$ for which a spin-orbit lock can be maintained for a
given orbit. The effect of this picture is that the dominant tidal wave for
longer orbital period systems at the weakest allowed tidal dissipation (largest
value of $Q_\star'$) has a higher frequency (shorter period) than the orbit, and
hence the tightest upper limit on $Q_\star'$ lies at a fraction of the orbital
period.

Moving sufficiently away from the tidal periods where the dissipation is best
constrained, we see that the highest quantile curve has a slope $\frac{d\log
Q_\star'}{dP_{tide}}\approx5$, matching the prior limits we placed on the
powerlaw index of the assumed tidal dissipation model (Eq.
\ref{eq:Q_prescription}). Thus, far away from the best constrained period range,
it is our prior, rather than data, that dominates the constraints. In each plot in
Fig. \ref{fig:first_individual_constraints} ---
\ref{fig:last_individual_constraints}, the region between the two vertical red
lines provides a rough idea of the range of periods where the constraints are
driven by the observational data, and hence is informative for understanding
tidal dissipation. The exact locations were defined by requiring that the
97.7th percentile of the distribution of $log_{10}Q_\star'$ does not exceed its
minimum by more than 0.7 dex. Only samples after the empirically determined
burn-in period (see Sec. \ref{sec:convergence_diagnostics}) were used when
building these distributions. We thus choose the largest burn-in period across
all quantiles and tidal periods between the red lines in the plots.

Additionally, as expected, for all systems where the observed present day
eccentricity is consistent with circular orbit, the 2.3\% quantile is determined
by the lower limit we assume in Eq. \ref{eq:Q_prescription}. For circularized
systems our method only gives an upper limit to $Q_\star'$. Only systems for
which observations can confidently exclude circular orbits produce a lower
limit on $Q_\star'$ (upper limit on the dissipation).

In particular, none of the M 35 binaries constrains $Q_\star'$ from below. This
is because, as Fig.~\ref{fig:m35_period_eccentricity} shows, the orbits of all
binaries in M 35 out to orbital periods of 10\,d are indistinguishable from
circular, and all longer period binaries are in the regime where tides have not
significantly impacted the orbit. The former as already discussed give only
upper limit to $Q_\star'$ (lower limit on the dissipation) and the latter
produce no useful constraints (e.g. the last 6 panels of
Fig.~\ref{fig:m35_individual_constraints}).

Tables \ref{tab:results_first} --- \ref{tab:results_last} list the 2.3\%,
15.9\%, 84.1\%, and 97.7\% quantiles of $\log_{10}Q_\star'$ for each system at
the tidal period where the 97.7-th quantile is the smallest (i.e. the bottom of
the V). The table also compares these individual constraints with the combined
constraint (see Sec. \ref{sec:combined_results}). We also provide in machine
readable format the quantile curves shown in each of the plots (see Sec.
\ref{sec:data_availability}).

One result immediately evident from Figures
\ref{fig:first_individual_constraints} --- \ref{fig:last_individual_constraints}
is that explaining the observed period-eccentricity distribution of the young M
35 binaries requires significantly more dissipation (by a factor of
approximately 30) than explaining the much older NGC 6819 and NGC 188 (note
different lower limits of the y-axes in the figures). The constraints based on
all M 35 binaries are consistent with each other, and many binaries require
$\log_{10}Q_\star'<5$.  In contrast binaries in NGC 6819 and NGC 188 produce
consistent constraints within and between the two clusters, with several
requiring values of $\log_{10}Q_\star'\gtrsim5.5$.

\begin{figure*}
    \includegraphics[width=\textwidth]{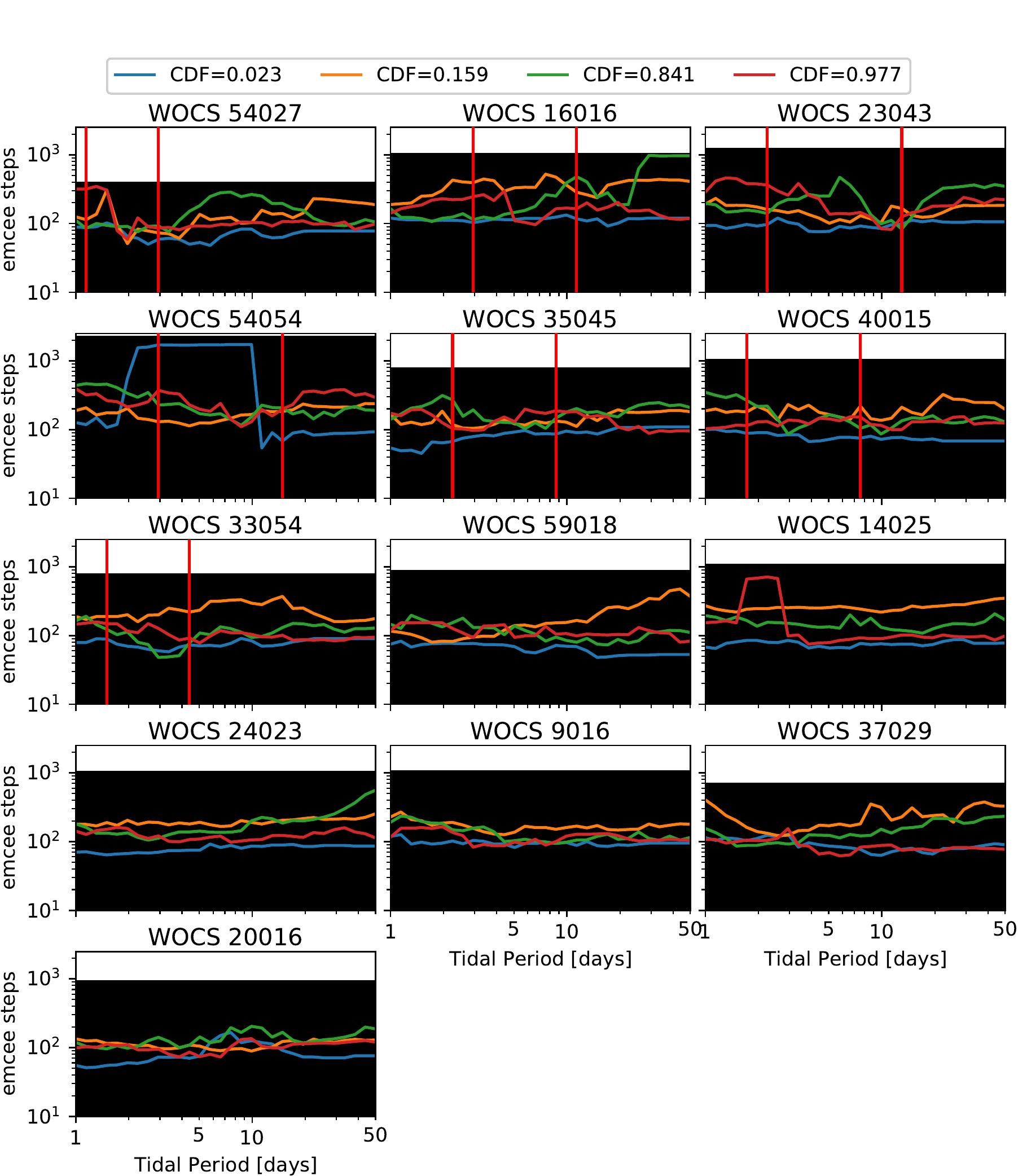}
    \caption{
        Diagnostics for the convergence of the MCMC chains. Each curve shows the
        estimate of the number of emcee steps required for subsequent samples to
        have the equilibrium probability of lying below or above a particular
        quantile (color lines). Those were used as the burn-in period when
        estimating posteriors (see text).  The black area of each plot shows the
        total number of steps generated during the sampling of this system. The
        vertical red lines match those in Fig.
        \ref{fig:m35_individual_constraints}, showing the range of periods where
        the constraints are driven by the data as opposed to priors.
    }
    \label{fig:m35_burnin}
    \label{fig:burnin_first}
\end{figure*}

\begin{figure*}
    \includegraphics[width=\textwidth]{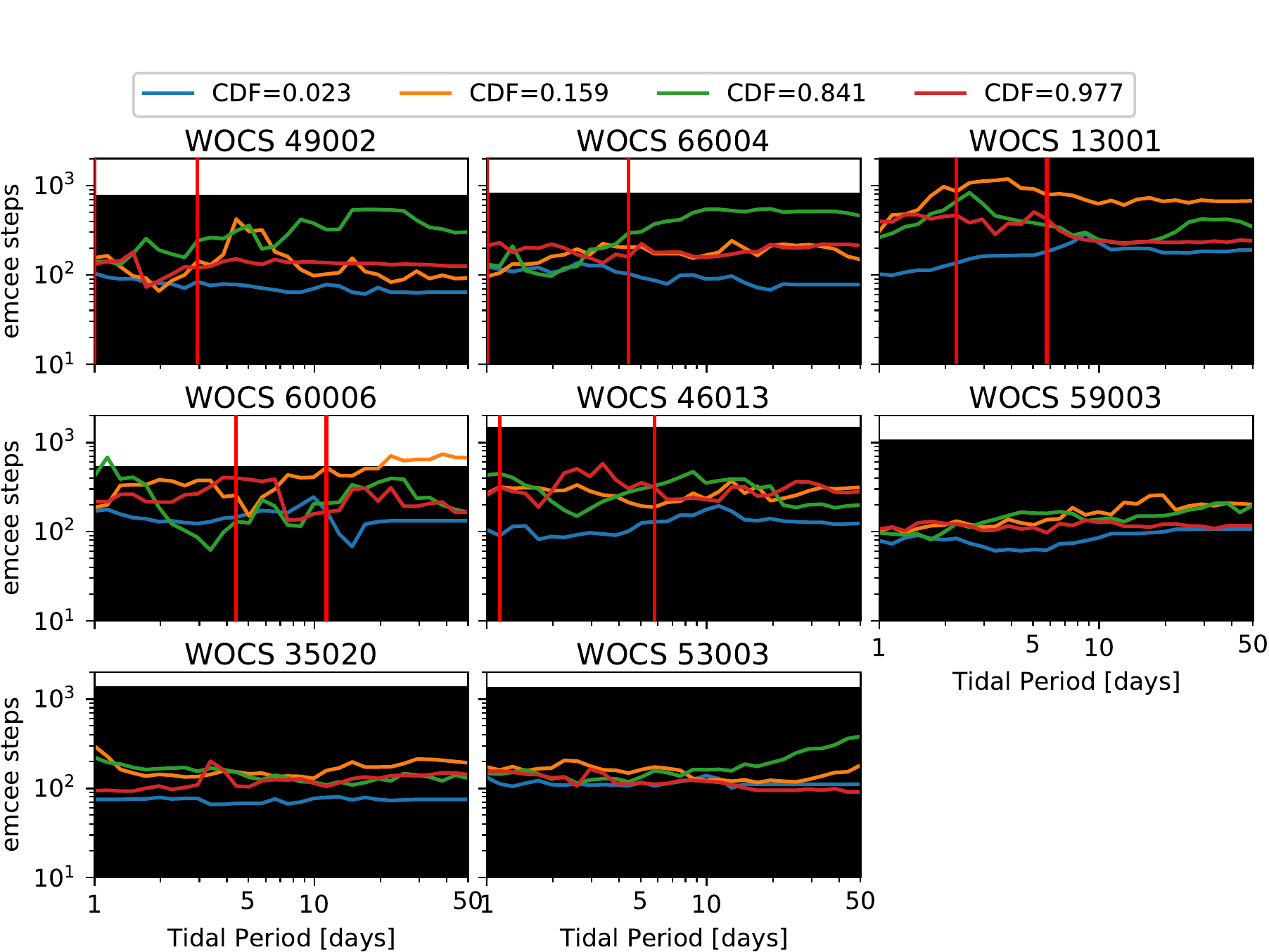}
    \caption{
        Same as Fig. \ref{fig:m35_burnin} but for the NGC 6819 binaries
        analyzed.
    }
    \label{fig:ngc6819_burnin}
\end{figure*}

\begin{figure*}
    \includegraphics[width=\textwidth]{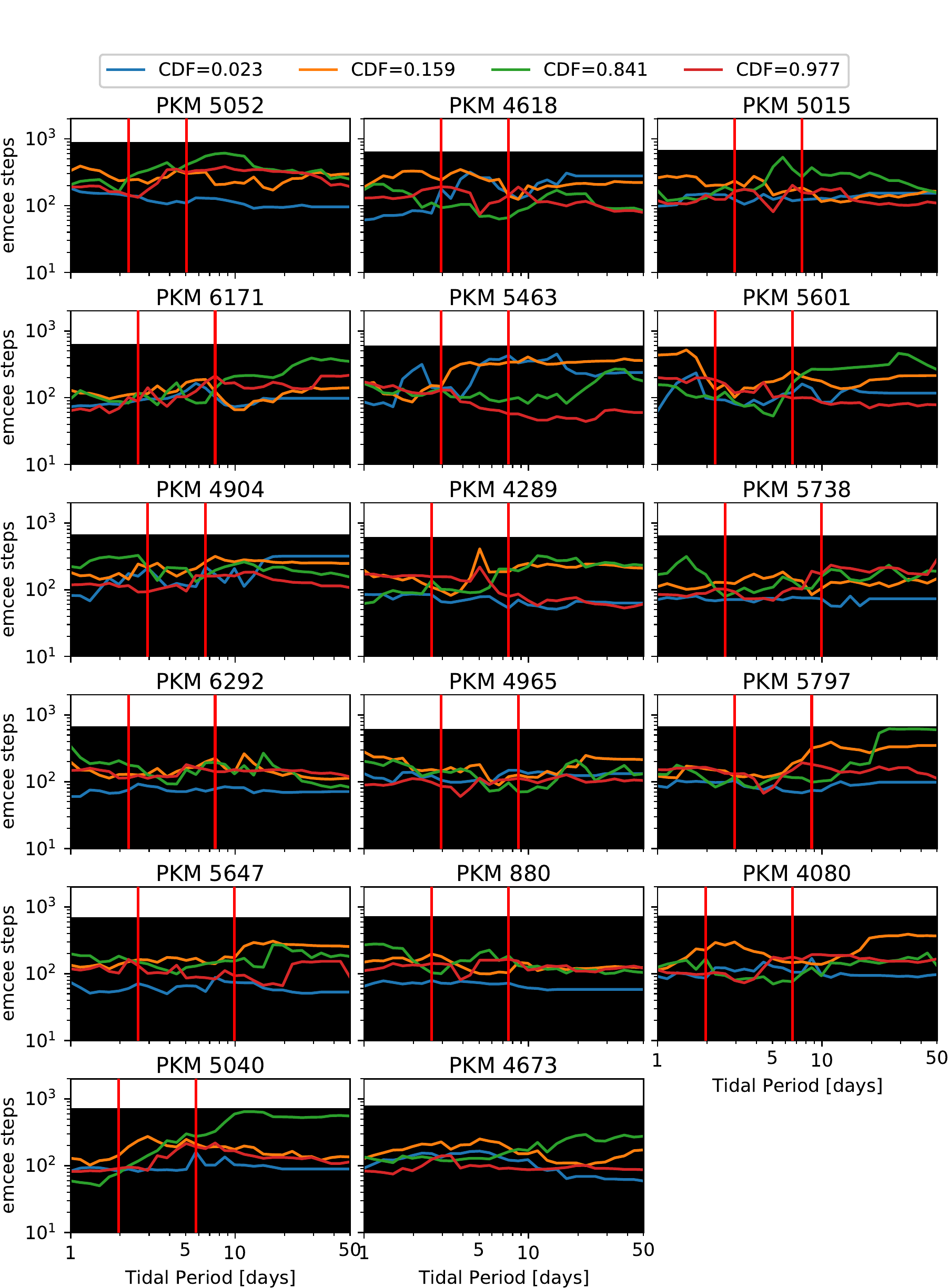}
    \caption{
        Same as Fig. \ref{fig:m35_burnin} but for the NGC 188 binaries
        analyzed.
    }
    \label{fig:ngc188_burnin}
    \label{fig:burnin_last}
\end{figure*}

\begin{figure*}
    \includegraphics[width=\textwidth]{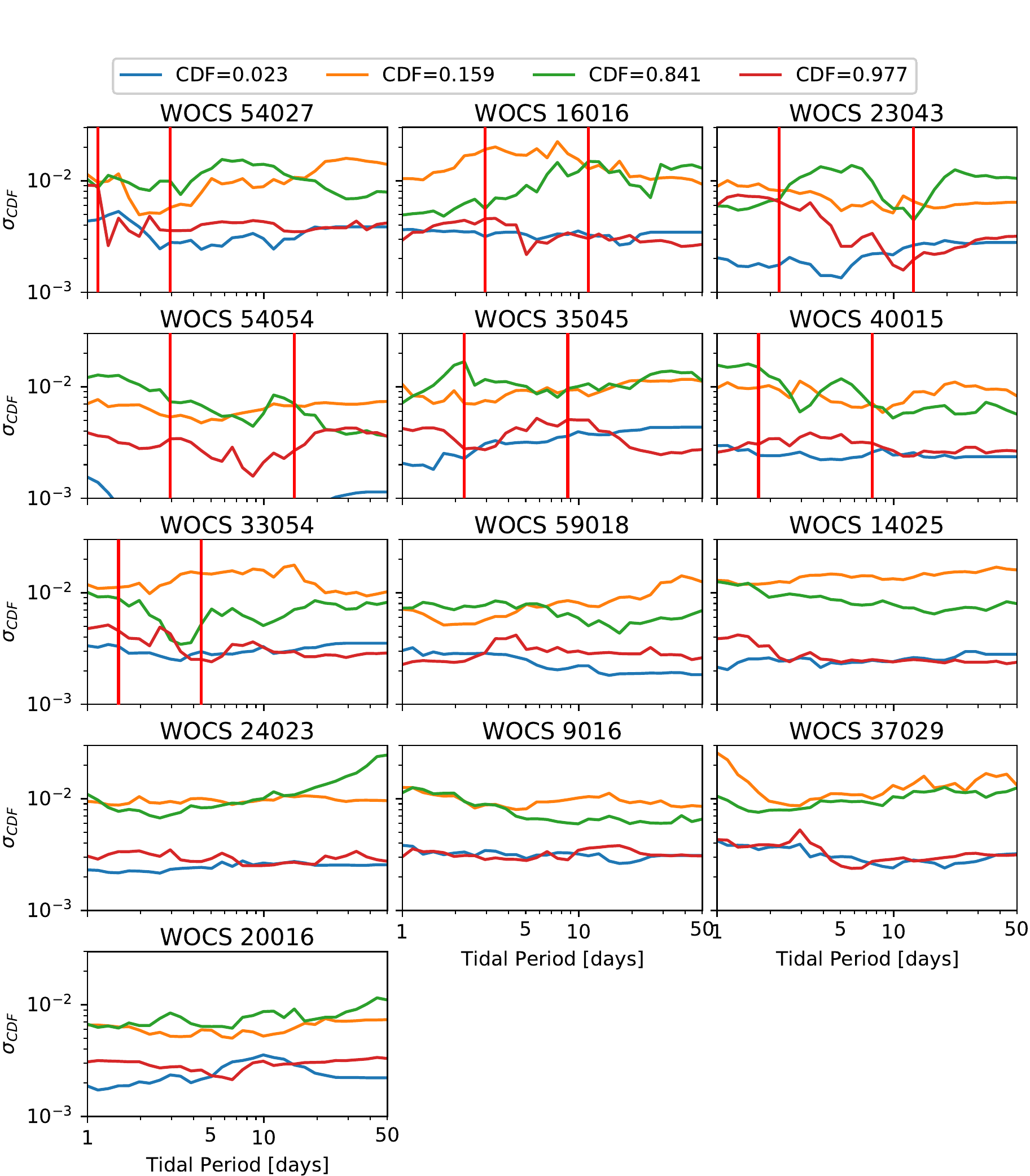}
    \caption{
        The estimated uncertainty in the fraction of samples below each of the 4
        target quantiles for the M 35 binaries analyzed. Each curve shows the
        estimate of the standard deviation of the cumulative distribution
        evaluated at a particular quantile (color lines). The red lines match
        those in Fig.  \ref{fig:m35_individual_constraints}.
    }
    \label{fig:m35_individual_cdfstd}
    \label{fig:individual_cdfstd_first}
\end{figure*}

\begin{figure*}
    \includegraphics[width=\textwidth]{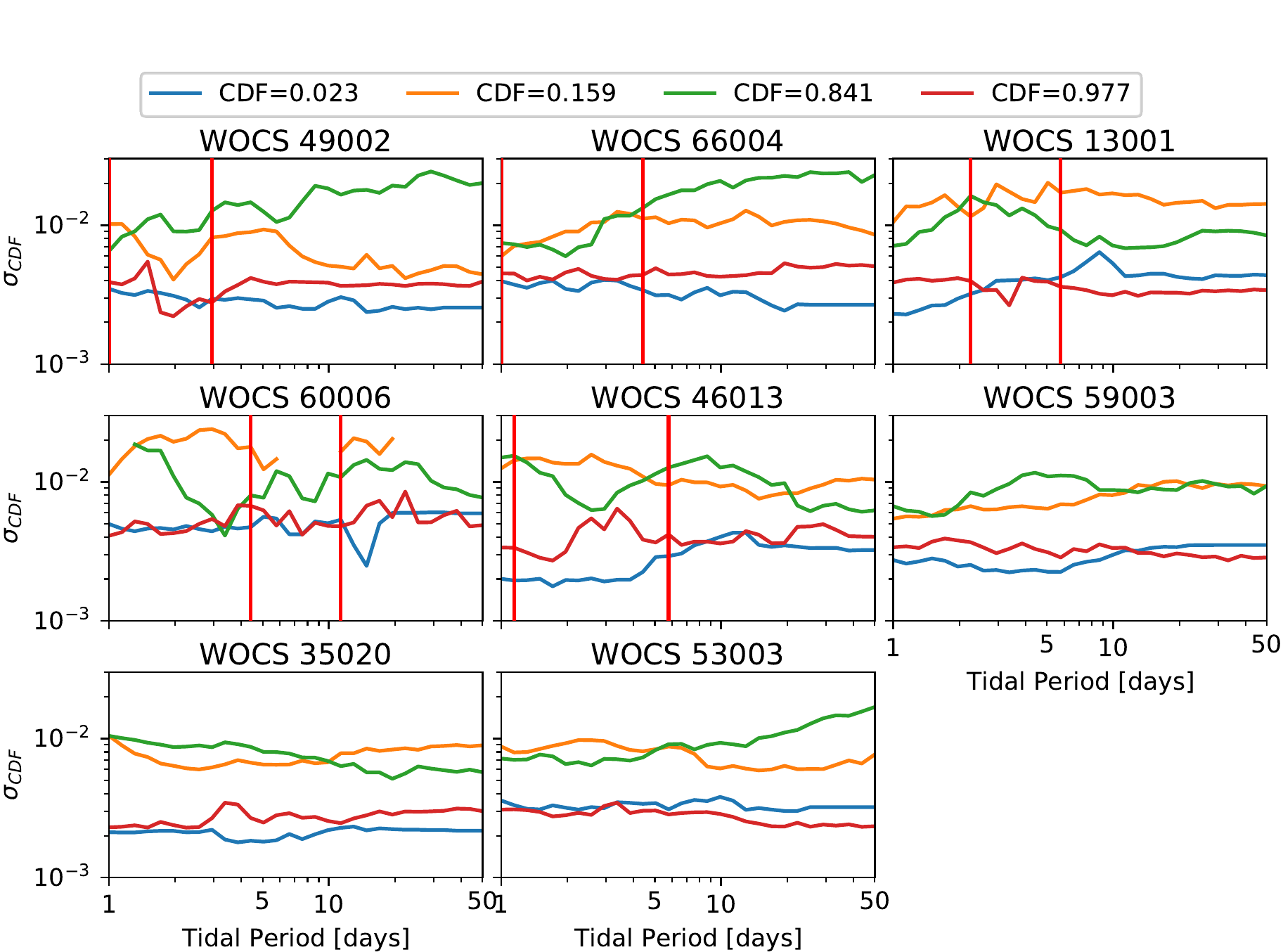}
    \caption{
        Same as Fig. \ref{fig:m35_individual_cdfstd} but for the NGC 6819
        binaries analyzed.
    }
    \label{fig:ngc6819_individual_cdfstd}
\end{figure*}

\begin{figure*}
    \includegraphics[width=\textwidth]{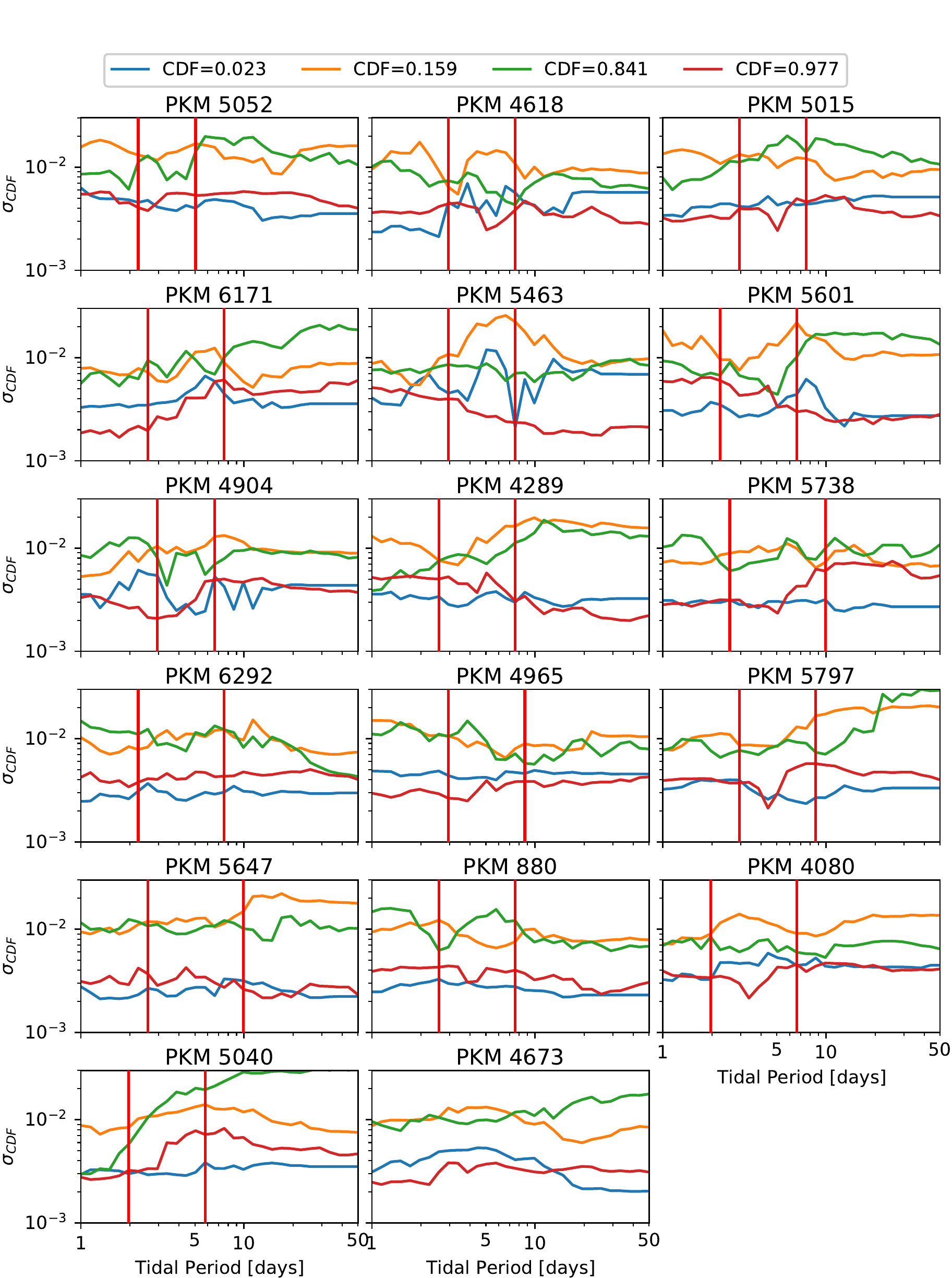}
    \caption{
        Same as Fig. \ref{fig:m35_individual_cdfstd} but for the NGC 188
        binaries analyzed.
    }
    \label{fig:ngc188_individual_cdfstd}
    \label{fig:individual_cdfstd_last}
\end{figure*}

Figures \ref{fig:burnin_first} --- \ref{fig:burnin_last} show the estimate of
the required burn-in period for the sampling to reach the equilibrium
distribution for a given quantile of $Q_\star'(P_{tide})$ (lines), calculated
per Section \ref{sec:convergence_diagnostics}. The black area of the plots show
the total number of samples actually generated for each system. Figures
\ref{fig:individual_cdfstd_first} --- \ref{fig:individual_cdfstd_last} show the
uncertainty (standard deviation) of the fraction of the posterior distribution
below each of the estimated quantiles based on samples after the corresponding
burn-in period. In each plot, the red-lines again delineate the range of tidal
periods where the dissipation is reliably constrained by observations.

We see from Fig. \ref{fig:burnin_first} --- \ref{fig:burnin_last} that all
chains contain sufficient samples to ensure that the 2.3\%, 15.9\%, 84.1\%, and
97.7\% quantiles of $\log_{10}Q_\star'$ have converged to their equilibrium
values for all tidal periods this analysis is sensitive to (between the red
vertical lines).  Furthermore, Fig.  \ref{fig:individual_cdfstd_first} ---
\ref{fig:individual_cdfstd_last} show that the estimated quantiles correspond to
the targeted CDF values to better than 10\% relative precision. For example, for
the M 35 binary WOCS 40015, Fig.  \ref{fig:m35_burnin} shows that about 200
steps are required for all four diagnostic quantiles to converge, while we
generated more than 1000 samples for that system. Similarly,
Fig.~\ref{fig:m35_individual_cdfstd} shows that for the diagnostic quantiles, in
the well-constrained range of tidal periods, the fraction of samples below the
corresponding estimated value is $(2.3\pm0.3)\%$, $(16\pm1)\%$, $(84\pm1)\%$,
and $97.7\pm0.4)\%$ respectively for the same binary.

\subsection{Combined Constraints}
\label{sec:combined_results}

\begin{figure}
    \includegraphics[width=\columnwidth]{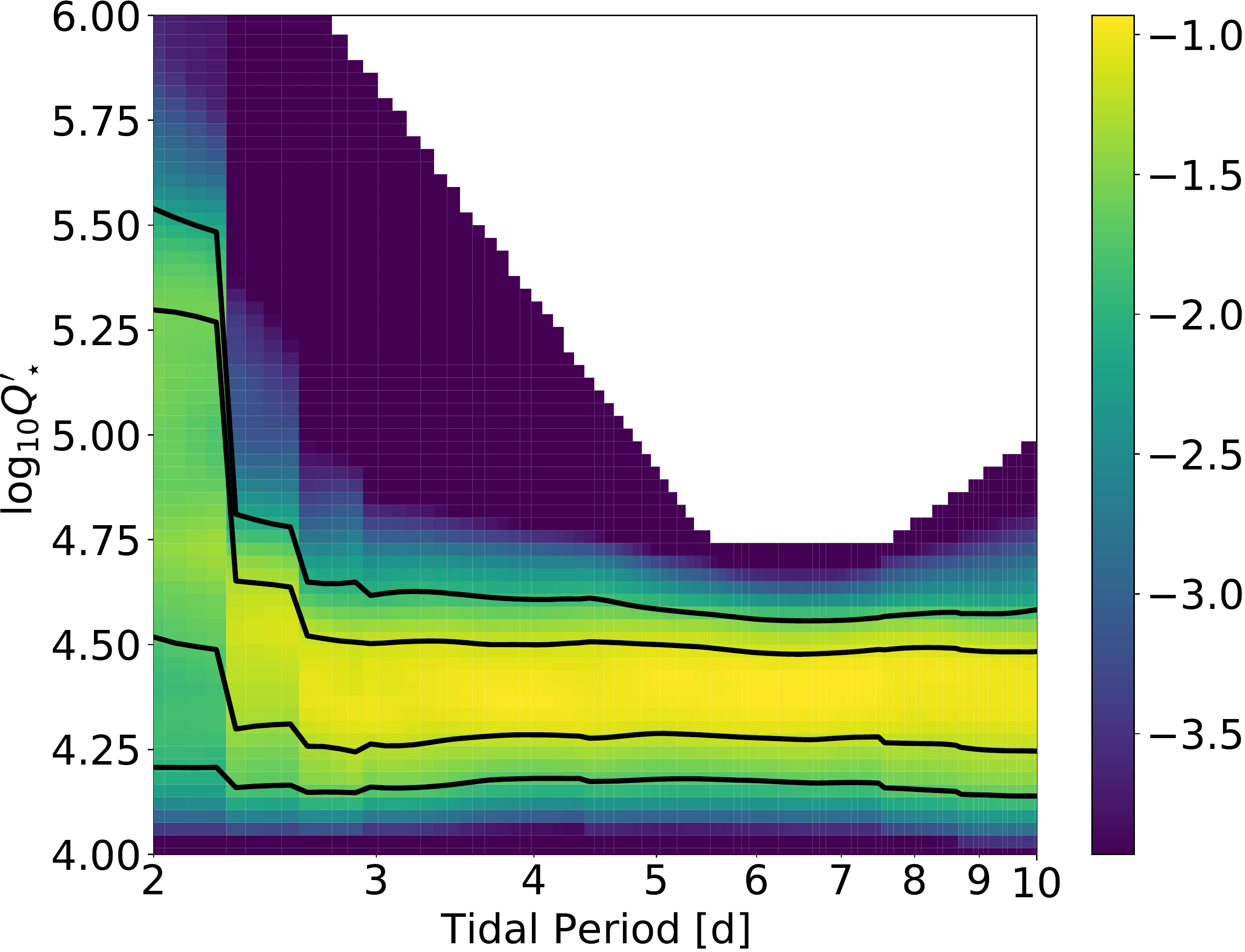}
    \caption{
        Combined constraint from all M 35 binaries constructed by multiplying
        together the individual constraints of Fig.
        \ref{fig:m35_individual_constraints} only in the regions between the two
        red lines for each system. The sudden jumps in period are due to
        contributions from various constraints sharply turning on and off
        at these period boundaries.
    }
    \label{fig:m35_combined_constraint}
\end{figure}

In order to find constraints that explain all binaries simultaneously, we
construct a combined probability density for the tidal dissipation in M 35 by
multiplying the posteriors of all 13 systems together, but only in the range we
identify as data dominated (i.e. between the red lines in Fig.
\ref{fig:first_individual_constraints} ---
\ref{fig:last_individual_constraints}). The result is shown in Fig.
\ref{fig:m35_combined_constraint}. Note that the y axis has been restricted to
$4<\log_{10}Q_\star'<6$, since higher values have negligible likelihood.

\begin{figure}
    \includegraphics[width=\columnwidth]{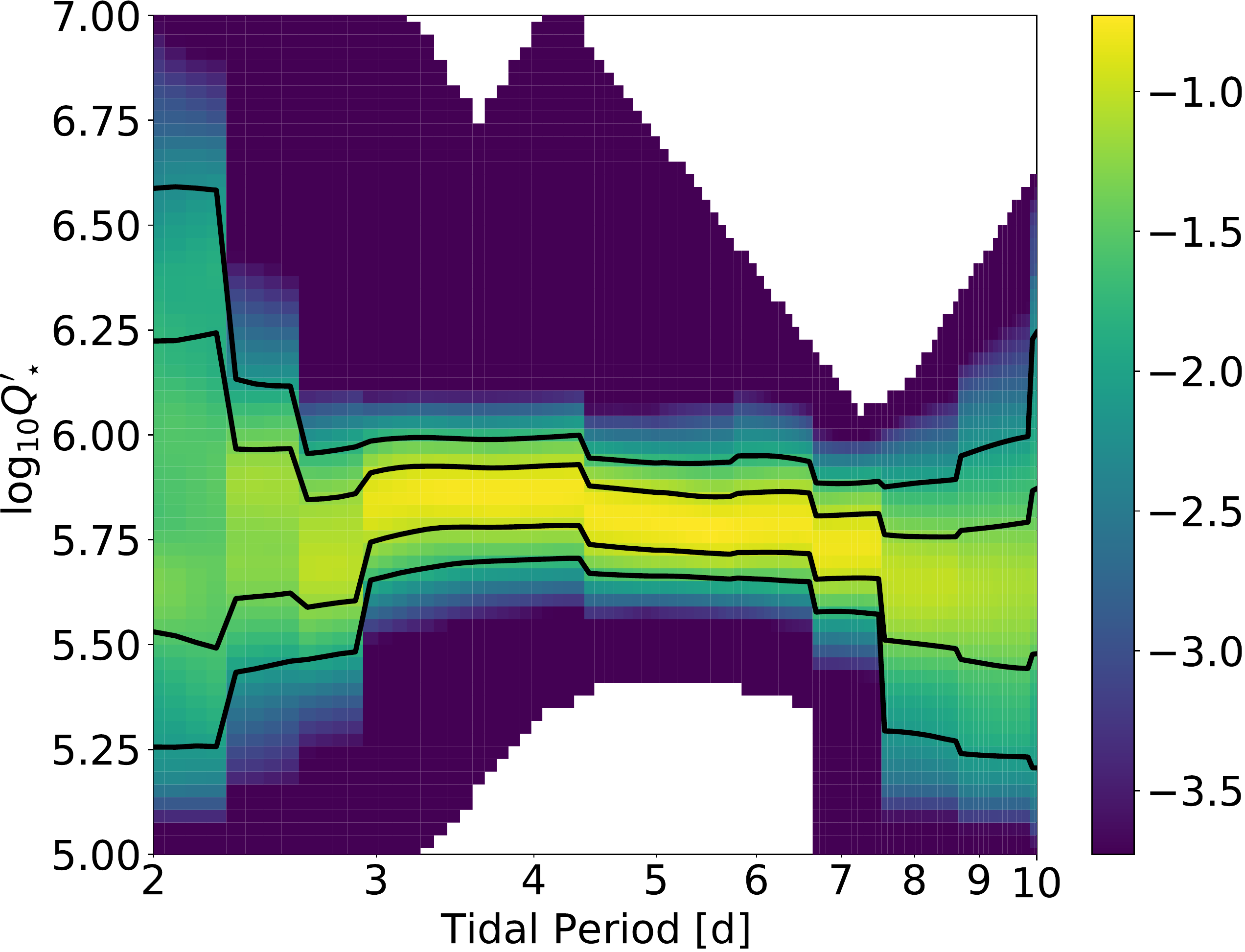}
    \includegraphics[width=\columnwidth]{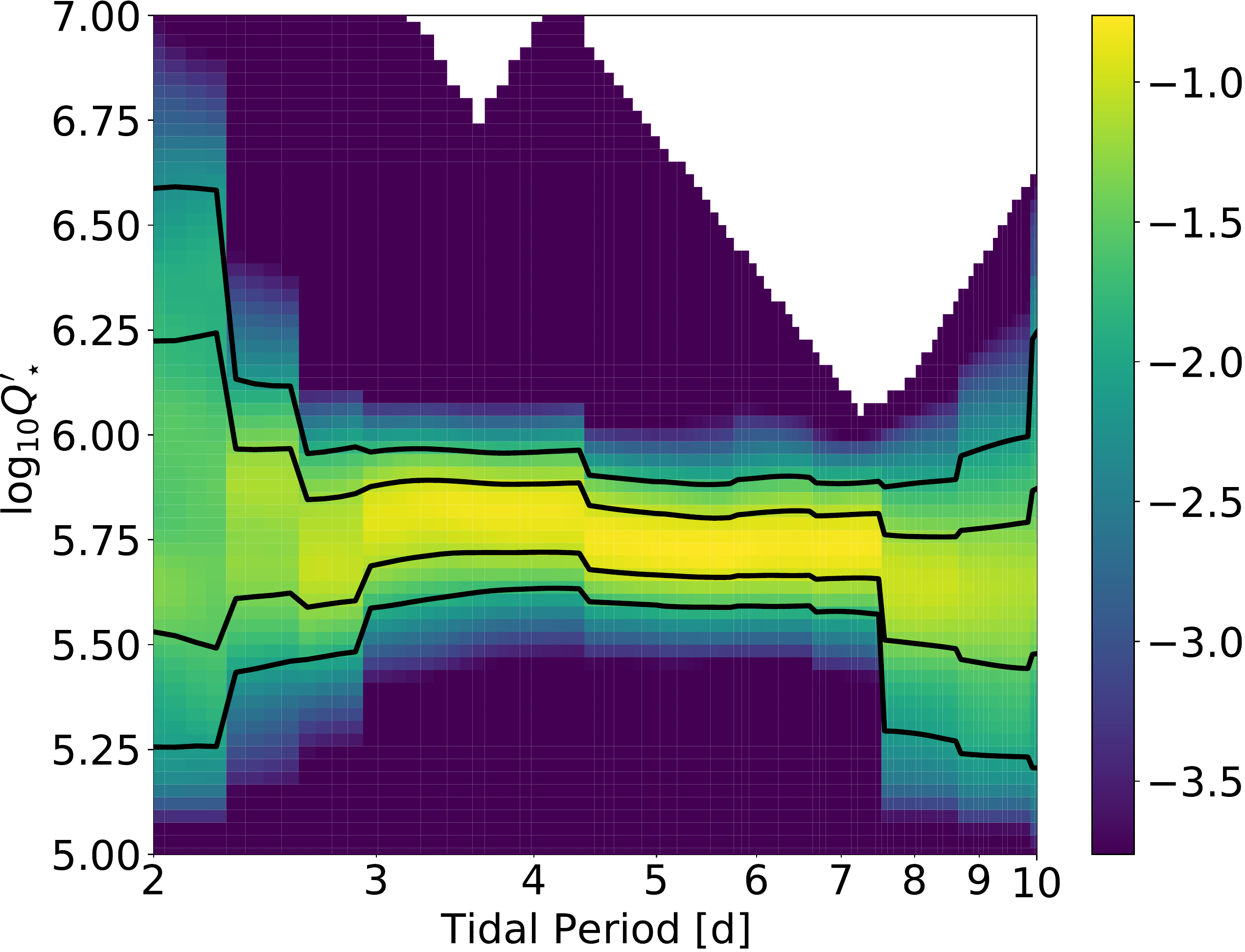}
    \caption{
        Top: Combined constraint from all NGC 6819 and NGC 188 binaries
        constructed by multiplying together the individual constraints of Fig.
        \ref{fig:m35_individual_constraints} only in the regions between the two
        red lines for each system. The sudden jumps in period are due to
        contributions from various constraints sharply turning on and off at
        these period boundaries. Bottom: same as top, but excluding NGC 188
        binary PKM 4904 due to its potentially suspicious location in the
        period-eccentricity diagram of the cluster.
    }
    \label{fig:ngc6819_ngc188_combined_constraint}
\end{figure}

Similarly, we combine all constraints of the 25 binaries in NGC 6819 and NGC 188
in Fig. \ref{fig:ngc6819_ngc188_combined_constraint}. In this case however, we
constructed two distribution, one with and one without the NGC 188 binary PKM
4904. As can be seen in the figure, the two constraints are virtually identical.
As noted in \ref{sec:ngc188_data}, PKM 5078 and PKM 4904 are in clearly
non-circular orbits but have orbital periods well into the range where all other
binaries in this cluster are circularized. As already discussed this could be an
indication that tides are not the only mechanism shaping the orbits of these
systems. PKM 5078 is a double line spectroscopic binary, and hence was already
excluded from our analysis (postponed to future publications). It was important
to check the possibility that PKM 4904 biases our results, but it does not
appear to be the case. Note that these binaries still affect the results by
shifting upward the period-eccentricity envelope used to define our posterior
likelihood function for the MCMC analysis. As discussed in
\ref{sec:ngc188_data}, raising (potentially erroneously) the envelope only
broadens the inferred posterior, while still including the constraints that would
be obtained had the envelope been defined without taking these binaries into
account.

\begin{figure}
    \includegraphics[width=\columnwidth]{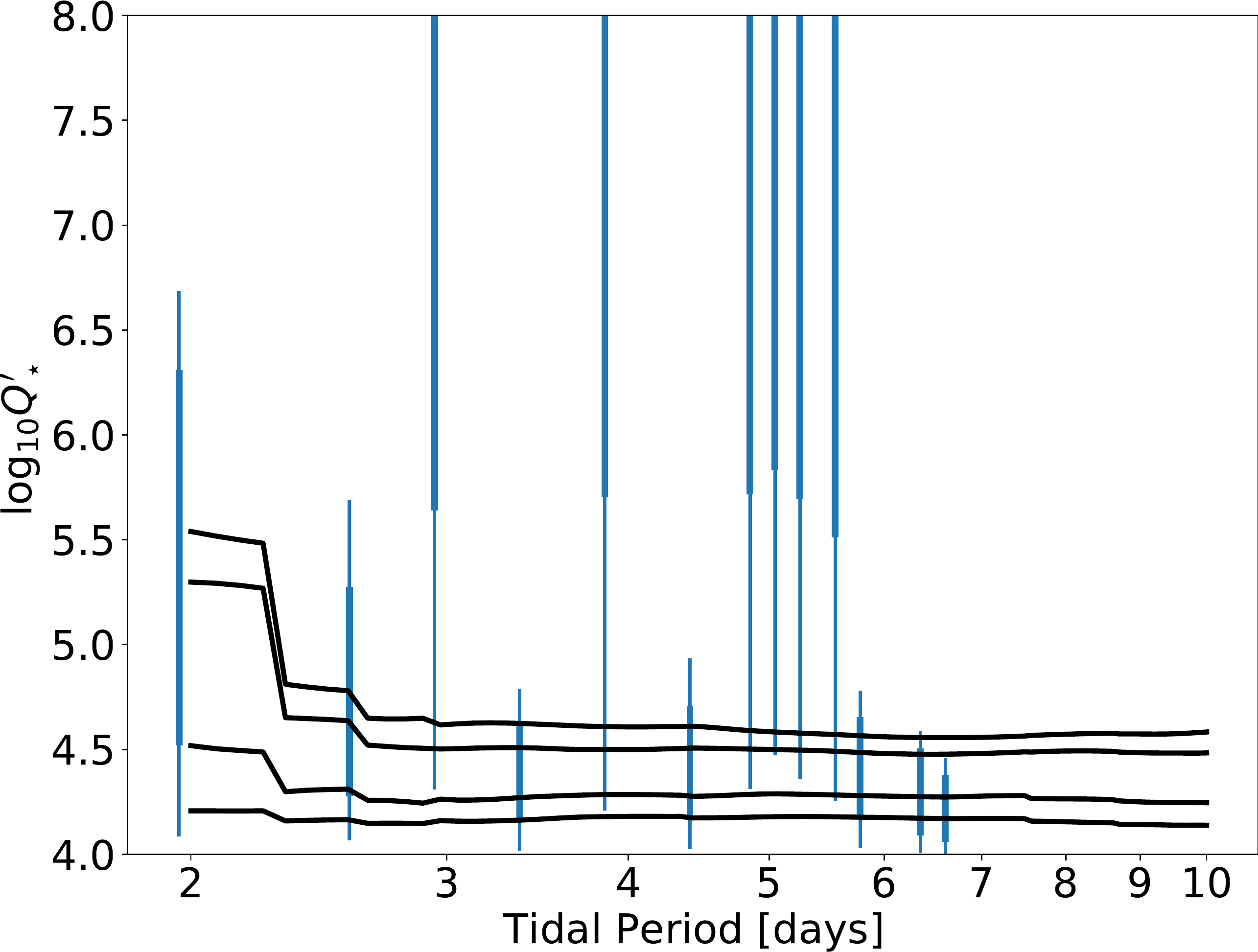}
    \caption{
        Comparison between the combined constraint of the tidal dissipation in M
        35 stars to the individual constraints for each binary. The lines show
        the 2.3\%, 15.9\%, 84.1\%, and 97.7\% quantiles of the combined
        constraint (same as in Fig. \ref{fig:m35_combined_constraint}), and the
        error bars show the same quantiles for individual systems at the tidal
        period where the dissipation is best constrained by observations. The
        thick part of the error bars spans from 15.9\% to 84.1\% of the
        posterior distribution and the thin error bar spans from 2.3\%, to
        97.7\%. The x coordinate of error bars extending past the top of the
        plot is not well defined.
    }
    \label{fig:m35_combined_individual_comparison}
\end{figure}

A reasonable question to ask of our combined constraints in Fig.
\ref{fig:m35_combined_constraint} and
\ref{fig:ngc6819_ngc188_combined_constraint} is whether they are able to
simultaneously reproduce the observed and envelope eccentricities of all
binaries included in this study. In other words, are the combined constraints in
agreement with the individual constraints of Sec. \ref{sec:individual_results}.
Fig. \ref{fig:m35_combined_individual_comparison} compares the combined
constraint from all M 35 binaries with the individual constraints used in
building the combined one. The curves in the figure show the same 1- and
2-$\sigma$ equivalent quantiles shown in Fig. \ref{fig:m35_combined_constraint}
and the error bars show the individual constraints for each system for the tidal
period with the lowest 97.7-th percentile (the bottom of the V in the individual
constraint plots). The thick/thin error bars shows the 1-/2-$\sigma$ equivalent
quantiles for each system. The error bars that extend past the top of the plot
correspond to long orbital period systems that provide no information on the
dissipation beyond our priors, so their x coordinates are not meaningful.

\begin{figure}
    \includegraphics[width=\columnwidth]{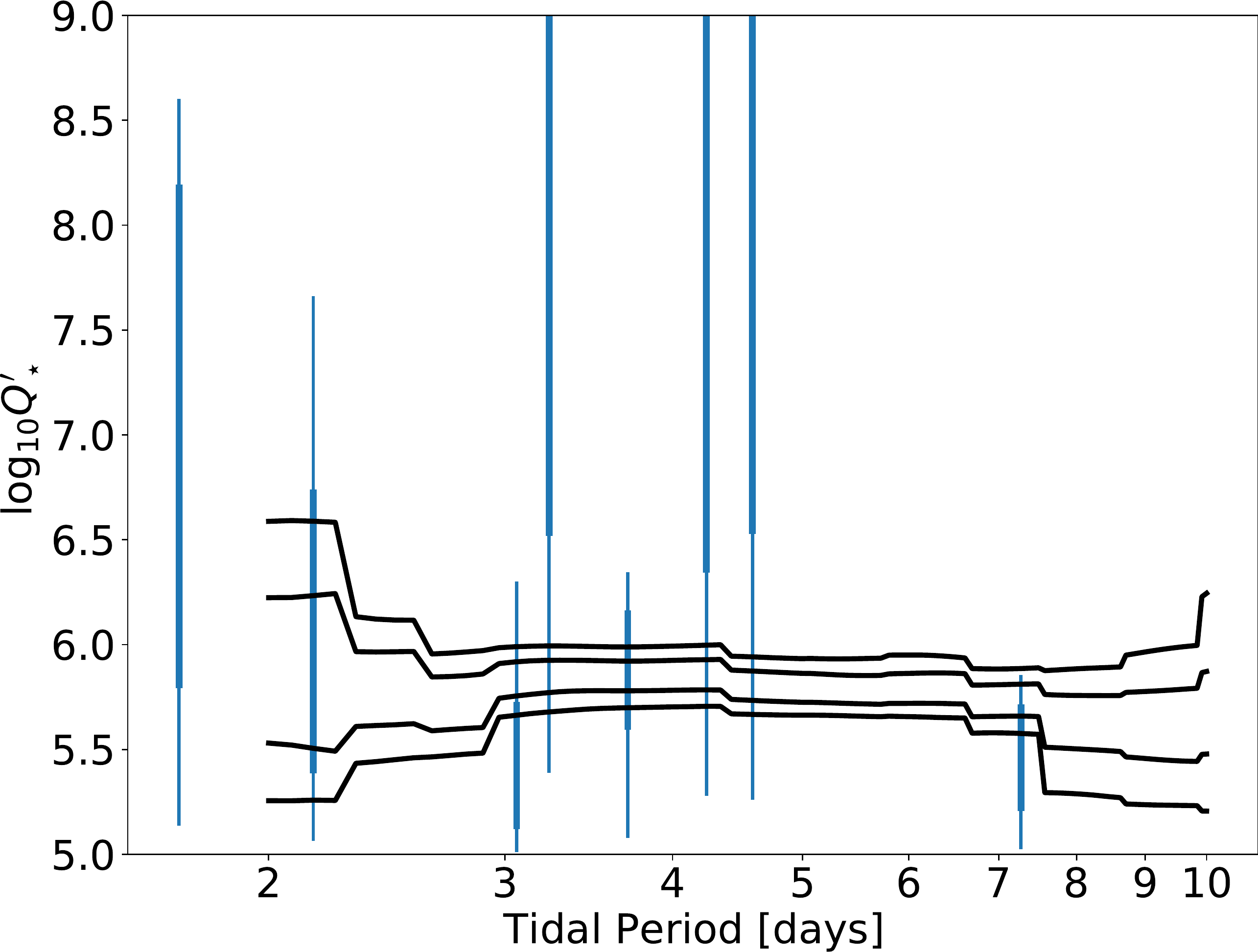}
    \includegraphics[width=\columnwidth]{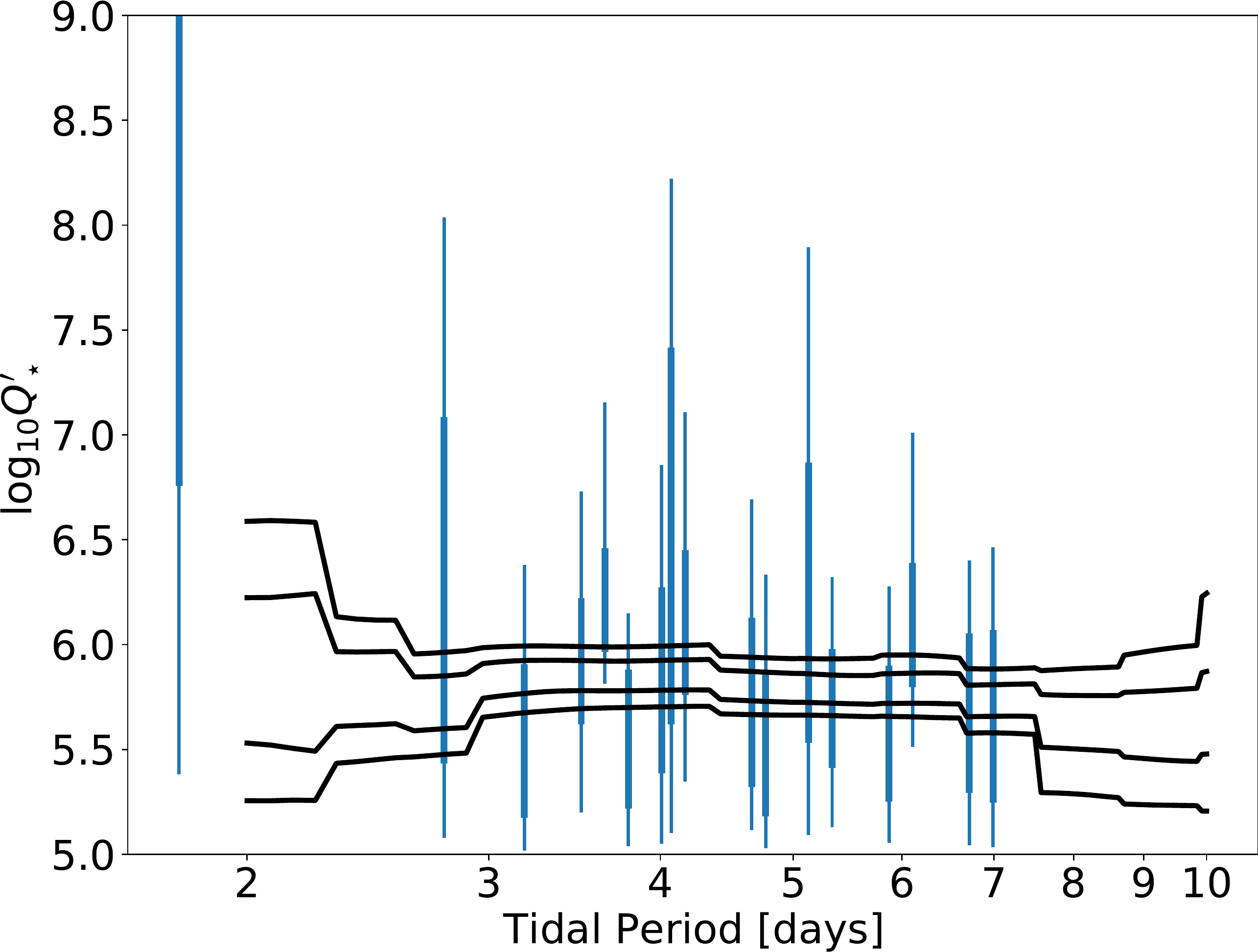}
    \caption{
        Comparison between the combined constraint of the tidal dissipation in
        NGC 6819 (top) and NGC 188 (bottom) stars to the individual constraints
        for each binary. The lines show the 2.3\%, 15.9\%, 84.1\%, and 97.7\%
        quantiles of the combined constraint for both clusters (same as in Fig.
        \ref{fig:ngc6819_ngc188_combined_constraint}), and the error bars show
        the same quantiles for individual systems at the tidal period where the
        dissipation is best constrained by observations. The thick part of the
        error bars spans from 15.9\% to 84.1\% of the posterior distribution and
        the thin error bar spans from 2.3\%, to 97.7\%. The x coordinate of
        error bars extending past the top of the plot is not well defined.
    }
    \label{fig:ngc6819_ngc188_combined_individual_comparison}
\end{figure}

Fig. \ref{fig:ngc6819_ngc188_combined_individual_comparison} shows the same
comparison between the individual constraints for NGC 6819 and NGC 188 binaries
against the combined constraint from both clusters. We have introduced small
offsets in tidal period for some of the points in order to avoid overlaps which
make the plot difficult to read.

The combined constraints are also included in Tables \ref{tab:m35_results} (M
35), \ref{tab:ngc6819_results} (NGC 6819), and \ref{tab:ngc188_results} (NGC
188) along the corresponding individual constraints.

\begin{table*}
    {
        \centering
        \caption{Comparison of the combined constraints based on all M\,35
        binaries with the individual constraints for each binary at the tidal
    period where 97.7-th percentile of the $log_{10}Q_\star'$ constraint is the
smallest.}
        \label{tab:m35_results} \label{tab:results_first}
        \begin{tabular}{c@{\quad\quad\quad}c@{\quad\quad\quad}cccc@{\quad\quad\quad}cc}
&&\multicolumn{4}{c@{\quad\quad\quad}}{\textbf{Individual CDF$^{-1}$}} & \multicolumn{2}{c}{\textbf{Combined CDF$^{-1}$}}\\
\multicolumn{1}{c@{\quad\quad\quad}}{\textbf{WOCS}} & \multicolumn{1}{c@{\quad\quad\quad}}{\textbf{Ptide}} & \multicolumn{1}{c@{}}{\textbf{2.3\%}} & \multicolumn{1}{c@{}}{\textbf{15.9\%}} & \multicolumn{1}{c@{}}{\textbf{84.1\%}} & \multicolumn{1}{c@{\quad\quad\quad}}{\textbf{97.7\%}} & \multicolumn{1}{c@{}}{\textbf{ 84.1\%}} & \multicolumn{1}{c@{}}{\textbf{ 97.7\%}} \\
\hline
\hline
54027 & 1.96 & 4.08 & 4.52 & 6.31 & 6.69 & --- & --- \\
16016 & 5.78 & 4.03 & 4.19 & 4.66 & 4.78 & 4.49 & 4.57 \\
23043 & 5.78 & 4.01 & 4.09 & 4.50 & 4.59 & 4.48 & 4.56 \\
54054 & 6.61 & 4.00 & 4.06 & 4.38 & 4.47 & 4.48 & 4.56 \\
35045 & 4.41 & 4.02 & 4.17 & 4.71 & 4.93 & 4.51 & 4.61 \\
40015 & 3.37 & 4.02 & 4.15 & 4.62 & 4.79 & 4.51 & 4.62 \\
33054 & 2.57 & 4.07 & 4.28 & 5.28 & 5.69 & 4.63 & 4.77 \\
59018 & 5.05 & 4.48 & 5.83 & 11.27 & 13.05 & 4.50 & 4.58 \\
14025 & 2.94 & 4.31 & 5.64 & 11.30 & 13.08 & 4.50 & 4.63 \\
24023 & 4.41 & 4.31 & 5.72 & 11.25 & 12.94 & 4.50 & 4.59 \\
9016 & 5.78 & 4.36 & 5.69 & 11.09 & 12.68 & 4.50 & 4.58 \\
37029 & 3.85 & 4.21 & 5.70 & 11.34 & 13.00 & 4.50 & 4.61 \\
20016 & 5.05 & 4.25 & 5.51 & 10.97 & 12.48 & 4.49 & 4.57 \\
\end{tabular}

    }
\end{table*}

\begin{table*}
    {
        \centering
        \caption{Comparison of the combined constraints based on all NGC 6819
            and NGC 188 binaries, except PKM 4904 with the individual
            constraints for each NGC 6819 binary at the tidal period where
            97.7-th percentile of the $log_{10}Q_\star'$ constraint is the
        smallest.}
        \label{tab:ngc6819_results}
        \begin{tabular}{c@{\quad\quad\quad}c@{\quad\quad\quad}cccc@{\quad\quad\quad}cccc}
&&\multicolumn{4}{c@{\quad\quad\quad}}{\textbf{Individual CDF$^{-1}$}} & \multicolumn{4}{c}{\textbf{Combined CDF$^{-1}$}}\\
\multicolumn{1}{c@{\quad\quad\quad}}{\textbf{WOCS}} & \multicolumn{1}{c@{\quad\quad\quad}}{\textbf{Ptide}} & \multicolumn{1}{c@{}}{\textbf{2.3\%}} & \multicolumn{1}{c@{}}{\textbf{15.9\%}} & \multicolumn{1}{c@{}}{\textbf{84.1\%}} & \multicolumn{1}{c@{\quad\quad\quad}}{\textbf{97.7\%}} & \multicolumn{1}{c@{}}{\textbf{ 2.3\%}} & \multicolumn{1}{c@{}}{\textbf{ 15.9\%}} & \multicolumn{1}{c@{}}{\textbf{ 84.1\%}} & \multicolumn{1}{c@{}}{\textbf{ 97.7\%}} \\
\hline
\hline
49002 & 1.72 & 5.14 & 5.79 & 8.19 & 8.60 & --- & --- & --- & --- \\
66004 & 1.96 & 5.06 & 5.39 & 6.74 & 7.66 & 5.26 & 5.51 & 6.23 & 6.59 \\
13001 & 3.37 & 5.08 & 5.60 & 6.16 & 6.35 & 5.70 & 5.78 & 5.92 & 5.99 \\
60006 & 6.61 & 5.02 & 5.21 & 5.72 & 5.85 & 5.58 & 5.66 & 5.81 & 5.89 \\
46013 & 3.37 & 5.01 & 5.12 & 5.73 & 6.30 & 5.66 & 5.76 & 5.92 & 5.99 \\
59003 & 2.94 & 5.39 & 6.52 & 11.40 & 12.63 & 5.68 & 5.77 & 5.93 & 5.99 \\
35020 & 5.05 & 5.26 & 6.53 & 11.47 & 13.08 & 5.67 & 5.73 & 5.87 & 5.94 \\
53003 & 3.85 & 5.28 & 6.34 & 11.18 & 12.87 & 5.71 & 5.78 & 5.93 & 6.00 \\
\end{tabular}

    }
\end{table*}

\begin{table*}
    {
        \centering
        \caption{Comparison of the combined constraints based on all NGC 6819
            and NGC 188 binaries, except PKM 4904 with the individual
            constraints for each NGC 188 binary at the tidal period where
            97.7-th percentile of the $log_{10}Q_\star'$ constraint is the
        smallest.}
        \label{tab:ngc188_results}
        \label{tab:results_last}
        \begin{tabular}{c@{\quad\quad\quad}c@{\quad\quad\quad}cccc@{\quad\quad\quad}cccc}
&&\multicolumn{4}{c@{\quad\quad\quad}}{\textbf{Individual CDF$^{-1}$}} & \multicolumn{4}{c}{\textbf{Combined CDF$^{-1}$}}\\
\multicolumn{1}{c@{\quad\quad\quad}}{\textbf{PKM}} & \multicolumn{1}{c@{\quad\quad\quad}}{\textbf{Ptide}} & \multicolumn{1}{c@{}}{\textbf{2.3\%}} & \multicolumn{1}{c@{}}{\textbf{15.9\%}} & \multicolumn{1}{c@{}}{\textbf{84.1\%}} & \multicolumn{1}{c@{\quad\quad\quad}}{\textbf{97.7\%}} & \multicolumn{1}{c@{}}{\textbf{ 2.3\%}} & \multicolumn{1}{c@{}}{\textbf{ 15.9\%}} & \multicolumn{1}{c@{}}{\textbf{ 84.1\%}} & \multicolumn{1}{c@{}}{\textbf{ 97.7\%}} \\
\hline
\hline
5052 & 3.37 & 5.10 & 5.62 & 7.42 & 8.22 & 5.70 & 5.78 & 5.93 & 5.99 \\
4618 & 5.05 & 5.51 & 5.80 & 6.39 & 7.01 & 5.66 & 5.72 & 5.86 & 5.95 \\
5015 & 4.41 & 5.05 & 5.39 & 6.27 & 6.86 & 5.70 & 5.78 & 5.93 & 5.99 \\
6171 & 4.41 & 5.13 & 5.41 & 5.98 & 6.32 & 5.66 & 5.72 & 5.86 & 5.93 \\
5463 & 5.05 & 5.35 & 5.76 & 6.45 & 7.11 & 5.70 & 5.78 & 5.93 & 6.00 \\
5601 & 3.85 & 5.20 & 5.62 & 6.22 & 6.73 & 5.69 & 5.78 & 5.92 & 5.99 \\
4904 & 4.41 & 5.81 & 5.97 & 6.46 & 7.15 & 5.70 & 5.78 & 5.92 & 5.99 \\
4289 & 3.85 & 5.12 & 5.32 & 6.13 & 6.69 & 5.67 & 5.73 & 5.87 & 5.94 \\
5738 & 5.78 & 5.03 & 5.25 & 6.07 & 6.46 & 5.58 & 5.66 & 5.81 & 5.88 \\
6292 & 4.41 & 5.06 & 5.25 & 5.90 & 6.28 & 5.66 & 5.72 & 5.86 & 5.95 \\
4965 & 5.05 & 5.04 & 5.29 & 6.05 & 6.40 & 5.58 & 5.66 & 5.81 & 5.89 \\
5797 & 5.78 & 5.03 & 5.18 & 5.88 & 6.33 & 5.66 & 5.73 & 5.87 & 5.94 \\
5647 & 5.05 & 5.04 & 5.22 & 5.88 & 6.15 & 5.70 & 5.78 & 5.92 & 5.99 \\
880 & 3.85 & 5.02 & 5.17 & 5.91 & 6.38 & 5.68 & 5.77 & 5.92 & 5.99 \\
4080 & 3.85 & 5.09 & 5.53 & 6.87 & 7.89 & 5.66 & 5.72 & 5.86 & 5.93 \\
5040 & 3.37 & 5.08 & 5.43 & 7.09 & 8.04 & 5.48 & 5.60 & 5.85 & 5.96 \\
4673 & 1.96 & 5.38 & 6.76 & 11.46 & 12.75 & --- & --- & --- & --- \\
\end{tabular}

    }
\end{table*}

As Fig. \ref{fig:m35_combined_individual_comparison} and
\ref{fig:ngc6819_ngc188_combined_individual_comparison} demonstrate, the
combined constraints are capable of reproducing the observed eccentricities for
all binaries analyzed, as well as the period-eccentricity envelopes of all three
clusters.

It is important to point out that the systems whose error bars stretch beyond
the top of the plots are simply consistent with a very broad range of $Q_\star'$
values, stretching over most of the prior range allowed with little preference
if any for some values of $Q_\star'$ over others. They should not be seen as
preferring a higher value of $Q_\star'$ than the combined constraint. This can
be seen from the corresponding plots in Fig.
\ref{fig:first_individual_constraints} ---
\ref{fig:last_individual_constraints}.  For example, the 6 systems that appear
as outliers in Fig.  \ref{fig:m35_combined_individual_comparison} are the 6
longest period systems we analyzed for M\,35 (those with WOCS identifiers 59018,
14025, 24023, 9016, 37029, 20016). As can be seen from
\ref{fig:m35_individual_constraints}, the $\log_{10}Q_\star'$ samples are very
close to uniformly distributed over the entire prior range, hence those systems
provide no useful information.  What is more, because the posteriors of these
systems are so broad, their x coordinates in the figures above are ill defined.
We only included those systems in the above plots for completeness.

Another conclusion that can be drawn from the combined constraints is that, at
the tidal periods probed by this analysis ($2\,d < P_{tide} < 10\,d$), there is
no evidence for frequency dependence  of $Q_\star'$, even though our model
explicitly allows for such a dependence. In fact, for sun-like stars on the main
sequence (NGC 6819 and NGC 188), we place tight limits on the dissipation of
$5.7 < \log_{10}Q_\star' < 6$ for $3\,d < P_{tide} < 7.5\,d$, implying that if
$Q_\star'\propto P_{tide}^\alpha$, then $|\alpha|<1$.  Note that for models,
like inertial wave coupling, predicting dramatic variations in $Q_\star'$ with
very small frequency changes, our result should be interpreted as constraints on
the appropriately frequency-smoothed dissipation predicted by such models.

For the young stars of M 35, we only measure an upper limit to $Q_\star' <
4\times10^4$ (lower limit on the dissipation), exceeding the dissipation of
main-sequence stars by an order of magnitude (see Appendix
    \ref{sec:disk_lifetime_effect} for a discussion of a small offset in the
constraint that's potentially introduced by ignoring the tidal evolution before
2\,Myrs).

\subsection{Assumptions and Caveats}

The results presented in this article depend on several critical assumptions
that bear consideration. First, our prescription for $Q_\star'$ only includes
frequency dependence, and we were able to split our sample into pre-MS dominated
and MS dominated circularization, even though age dependent $Q_\star'$ was not
explicitly included in our modeling. However, depending on the exact dissipation
mechanism, tidal frequency and age are far from the only factors that can affect
the dissipation.

For instance, some dynamical tide models invoke tidal excitation of g-modes at
the convective-radiative boundary, which then travel inward.
\citep{Barker_Ogilvie_10} find that if these waves achieve sufficient amplitude
to break, $Q_\star'\propto M_\star^2 P_{tide}^{8/3}$. Wave breaking does not
occur for weak tides, or if the star has a significant convective core
($M_\star\gtrsim 1.1M_\odot$) preventing waves from reaching the center.  For
weakly non-linear g-modes (non-breaking) \citep{Essick_Weinberg_16} argue the
non-linear excitation of (grand)daughter modes leads to $Q_\star'\propto
M_\star'^{0.5} P_{tide}^{2.4}$, where $M_\star'$ is the mass of the companion
star. Finally, if g-modes remain linear, \citep{Ma_Fuller_21} predict
$Q_\star'\propto M_\star^{-8/3} M_\star' P_{tide}^{-13/3}$.

Another flavor of theoretical models invoke tidal excitation of inertial modes
in the convective envelopes of stars, predicting dramatically increased
dissipation for low mass stars if $P_{spin} < 2 P_{tide}$.  Detecting this edge
at the correct location would be a very strong indication of this mechanism
dominating the dissipation. However, the angular momentum in the orbit is many
orders of magnitude larger than the angular momentum in the spin. Consequently,
tides cause stars to spin pseudo-synchronously on a much shorter timescale than
they circularize the orbit. As a result, tidal circularization occurs almost
exclusively with $P_{spin} < 2 P_{tide}$. Therefore, in the context of inertial
mode coupling, our results should be interpreted as measuring a frequency
smoothed version of the enhanced dissipation. Detecting inertial mode
enhancement thus requires studying the effects of tides on other system
properties, like tidal synchronization, or looking for signs of the expected
mass and/or spin dependence of the predicted enhanced dissipation.

Allowing for additional degrees of freedom in the tidal dissipation
prescription, like explicit age, spin, and amplitude dependence, will be
investigated in further analyses we are currently pursuing, which include much
larger samples of systems and hence have a chance of disentangling some of these
dependencies.

The g-mode class of models also point out a second caveat in our analysis. We
assumed that the tides couple to the surface convective zone of the stars in our
sample, while these models predict coupling to the radiative core. This
distinction makes a difference only if significant differential rotation can
develop between the core and the envelope. If stars spin like solid bodies, one
can always find an effective $Q_\star'$ regardless of which zone couples to
tides. Furthermore, since the orbit  dominates the angular momentum budget of
the system by many orders of magnitude, even large differential rotation will
make little difference for the results presented here.

An additional critical assumption we make for this analysis is that tides are
the only thing driving orbital evolution since early in the binary's history. It
is not uncommon to find triple and higher multiplicity systems, especially in
clusters. If additional, undetected, companions are present in the binaries we
analyze, they may excite the orbital eccentricity, causing us to under-estimate
the amount of tidal dissipation present. In fact, we already mentioned this
possibility as a possible explanation for the apparent outlier eccentricities
of NGC 188 members PKM 4904 and PKM 5078, and even excluded these binaries when
calculating the combined constraint in Fig.
\ref{fig:ngc6819_ngc188_combined_constraint}. One way such companions would show
up in the data is through deviations of the RV measurements from strict
periodicity (i.e. long term trends). None of the binaries included in our
sample show evidence of this. Alternatively, we would expect binaries strongly
affected by tertiary companions to produce anomalously large $Q_\star'$
constraints, which again we see no evidence of. However, neither of these
non-detections is a 100\% guarantee that companions are not present.

Another potential caveat to our approach was recently pointed out by
\citet{Zanazzi_22} who find that the relatively small number of binary systems
present in an individual cluster may miss important features of the
period-eccentricity distribution. In particular, these authors argue for the
presence of what they dub a ``cold core'' --- a population of binaries with
circular orbits out to much longer orbital periods than the envelope of the
period-eccentricity distribution. The relevant finding for this effort is that
for sample sizes of several dozen systems, as we have for each cluster, it is
not entirely unlikely that binaries close to the period-eccentricity envelope
are not present, thus systematically biasing the envelope down (or to the
right). This would in turn bias the upper limit of $Q_\star'$ we infer downward.
That said, the period-eccentricity envelope we use for NGC\,188 is consistent
with what \citet{Zanazzi_22} argue is the true envelope based on two large
samples of binaries --- one combining the binaries of all open clusters, and one
based on TESS and \textit{Kepler} eclipsing binaries. As can be seen from Fig.
\ref{fig:ngc188_individual_constraints} and
\ref{fig:ngc6819_ngc188_combined_individual_comparison} the upper limit on
$Q_\star'$ for the two older clusters is mostly determined by NGC\,188, and thus
not impacted by potentially misidentifying the envelope due to small sample
size. This is however a potentially important caveat to the M\,35 constraints.

\subsection{Comparison to Previous Empirical Studies of Stellar Tidal
Dissipation}

For main-sequence stars, our results agree with the previous analysis of open
cluster circularization \citep{Meibom_Mathieu_05, Milliman_et_al_14}, but
achieve higher precision, based only on a small fraction of the data, and
properly account for observational uncertainties and variable tidal dissipation.
Furthermore, the constraints \citet{Meibom_Mathieu_05} and
\citet{Milliman_et_al_14} extract for different clusters show significant mutual
disagreements \citep[see for example Fig. 12 of][]{Milliman_et_al_14}. In
contrast, we show that our combined constraints are consistent with each binary
of each cluster.

Recently, \citet{Justesen_Albrecht_21} showed that for stars with surface
convective zones, the purely pre-MS circularization predicted by the
\citet{Zahn_Bouchet_89} equilibrium tide prescription for $Q_\star'$ fits very
well with the observed eccentricities of a larger collection of approximately
equal mass eclipsing binaries with effective temperatures below 6250K. However,
for the sample of binaries used in that study, the ages were not well
constrained. As a result, it is entirely possible that the
\citep{Meibom_Mathieu_05} picture is just as consistent with the observations.

Our results are also within the wide error bars of the dissipation
that \citet{Jackson_Greenberg_Barnes_08a} find is required to explain the
decreased eccentricities of short period HJs relative to longer period ones.

\begin{figure}
    \includegraphics[width=\columnwidth]{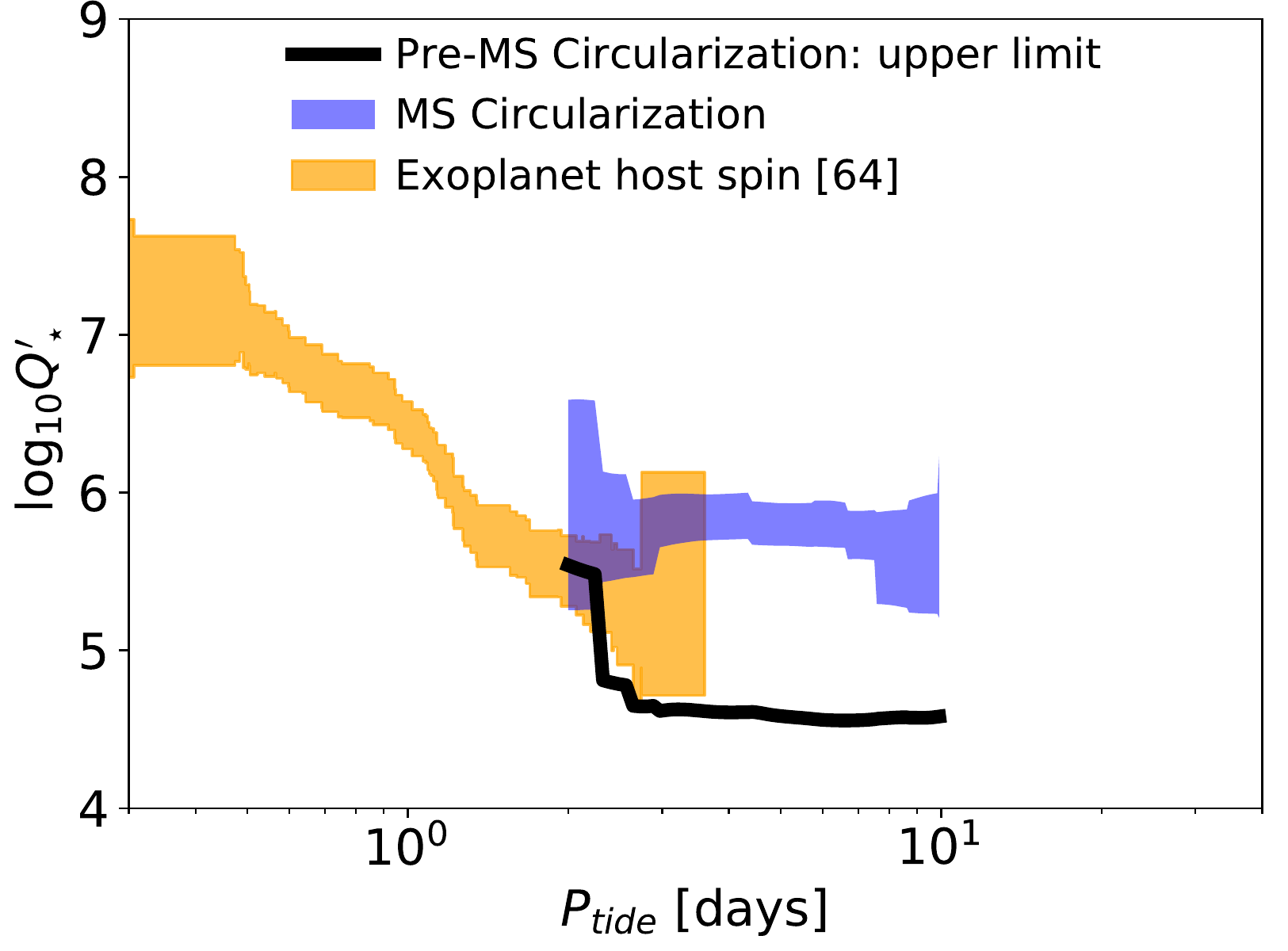}
    \caption{
        The constraints from this work and previous work from our group based on
        tidal spin-up of hot Jupiter host stars.
    }
    \label{fig:compare_to_hj_hosts}
\end{figure}

This result also smoothly extends the previous tidal dissipation calibration
from our group \citep{Penev_et_al_18}, based on tidal spin-up of hot Jupiter
(HJ) host stars. In that previous work, we found that in order to reproduce the
observed spins of HJ hosts, tidal dissipation must be quite small at short tidal
periods and grow rapidly with tidal period:
$Q_\star(P_{tide}=0.5\,d)'\gtrsim10^7$ and $Q_\star'\propto P_{tide}^{3.1}$. The
results presented here match well the dissipation found for HJ hosts at the
longest periods probed by \citet{Penev_et_al_18}, but seem to show tidal
dissipation saturating at periods of several days. The two constraints are
plotted together in Fig. \ref{fig:compare_to_hj_hosts}. The yellow area shows
the \citet{Penev_et_al_18} results, the blue area shows the main-sequence
dissipation found here, and the black line shows the 95\% upper limit of the
pre-main-sequence circularization found here. For \citet{Penev_et_al_18}, the
vertical extent at a tidal period $P$ was calculated by combining the
constraints from all systems with tidal periods in the range from $P/1.5$ to
$1.5\times P$.

Two additional extensive studies of tidal dissipation in exoplanet systems are
\citet{Hansen_10} and \citet{Bonomo_et_al_17}.

At first glance, the \citet{Hansen_10} results seems inconsistent with the
results presented here and by some of the other studies mentioned above. In
fact, the authors themselves point that out. However, the \citep{Hansen_10}
dissipation estimates include systems with hosts stars that lack significant
surface convective zones. In fact, one of the key systems which sets the stellar
dissipation in their analysis is HAT-P-2, whose host star has a surface
temperature of 6300K, slightly above the boundary where the surface convective
zone becomes negligible. Furthermore, \citet{Hansen_10} parametrize tides with a
single parameter that is not uniquely related to $Q_\star'$ and do not allow for
the possibility that their tidal parameter is different for different systems or
changes as systems evolve. This makes it impossible to isolate the dissipation
their analysis would require for stars with structure similar to the ones we
analyze here.

\citet{Bonomo_et_al_17} calibrate planetary and stellar tidal dissipation in HJ
systems based on a similar basic idea to this study. For stellar dissipation in
particular, they require that the circularization timescale for eccentric
systems must be larger than the system age, and apply that limit one system at a
time. They find lower limits on $Q_\star'$ varying from $10^9$ to under $10^5$,
correlated with the semimajor axis. The correlation with star-planet separation
simply reflects the degeneracy between $Q_\star'$ and semi-major axis when
predicting how effective tides are at changing the orbit. These authors also
make no effort to distinguish between stars with or without significant surface
convection. However, since the authors treated each system separately we are
able to check for that. Of the 8 shortest period systems found by
\citet{Bonomo_et_al_17} to require $Q_\star'>10^6$, half are hot (negligible
surface convection): WASP-18, WASP-14, XO-3, and HAT-P-14; and the other half
are cool (significant surface convection): HAT-P-16, HAT-P-20, WASP-10, and
WASP-89. Furthermore, except for WASP-89, these systems have small
eccentricities, meaning that the dominant tidal wave will have a period equal to
half the period with which the planet goes around the star in a reference frame
rotating with the star. Since these are not fast rotating stars, these periods
are several days, placing the 4 cool systems in direct conflict with the results
here and of some of the other studies mentioned above. One tempting explanation
is that this difference is real, perhaps due to the hugely different amplitudes
of the tidal waves a companion star would raise compared to a companion planet.
To really compare the two results, however, one would need to properly model the
entire evolution of these star-planet systems, perhaps following a similar
approach to what we did here.

At first glance, our results appear to be inconsistent with
\citet{Collier_Cameron_Jardine_18}, who found that significantly less
dissipation than we find is required to explain the observed upper boundary of
the HJ population in the mass ratio --- orbital separation diagram. In
particular, they find that for systems expected to experience enhanced
dissipation due to inertial waves (all our systems are in this regime)
$\log_{10}Q_\star'=7.3\pm0.4$ is required for tidal inspiral to produce the
observed edge in the distribution. However, as is evident from their Fig. 3,
signatures of tidal inspiral are only detectable for orbital semimajor axes less
than 3 times the Roche limit. For a Jupiter mass planet around a solar mass star
this corresponds to orbital periods less than 2.5\,days or tidal periods less
than $\sim2$\,days, even allowing for a correction due to the stellar spin:
$P_{tide}=0.5 \left(P_{orb}^{-1} - P_\star^{-1}\right)^{-1}$ ($P_\star$ is the
spin period of the star). Since the majority of the signal in the
\citet{Collier_Cameron_Jardine_18} study comes at even shorter periods, their
constraints are sensitive to the tidal dissipation below the frequnecy range
probed here. These would be more appropriately compared to the
\citet{Penev_et_al_18} results.  Since the authors did not test for, or allow
for frequnecy dependent dissipation, it is entirely possible that the two
constraints are consistent. A reliable test would require repeating the
\citet{Collier_Cameron_Jardine_18} analysis with a more flexible prescription
for $Q_\star'$.

\section{Conclusions}
\label{sec:conclusions}

We used available measurements of the orbital eccentricities of 38 single lined
spectroscopic binaries from 3 different open clusters to infer the tidal
dissipation required to simultaneously explain the observed period-eccentricity
envelope in those clusters, while at the same time allowing at least the
measured present day eccentricity of each system to be reproduced for
reasonable choices of initial conditions.

Our analysis allowed for, but did not find evidence for powerlaw frequency
dependence of the tidal dissipation; in fact, for any tidal period between 2
and 10 days, the required tidal quality factor for main-sequence stars was found
to satisfy $5.5 < \log_{10}Q_\star' < 6$, with the majority of the range
constrained significantly tighter than that.  However, we find significantly
higher, by more than an order of magnitude, dissipation for binaries in the
cluster for which tidal circularization predominantly happens during the
pre-main-sequence, compared to the two clusters dominated by main-sequence
circularization.

We used Bayesian analysis to properly account for observational uncertainties,
and a flexible tidal model to allow for frequency dependent tidal dissipation.
Our orbital evolution simulations accounted for evolving stellar structure, loss
of angular momentum to stellar winds, and the internal redistribution of angular
momentum between the surface and the interior of stars. Furthermore, we properly
simulated tides for stars in eccentric orbits, by splitting the tidal potential
into its Fourier components, and allowing for separate dissipation for each
tidal wave under a common frequency dependent prescription. The only studies
that account for all of these complexities \citep[e.g.][]{Bolmont_Mathis_16,
Benbakoura_et_al_19} assume a particular theoretical model for the dissipation
and only treat circular orbits with no spin-orbit misalignment, making them
incapable of tackling the analysis presented here and preventing comparison with
other tidal dissipation models than the ones assumed.  The improved modeling
used here resolves the internal inconsistencies of previous constraints based on
these same binaries \citep{Meibom_Mathieu_05, Milliman_et_al_14}. For example,
the circularization period \citet{Milliman_et_al_14} found for NGC\,6819 was
well below the claimed evolution of circularization with age, below that of even
younger clusters. In the \citet{Meibom_Mathieu_05} interpretation that would
suggest this cluster has undergone less circularization than the others, hence
requiring larger $Q_\star'$.  This should have been particularly pronounced when
comparing to NGC\,188, since that is the cluster with largest leverage in
defining the trend of increasing circularization period on the MS. As we show
here, once the particular characteristics of the individual binaries are
properly accounted for, the same tidal dissipation is able to explain both
NGC\,188 and NGC\,6819 circularization.

\section*{Acknowledgements}

This research was supported by NASA grant 80NSSC18K1009. The authors acknowledge
the Texas Advanced Computing Center (TACC) at The University of Texas at Austin
for providing high performance computing resources that have contributed to the
research results reported within this paper. \url{http://www.tacc.utexas.edu}.
The authors also acknowledge use of additional high performance computing
resources provided by the Ganymede HPC cluster at the University of Texas at
Dallas. The authors would like to thank Dr. Adrian Barker for providing
constructive and detailed feedback which significantly improved this
manuscript.

\section*{Data Availability}
\label{sec:data_availability}

All tables of input parameters presented in this article are derivative of
published tables already available in machine readable form through the VizieR
online service. They are reproducible by applying simple filtering to those
tables, and thus do now warrant creating and publishing independent copies.

We have created a zenodo archive \citep{Penev_Schussler_22b} to accompany this
article, which provides machine readable tables of the following:

\begin{itemize}
    \item The generated MCMC samples for each binary in the original HDF5 format
        produced by the emcee package \citep{Foreman-Mackey_et_al_13} (see
        \url{https://emcee.readthedocs.io/en/stable/user/backends/}).
    \item The 2.3\%, 15.9\%, 84.1\%, and 97.7\% quantiles of $\log_{10}Q_\star'$
        for the given binary as a function of tidal period for each binary.
        (i.e. the coordinates defining the quantile curves in
            Fig.~\ref{fig:first_individual_constraints} ---
        \ref{fig:last_individual_constraints}).
    \item The burn-in period for each quantile vs tidal period (i.e. the
        coordinates defining the burn-in curves in Fig.~\ref{fig:burnin_first}
    --- \ref{fig:burnin_last})
    \item The estimated standard deviation of the fraction of samples below each
        of the 4 target quantiles for each binary (i.e. the coordinates defining
        the curves in Fig.~\ref{fig:individual_cdfstd_first} ---
    \ref{fig:individual_cdfstd_last})
    \item The 84.1\%, and 97.7\% quantiles of the combined $\log_{10}Q_\star'$
        constraint from all M 35 binaries vs tidal period. Note that the lower
        quantiles are not constrained by the data  but rather come from the
        assumed priors for the $log_{10}Q_\star'$ model parameters.
    \item The 2.3\%, 15.9\%, 84.1\%, and 97.7\% quantiles of the combined
        $\log_{10}Q_\star'$ constraint from all NGC 6819 and NGC 188 binaries vs
        tidal period, excluding NGC 188 binary PKM 4904 (i.e. the coordinates of
            the quantile curves in the bottom panel of
        Fig.~\ref{fig:ngc6819_ngc188_combined_constraint}).
\end{itemize}

The version of POET used for calculating the orbital evolution is available
through zenodo archive \citet{Penev_Schussler_22a}.



\bibliographystyle{mnras}
\bibliography{bibliography} % if your bibtex file is called example.bib

%%%%%%%%%%%%%%%%%%%%%%%%%%%%%%%%%%%%%%%%%%%%%%%%%%

%%%%%%%%%%%%%%%%% APPENDICES %%%%%%%%%%%%%%%%%%%%%

\appendix

\section{Tidal Locking}

Numerical expediency forces us to introduce one more wrinkle to the tidal
dissipation model described in Section \ref{sec:tidal_dissipation_model}. In
particular, \poet{} is designed to efficiently handle discontinuities in the
frequency dependence of $\Delta_{m,m'}$ (or $Q'_{m,m'}$) when $\Omega_{m,m'}=0$.
This occurs if $\Delta_{m,m'}$ does not approach zero as $\Omega_{m,m'}
\rightarrow 0$. In such cases, when a particular tidal term reaches zero
frequency during the orbital evolution, it may happen that the tidal evolution
equations, assuming $\Omega_{m,m'}$ approaches zero from above and below,
predict that the time derivative $\dot{\Omega}_{m,m'}$ is negative and positive
respectively. In that case, at least for some time, the spin of the zone will
be locked in some integer ratio with the orbital frequency. If that situation is
detected, \poet{} eliminates the spin of the zone as an independently evolved
variable.  This allows the subsequent evolution to continue taking large time
steps without losing accuracy or precision, but with a check of whether the lock
is maintained at each subsequent step.

In the case of $\alpha<0$ in Eq. \ref{eq:Q_prescription}, such locking behavior
is frequent, and the described scheme dramatically speeds up the calculations.
When $\alpha>0$, exact spin-orbit locks don't occur.  However, it still
frequently happens that during the evolution stable equilibrium points are
encountered when some tidal term ends up with very small frequency
($\Omega_{m,m'} = m\Omega_\star - m'\Omega_{orb} \ll \Omega_{orb}$). Because the
equilibrium tidal frequency evolves as the orbit evolves, it is not possible to
handle these cases by eliminating one of the variables. If nothing is done in
those cases however, maintaining the desired precision in calculating the
evolution results in impractically small time steps. In order to avoid such
extremely time consuming calculations, we tweak the tidal prescription slightly
to ensure $\Delta_{m,m'}$ remains non-zero (though small) as
$\Omega_{m,m'} \rightarrow 0$ no matter the value of $\alpha$. To be precise,
when $\alpha<0$ we use Eq. \ref{eq:Q_prescription}, however when $\alpha>0$ we
instead use:

\begin{equation}
    Q'_{m,m'} = Q_0
    \begin{cases}
        \left(\frac{P_{max}}{P_0}\right)^\alpha
        & \text{if } P_{m,m'}>P_{max}
        \\
        \left(\frac{P_{m,m'}}{P_0}\right)^\alpha
        & \text{if } P_{max} > P_{m,m'} > P_0
        \\
        1 & \text{if } P_{m,m'} < P_0
    \end{cases}
    \label{eq:Q_prescription_alt}
\end{equation}

\begin{figure}

    \includegraphics[width=\columnwidth]{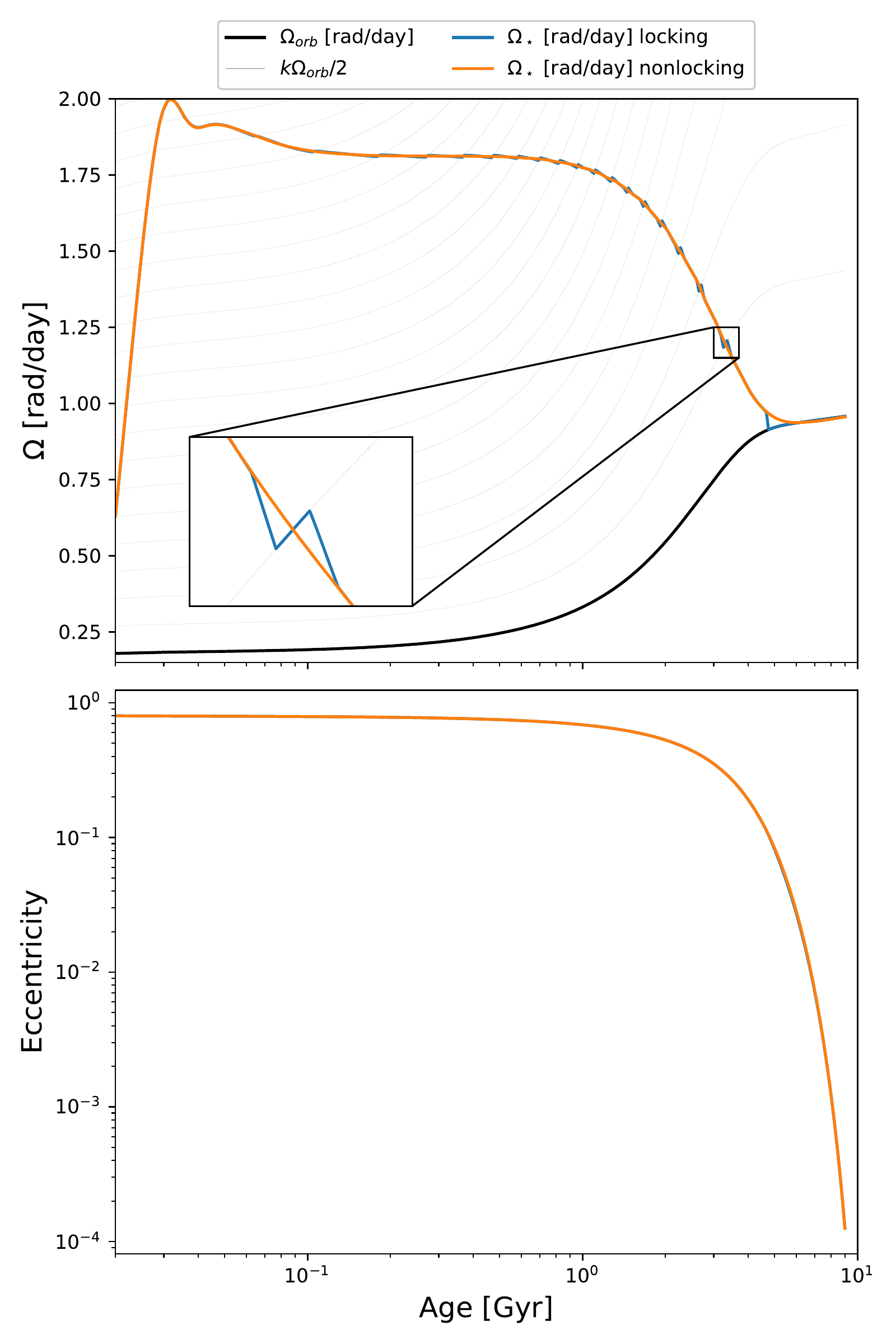}

    \caption{
        Top: Comparison of the orbital and stellar convective zone spin
        frequency evolutions calculated using Eq. \ref{eq:Q_prescription}
        (non-locking) and Eq. \ref{eq:Q_prescription_alt} (locking) for a binary
        of $1\,M_\odot$ + $0.7\,M_\odot$ stars, assuming $Q_0=10^6$,
        $P_0=5$\,days, and $\alpha=1$. The thin grey lines indicate possible
        spin-orbit locks (i.e.  $2 \Omega_\star = k \Omega_{orb}$ for some
            integer $k$. The inset zooms in near one such lock to show the
            effect of Eq.  \ref{eq:Q_prescription_alt}.
            Bottom: The associated eccentricity evolution. Evolution with both
            assumptions of tidal dissipation prescription are plotted but are
            indistinguishable in the plot.
        }

    \label{fig:forced_locking_demo}
\end{figure}

The value of $P_{max}$ ($P_{max}=50$\,days) was chosen to ensure that the
calculated eccentricity and semimajor axis evolution are indistinguishable from
those produced under Eq. \ref{eq:Q_prescription}, except that in some
cases calculations are between 3 and 4 orders of magnitude faster using the
prescription above. The effect of this tweak is shown in Fig.
\ref{fig:forced_locking_demo}, where we compare the evolution of a fiducial
binary calculated assuming exactly Eq. \ref{eq:Q_prescription} and the modified
prescription of Eq. \ref{eq:Q_prescription_alt}. As the inset in the figure
shows, the only effect is to add brief periods of spin-orbit locks each time the
system evolves through one of these co-rotation resonances.

\section{Convergence Diagnostics/Burn-in}
\label{sec:convergence_derivation}

We wish to determine how many MCMC steps must be taken in order to reliably
estimate a given quantity. This question will have a different answer depending
on the exact quantity being estimated. In this case, we wish to know the 2.3\%,
15.9\%, 84.1\%, 97.7\% percentiles of the distribution. We must determine how
many steps are required for the starting positions to be forgotten, and how many
further steps are required to get precise estimates of these quantiles.

We begin by selecting a set of 30 diagnostic tidal periods ($P_p$), distributed
uniformly in log-space between 1 and 50 days, and quantiles levels ($q_q \in
\{2.3\%, 15.9\%, 84.1\%, 97.7\%\}$), matching 1- and 2-$\sigma$ quantiles of the
normal distribution.  Then for each sample in the MCMC chain for a particular
binary system, we evaluate the tidal model (Eq.  \ref{eq:Q_prescription}) at
each of those periods, getting a tidal period specific chain of
$\log_{10}Q_\star'$ values. Let $x_{p,w,t}$ ($t \in \left\{1\ldots T\right\}$,
$w \in \left\{1 \ldots W\right\}$) be the value of $\log_{10}Q_\star'(P_p)$ for
the $t$-th steps of the $w$-th walker.

Using kernel density estimation (with Epanechnikov kernel) we estimate
the quantile $u_{p,q}$ of $\log_{10}Q_\star'$ at each of the $q_q$ levels for
each of the $P_p$ tidal periods. Our goal is to find a burn-in period for
$u_{p,q}$, as well as an estimate of the variance of the estimated cumulative
distribution of $\log_{10}Q_\star'$ evaluated at $u_{p,q}$.

For each diagnostic tidal period and quantile, we begin by constructing a
discrete sequence of the number of walkers at each step below a given quantile:
\begin{equation}
    z_{p,q,t}
    =
    \sum_{w}
    \begin{cases}
        1 & \mathrm{if}\ x_{p,w,t} < u_{p,q}\\
        0 & \mathrm{otherwise}
    \end{cases}
\end{equation}

As \citet{Raftery_Lewis_91} point out, while $z_{p,q,t}$ is not necessarily a
Markov process, the sufficiently thinned chain $\{z_{p,q,1+k(t-1)}\}$ will be
well approximated by one if $k$ is large enough. To find $k$
\citet{Raftery_Lewis_91} suggest using a Bayesian information criterion to
determine the smallest value of $k$ at which a first order Markov process is
preferred over a second order one, i.e. the first $k$ for which:
\begin{equation}
    \log \frac{ML_2(\{z_{p,q,1+k(t-1)}\})}{ML_1(\{z_{p,q,1+k(t-1)}\})}
    -
    s (s-1)^2 \log\lfloor \frac{T}{k} \rfloor < 0
\end{equation}
where $ML_1$ and $ML_2$ are the maximum likelihoods fitting
$\{z_{p,q,1+k(t-1)}\}$ with a first and second order Markov processes
respectively, $s$ is the number of distinct values present in $\{z_{p,q,t}\}$
(i.e. number of states the approximate Markov process samples over), and $T$ is
the number of steps in the original emcee chain.

Given the matrix of maximum likelihood estimates ($P_{i,j}$) of the transition
probabilities from state $i$ to state $j$ of the first order Markov process over
the thinned chain, we then find the equilibrium distribution $\pi_i$ by solving
$P \pi = \pi$. The burn-in period is then $M k$, where $M$ is the smallest
number such that the probability of being in each of the $s$ accessible states
after $M$ Markov steps is within some tolerance ($\varepsilon$) of the
equilibrium distribution, for all initial states:

\begin{equation}
    (P^M)_{i,j} - \pi_j \leq \varepsilon
\end{equation}

We choose $\varepsilon=10^{-3}$ for this study.

Finally, in order to estimate the variance in the value of the cumulative
distribution at $u_{p,q}$, we discard the burn-in samples found above and
generate $10^5$ independent realizations of the best-fit Markov process on
$\{z_{p,q,1+k(M + t-1)}\}$, and calculate their sample variance.

\begin{figure}
    \includegraphics[width=\columnwidth]{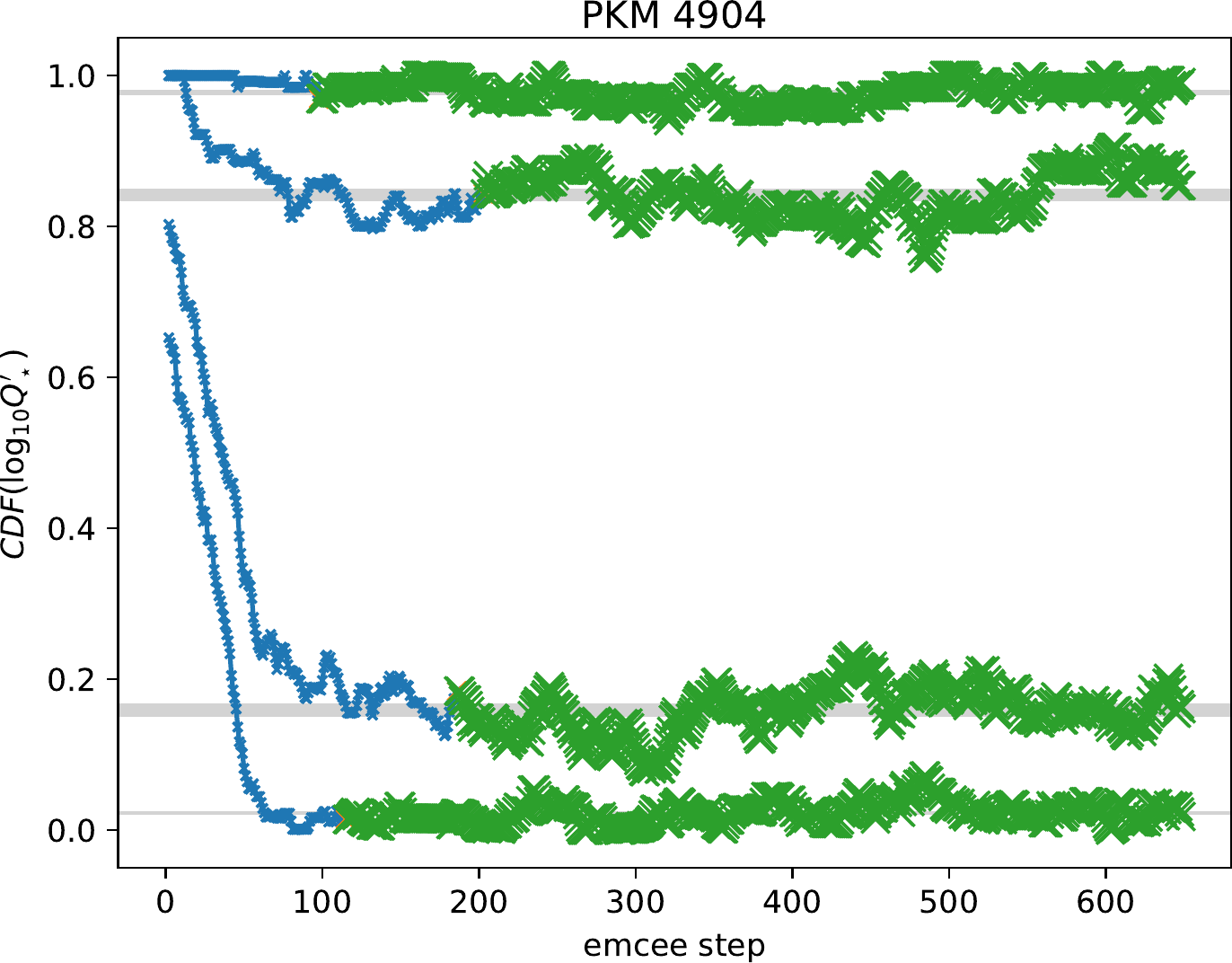}
    \caption{
        Demonstration of the convergence estimation algorithm. The sequence of
        ``x'' symbols shows the fraction of samples below the estimated value of
        each of our target quantiles vs the emcee step number. The smaller blue
        samples are from the estimated burn-in period and the larger green
        crosses are the samples we use in determining the constraints. The grey
        areas show the estimated standard deviation of the CDF around the target
        value (2.3\%, 15.9\%, 84.1\%, 97.7\% respectively) computed from the
        post burn-in samples.
    }
    \label{fig:quantile_convergence}
\end{figure}

Fig. \ref{fig:quantile_convergence} shows this algorithm applied to the binary
that showed the largest drift of the estimated quantiles: NGC 188 binary PKM
4904. We see the distribution narrowing dramatically for the first 100 or so
samples (the fraction of the distribution between the upper and lower boundary
becomes much larger). The 15.9\% percentile exhibits the largest
auto-correlation and as a result the estimated burn-in period is the longest.
The grey areas also show the estimated uncertainties in the values of the
cumulative distribution function (the middles of these areas are the targeted
quantiles) based on the post-burn-in samples.

\section{Effect of Binary Evolution Starting Age}
\label{sec:disk_lifetime_effect}

\begin{figure*}
    \includegraphics[width=\columnwidth]{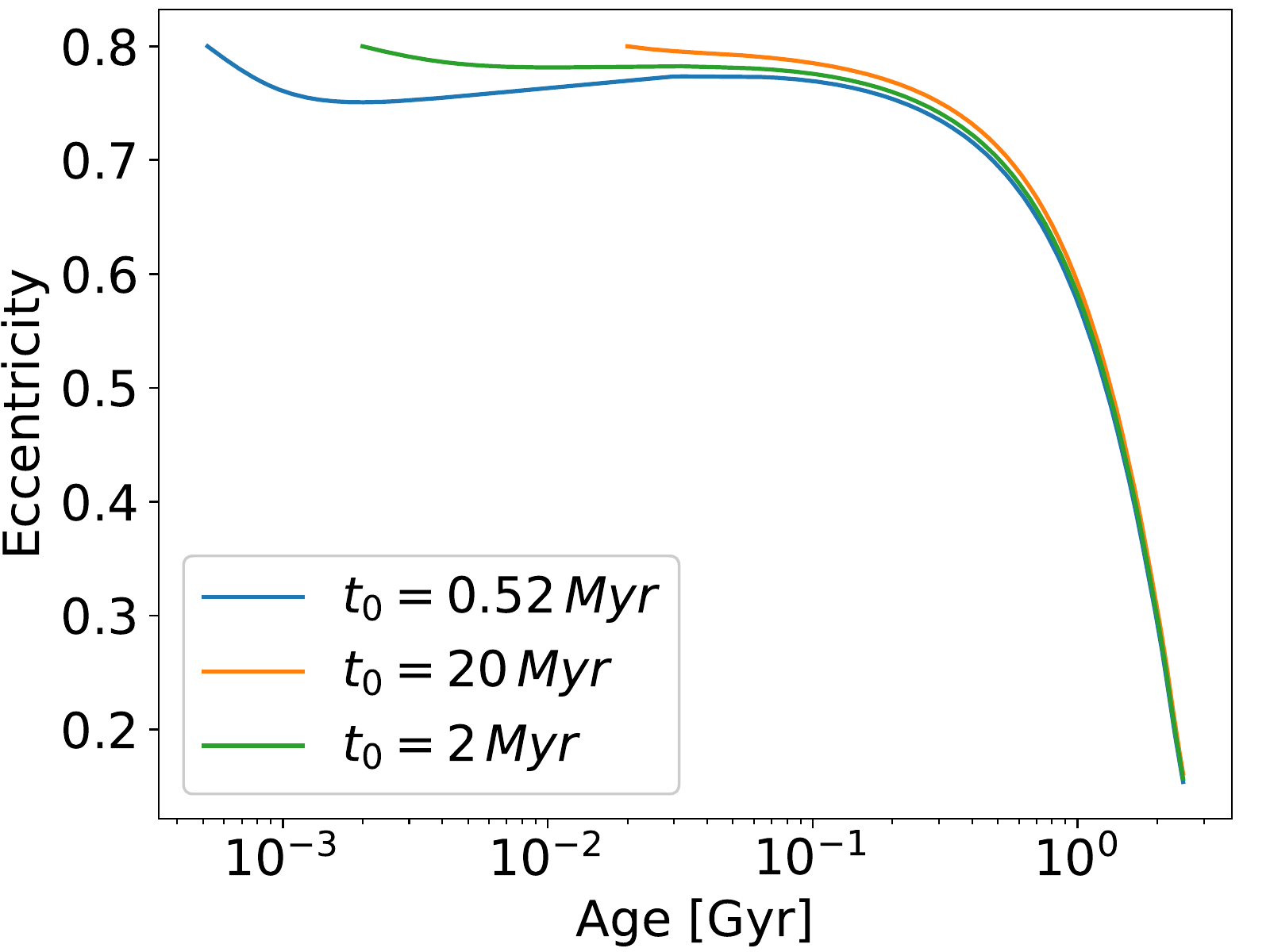}
    \includegraphics[width=\columnwidth]{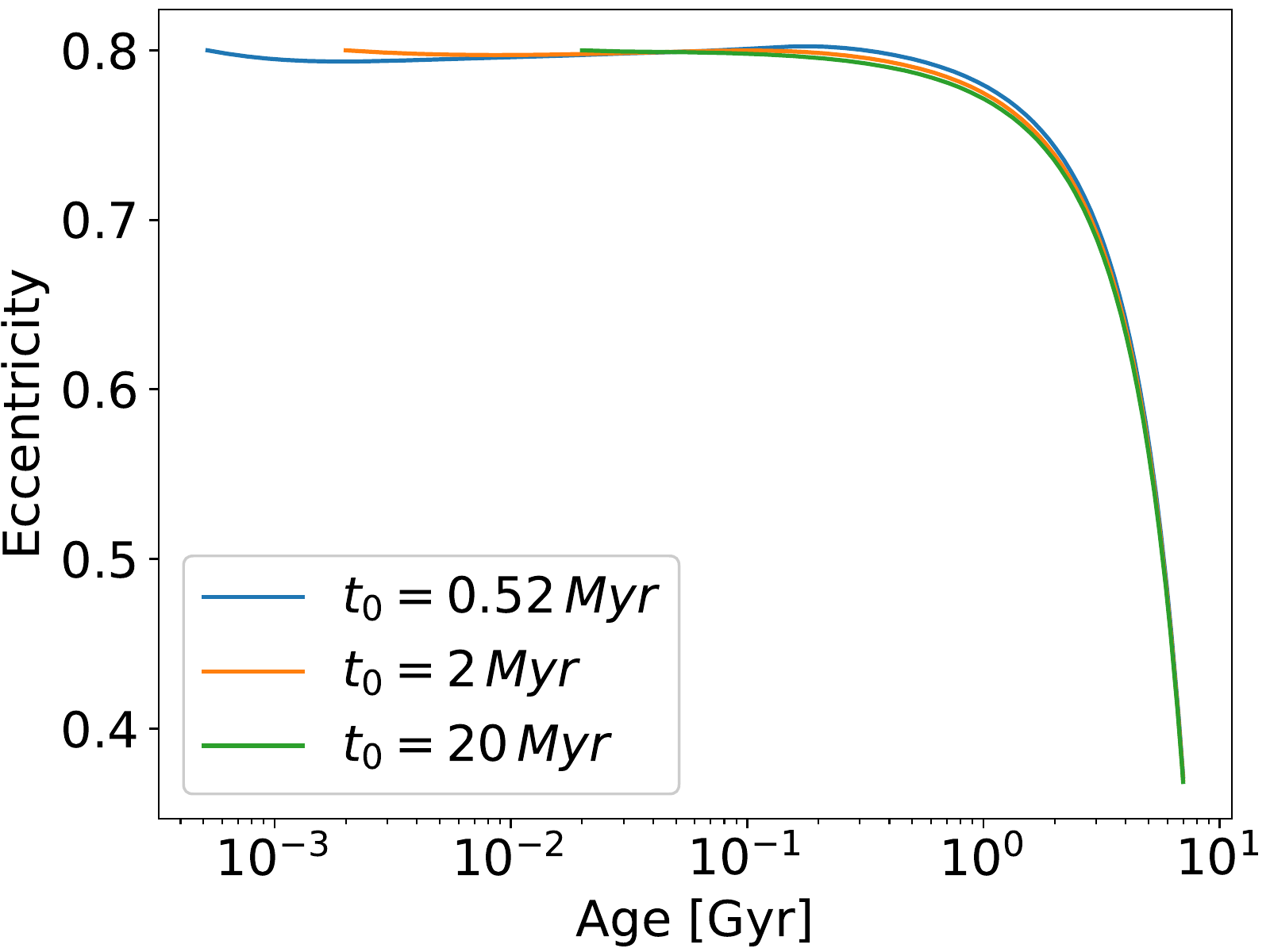}
    \caption{
        The negligible effect of the age at which binary evolution starts for
        a binary with two $1\,M_\odot$ stars with parameters suitable for
        NGC\,6819 (left) and NGC\,188 (right).
    }
    \label{fig:starting_age_effect_old}
\end{figure*}

In Sec. \ref{sec:orbital_evolution} we outlined that we start calculating tidal
evolution from an age of $t_0=20\,Myr$ for NGC 6819 and NGC 188 and $t_0=2\,Myr$
for M 35. The choice of starting age would justifiably raise concerns with
expert readers who are familiar with findings in the literature that pre-MS
evolution appears to dominate tidal circularization for the first few hundred
Myrs \citep[c.f.][]{Meibom_Mathieu_05}. In this appendix we demonstrate that
this choice did not significantly alter the final results.

For the older clusters, \citet{Meibom_Mathieu_05,Milliman_et_al_14} show that
the tidal evolution is dominated by the main sequence. As a result, the exact
details of what happens during the first few tens of Myrs are largely irrelevant
for establishing the eccentricity envelope at Gyr ages. More precisely, for
systems with long present day orbital period, tides hardly affect the orbit both
on the MS and during PMS. At short orbital periods tides would have circularized
the system by our starting age of 20\,Myrs, but since MS circularization is even
more important for theses systems, even omitting that initial phase, the orbit
is circularized by the time we reach the age of NGC\,6819 and NGC\,188. The only
possible concern then is with systems that are significantly, but not fully,
circularized by the corresponding age of the cluster being analyzed.

Fig.  \ref{fig:starting_age_effect_old} compares the eccentricity evolutions of
a fiducial system consisting of two identical solar mass stars, assuming a
constant tidal quality factor ($\log_{10}Q_\star'=5.8$), starting at ages of
$0.5\,Myr$, $2\,Myrs$, and $20\,Myrs$ respectively with an initial eccentricity
of 0.8 and final age and orbital period corresponding to the age and tidal
circularization period \citep[per][]{Meibom_Mathieu_05} of NGC\,6818 (left) and
NGC\,188 (right). In the left panel, the system is evolved to an age of
2.5\,Gyrs with a final orbital period of 8\,d, and in the right panel the final
age is 7\,Gyrs and the final orbital period is 14.5\,d. In all cases, the
evolution assumed $\log_{10}Q_\star'=5.8$ (the value found in this study). As
can be seen, there is virtually no difference between the three assumed starting
ages. This can be understood by the fact that tides do not only circularize, but
also shrink the orbit over time. As a result, a system that is significantly
circularized must have started with a much longer orbital period than it has
today. Thus, the effect of tides accelerates over time, and must have been much
smaller in the past. In other words, if MS tidal evolution dominates over the
PMS, and the system is only partially circularized, it must be that initially
tides were just barely effective in changing the system (approximately flat
portion of the curves in Fig.  \ref{fig:starting_age_effect_old}). Note that the
slight increase in eccentricity seen in the figure corresponds to the star
spinning super-synchronously for a short period of time.  This occurs regardless
of the initial spin period chosen for the stars, since as soon as tides become
important they deliver sufficient angular momentum to the stars that, as they
shrinks toward the zero-age main sequence, they end up spinning faster than
the orbit.

Note that if the tidal dissipation is a sufficiently strong function of age, the
PMS circularization may dominate over MS circularization regardless of the
binary age. In this case, the age at which tidal evolution turns on will be
important. However, that contradicts the \citet{Meibom_Mathieu_05} and
\citet{Milliman_et_al_14} observation that the circularization period gradually
increases over the MS. Furthermore, since our prescription for $Q_\star'$ (Eq.
\ref{eq:Q_prescription}) does not incorporate an explicit age dependence,
repeating the analysis with a different starting age will not change our
results.

\begin{figure}
    \includegraphics[width=\columnwidth]{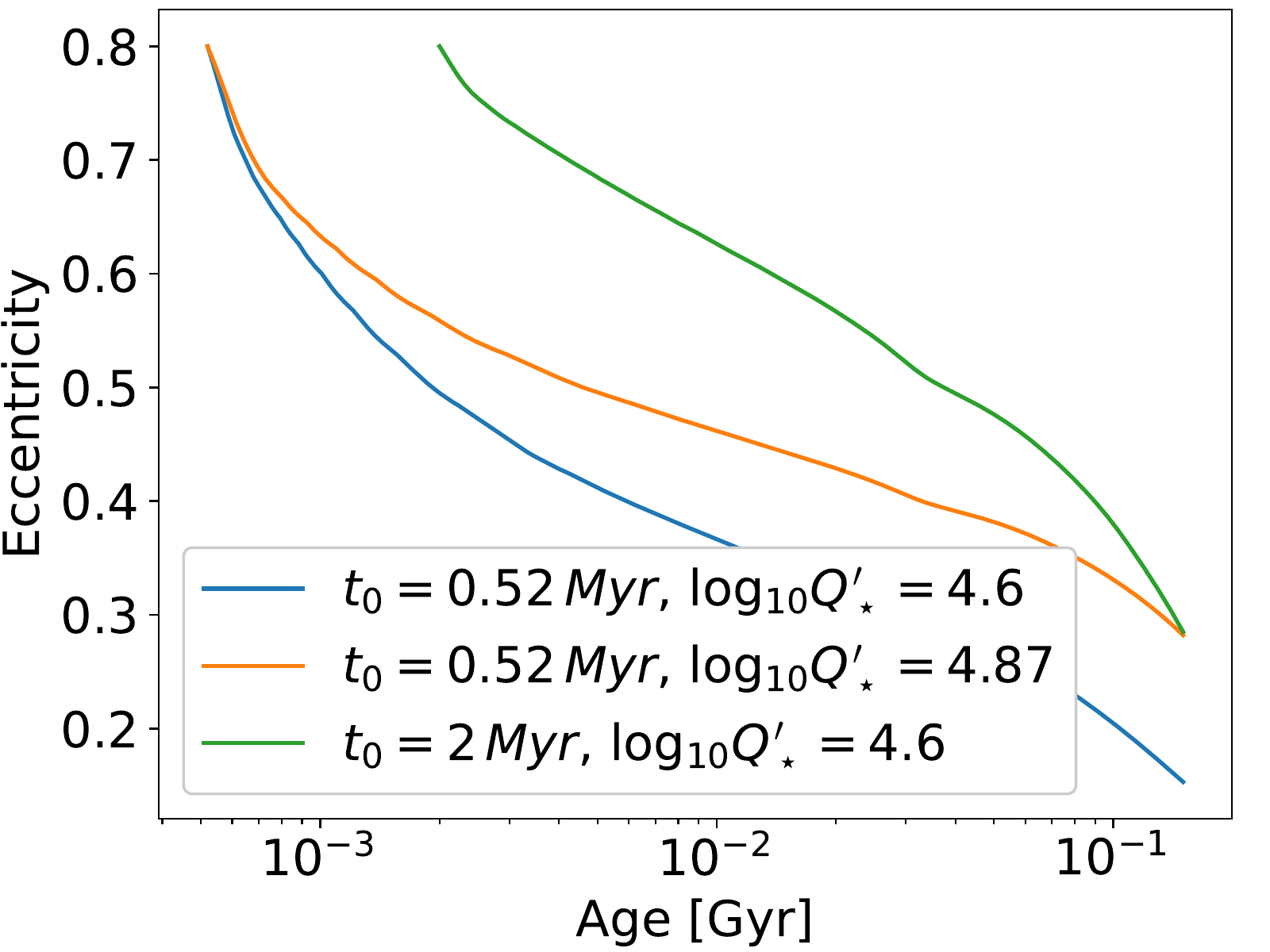}
    \caption{
            The effect of the age at which binary evolution starts for a binary
            with two $1\,M_\odot$ stars with parameters suitable for M\,35.
    }
    \label{fig:starting_age_effect_M35}
\end{figure}

The situation is a little less straightforward for M\,35 than for the older
clusters. The \citet{Meibom_Mathieu_05} circularization period for this cluster
is consistent with that of all younger clusters. This is usually interpreted to
imply that the circularization for such young systems is dominated by the
evolution at young ages when the star is large and hence subject to much
stronger tides.  This is why we use a much smaller starting age (2\,Myrs) when
constraining the dissipation in the 150\,Myr old M\,35. This was chosen as a
compromise.  At much younger ages, the rapid evolution of the radius and
internal structure of Sun-like stars caused numerical instabilities in the
calculated orbital evolution for some systems. At the same time, the effect on
the inferred tidal dissipation of starting at 2\,Myrs versus a much younger age
is sufficiently small to not compromise the results. Fig.
\ref{fig:starting_age_effect_M35} compares the eccentricity evolutions of a
binary consisting of two $1\,M_\odot$ stars started with an initial eccentricity
of 0.8 and initial orbital period such that at an age of 150\,Myrs the system
evolves to have an orbital period of 10.2\,days \citep[the tidal circularization
period][report for M\,35]{Meibom_Mathieu_05}. For the top and bottom curves,
evolution was calculated assuming $\log_{10}Q_\star'=4.6$ (the upper limit found
here).  Clearly, in this case, changing the starting age has a non-negligible
effect on the final eccentricity.  However, the effect on the inferred value of
$\log_{10}Q_\star'$ is quite limited. The middle curve shows the evolution
calculated with the smaller starting age, but with $\log_{10}Q_\star'=4.87$,
which reproduces the same final eccentricity as the evolution started at
$2\,Myrs$. As discussed above, choosing a system that is significantly, but not
completely, circularized is the worst-case scenario for the effect of the
starting age. As a result, we can conclude that pushing the starting age to
$2\,Myr$ can at most change the inferred value of $\log_{10}Q_\star'$ by no more
than 0.3\,dex.

\section{Effect of Initial Stellar Spin}
\label{sec:disk_period_effect}

\begin{figure}
    \includegraphics[width=\columnwidth]{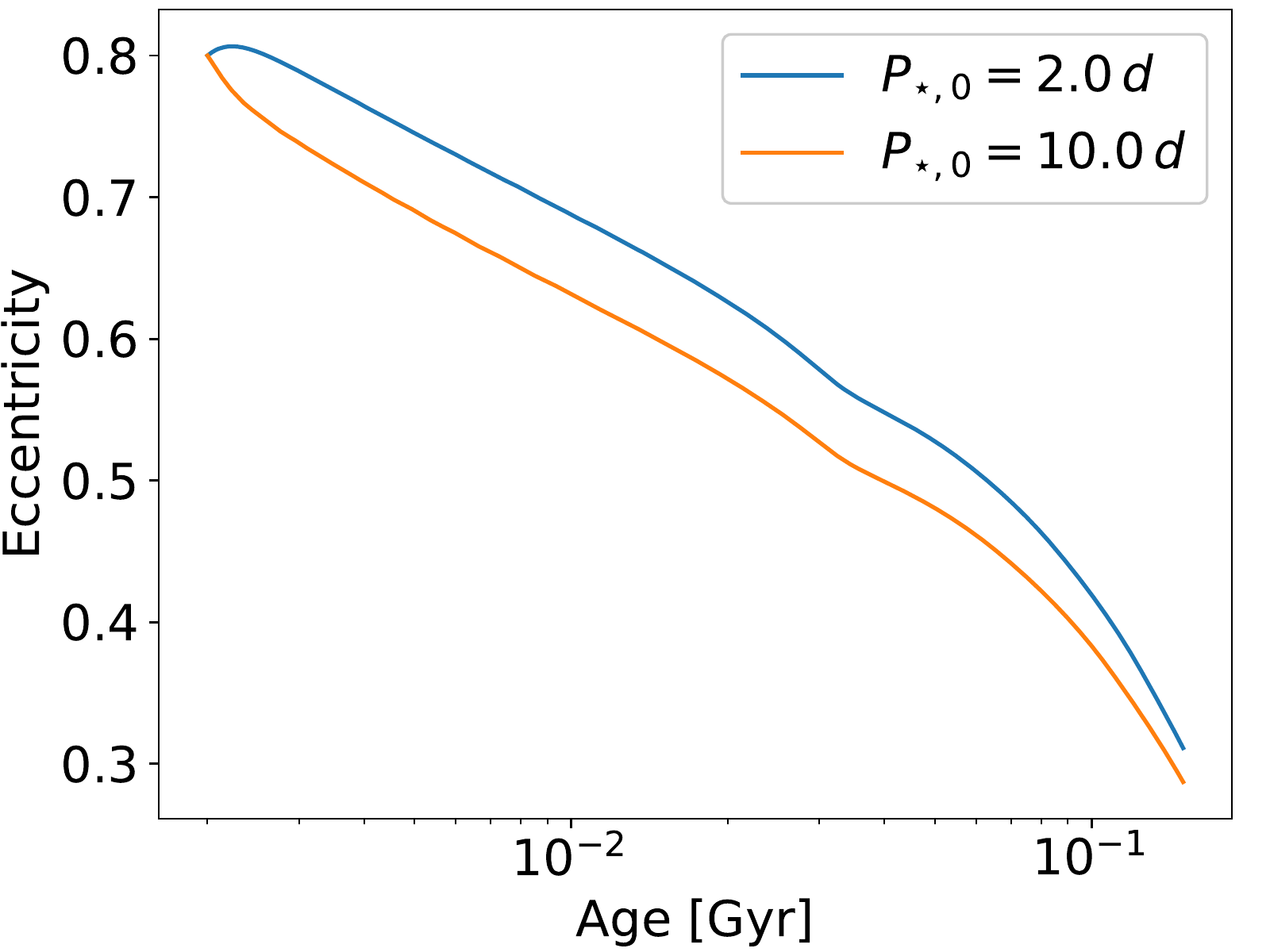}
    \caption{
            The effect of the initial stellar spin for the binary evolution of
            two $1\,M_\odot$ stars with parameters suitable for M\,35.
    }
    \label{fig:initial_spin_effect_M35}
\end{figure}

\begin{figure}
    \includegraphics[width=\columnwidth]{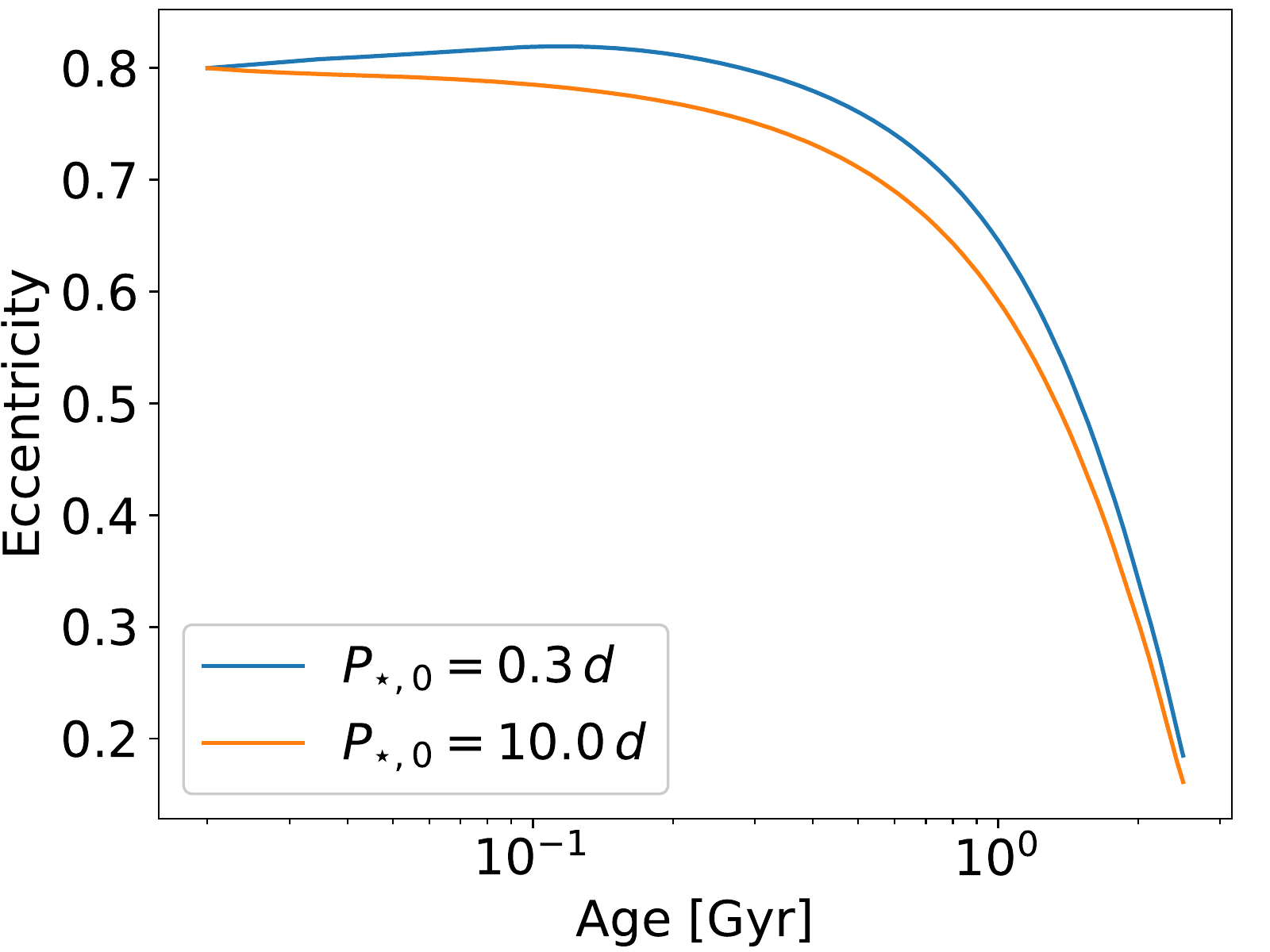}
    \caption{
            The effect of the initial stellar spin for the binary evolution of
            two $1\,M_\odot$ stars with parameters suitable for NGC\,6819.
    }
    \label{fig:initial_spin_effect_NGC6819}
\end{figure}

\begin{figure}
    \includegraphics[width=\columnwidth]{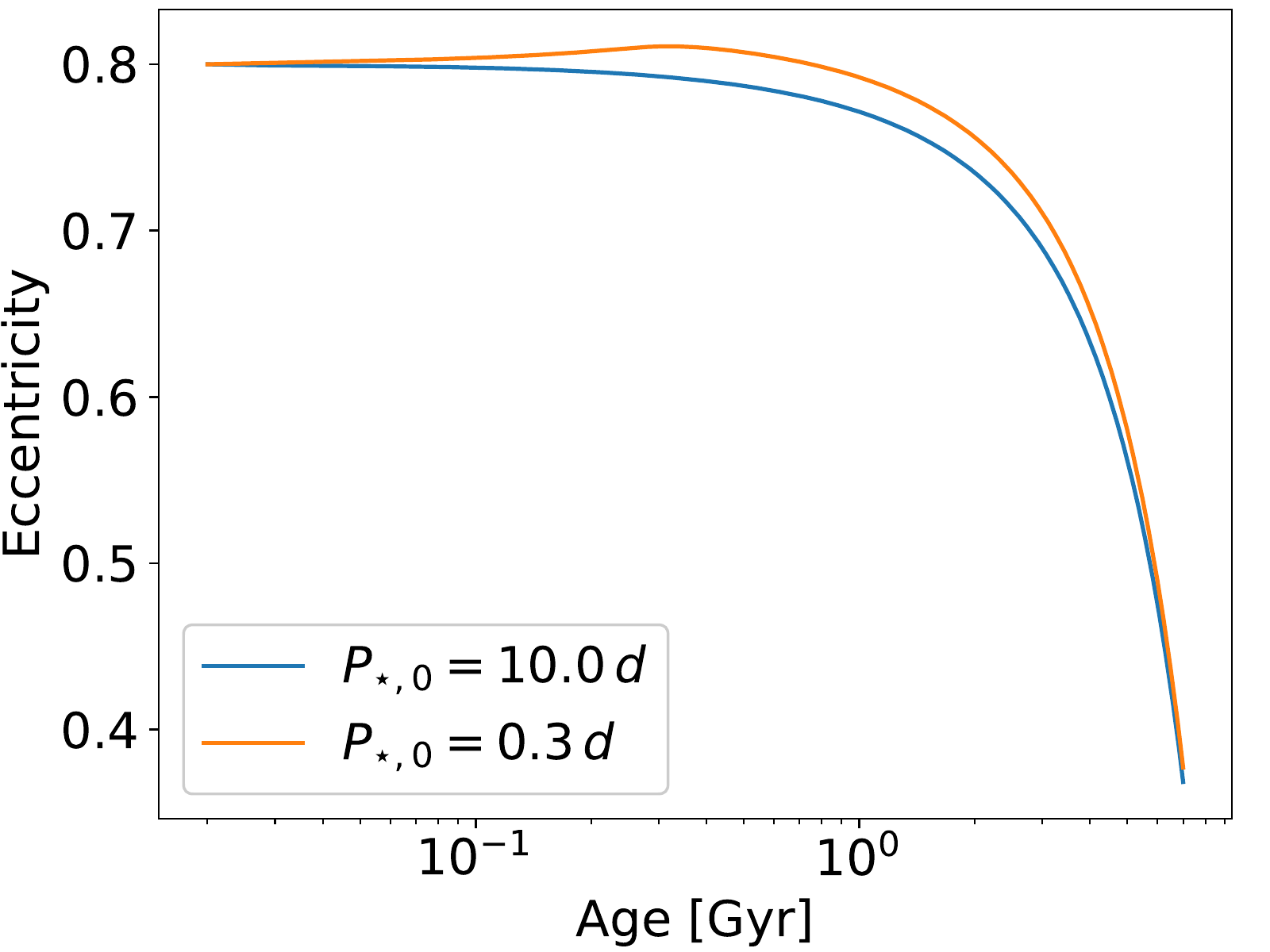}
    \caption{
            The effect of the initial stellar spin for the binary evolution of
            two $1\,M_\odot$ stars with parameters suitable for NGC\,188.
    }
    \label{fig:initial_spin_effect_NGC188}
\end{figure}

Another reasonable concern with assumptions made in our modeling has to do with
the initial spin assumed for the stars. In particular, we assume binary
evolution always starts with both stars having a surface spin period of 10\,d.
Observations of young stars show a range of initial rotation periods. In
particular, \citet{Gallet_Bouvier_15} report spin periods ranging from about
2\,d to about 8\,d for single stars with ages of a few Myr (relevant for our
M\,35 simulations). As stars shrink toward the main sequence from these initial
spins, the range shifts to 0.3 --- 7\,d for ages of 20\,Myrs (the assumed
starting age for NGC\,6819 and NGC\,188 in this analysis). In Fig.
\ref{fig:initial_spin_effect_M35}, \ref{fig:initial_spin_effect_NGC6819}, and
\ref{fig:initial_spin_effect_NGC188} we show that the assumed initial spin has
only a negligible effect on the eccentricity evolution, as long as it remains
within the observed ranges. This also makes intuitive sense, because stellar
spin accounts for only a very small fraction of the total angular momentum in
the system, so exchanging part of that with the orbit can only have a small
effect on the final eccentricity. The impact of different assumed values for the
initial spin are further suppressed by the fact that faster spinning stars lose
angular momentum to magnetic winds faster \citep{Gallet_Bouvier_15}.

The effect of the initial spin may be somewhat larger than in Fig.
\ref{fig:initial_spin_effect_M35}, \ref{fig:initial_spin_effect_NGC6819}, and
\ref{fig:initial_spin_effect_NGC188} if the tidal dissipation were to increase
sufficiently quickly with increasing stellar spin. However, even in that case,
the effect will be limited as the orbital angular momentum far exceeds the spin
angular momentum of the individual stars even for the fastest observed initial
spins. Since the parametrization used in this analysis (Eq.
\ref{eq:Q_prescription}) does not include explicit spin dependence, our results
would not significantly change if we were to allow the initial spin to vary.

%%%%%%%%%%%%%%%%%%%%%%%%%%%%%%%%%%%%%%%%%%%%%%%%%%

% Don't change these lines
\bsp	% typesetting comment
\label{lastpage}
\end{document}